\documentclass[journal,comsoc]{IEEEtran}

\usepackage{amsmath,amsfonts}
\usepackage{algorithmic}
\usepackage{array}
\usepackage{textcomp}
\usepackage{stfloats}
\usepackage{url}
\usepackage{verbatim}
\usepackage{graphicx}
\def\BibTeX{{\rm B\kern-.05em{\sc i\kern-.025em b}\kern-.08em
    T\kern-.1667em\lower.7ex\hbox{E}\kern-.125emX}}
\usepackage{balance}

\usepackage{enumitem}
\usepackage[nolist]{acronym}
\usepackage{scalerel}
\usepackage{tikz}
\usetikzlibrary{calc}
\usepackage{pgfplots}
\usepgfplotslibrary{groupplots}

\usepackage{float}
\pgfplotsset{compat=newest}
\usepackage{nicefrac}
\usepackage{amsmath, amsbsy, amssymb, amsthm}
\usepackage{tikzscale}
\usetikzlibrary{arrows}
\usepackage{color}
\usepackage{colortbl}
\usepackage{epigraph}
\usepackage{circuitikz}
\usepackage{multirow}
\usepackage{booktabs}
\usepackage[plainruled]{algorithm2e}
\usepackage[group-separator={,}]{siunitx}
\definecolor{darkgreen}{rgb}{0.125,0.5,0.169}
\usetikzlibrary{shapes,arrows}
\usetikzlibrary{positioning}
\usetikzlibrary{patterns}
\usetikzlibrary{decorations.pathreplacing}
\usetikzlibrary{trees}
\usetikzlibrary{svg.path}
\usepackage{marvosym}
\usepackage{bbm}
\usepackage{mathtools}
\usepackage{xfrac}
\usepackage{booktabs}
\usepackage{array}

\definecolor{orcidlogocol}{HTML}{A6CE39}
\tikzset{
  orcidlogo/.pic={
    \fill[orcidlogocol] svg{M256,128c0,70.7-57.3,128-128,128C57.3,256,0,198.7,0,128C0,57.3,57.3,0,128,0C198.7,0,256,57.3,256,128z};
    \fill[white] svg{M86.3,186.2H70.9V79.1h15.4v48.4V186.2z}
                 svg{M108.9,79.1h41.6c39.6,0,57,28.3,57,53.6c0,27.5-21.5,53.6-56.8,53.6h-41.8V79.1z M124.3,172.4h24.5c34.9,0,42.9-26.5,42.9-39.7c0-21.5-13.7-39.7-43.7-39.7h-23.7V172.4z}
                 svg{M88.7,56.8c0,5.5-4.5,10.1-10.1,10.1c-5.6,0-10.1-4.6-10.1-10.1c0-5.6,4.5-10.1,10.1-10.1C84.2,46.7,88.7,51.3,88.7,56.8z};
  }
}

\newcommand\orcidicon[1]{\href{https://orcid.org/#1}{\mbox{\scalerel*{
\begin{tikzpicture}[yscale=-1,transform shape]
\pic{orcidlogo};
\end{tikzpicture}
}{|}}}}

\usepackage{hyperref}

\makeatletter
\let\MYcaption\@makecaption
\makeatother

\usepackage[font=footnotesize]{subcaption}

\makeatletter
\let\@makecaption\MYcaption
\makeatother

\tikzset{>=latex}

\begin{acronym}
 \acro{R2M}{raw 2\textsuperscript{nd} moment}
 \acro{CSI}{channel state information}
 \acro{UE}{user equipment}
 \acro{EM}{electromagnetic}
 \acro{UL}{uplink}
 \acro{BS}{base station}
 \acro{TDD}{time division duplex}
 \acro{FDD}{frequency division duplex}
 \acro{ECC}{error-correcting code}
 \acro{MLD}{maximum likelihood decoding}
 \acro{HDD}{hard decision decoding}
 \acro{IF}{intermediate frequency}
 \acro{RF}{radio frequency}
 \acro{SDD}{soft decision decoding}
 \acro{NND}{neural network decoding}
 \acro{CNN}{convolutional neural network}
 \acro{ML}{maximum likelihood}
 \acro{GPU}{graphical processing unit}
 \acro{BP}{belief propagation}
 \acro{LTE}{Long Term Evolution}
 \acro{BER}{bit error rate}
 \acro{SNR}{signal-to-noise-ratio}
 \acro{ReLU}{rectified linear unit}
 \acro{BPSK}{binary phase shift keying}
 \acro{QPSK}{quadrature phase shift keying}
 \acro{AWGN}{additive white Gaussian noise}
 \acro{MSE}{mean squared error}
 \acro{LLR}{log-likelihood ratio}
 \acro{MAP}{maximum a posteriori}
 \acro{NVE}{normalized validation error}
 \acro{BCE}{binary cross-entropy}
 \acro{CE}{cross-entropy}
 \acro{BLER}{block error rate}
 \acro{SQR}{signal-to-quantisation-noise-ratio}
 \acro{MIMO}{multiple-input multiple-output}
 \acro{mMIMO}{massive multiple-input multiple-output}
 \acro{OFDM}{orthogonal frequency division multiplex}
 \acro{RF}{radio frequency}
 \acro{LOS}{line of sight}
 \acro{NLoS}{non-line of sight}
 \acro{NMSE}{normalized mean squared error}
 \acro{CFO}{carrier frequency offset}
 \acro{SFO}{sampling frequency offset}
 \acro{IPS}{indoor positioning system}
 \acro{TRIPS}{time-reversal IPS}
 \acro{RSSI}{received signal strength indicator}
 \acro{MIMO}{multiple-input multiple-output}
 \acro{ENoB}{effective number of bits}
 \acro{AGC}{automatic gain control}
 \acro{ADC}{analog to digital converter}
 \acro{ADCs}{analog to digital converters}
 \acro{FB}{front bandpass}
 \acro{FPGA}{field programmable gate array}
 \acro{JSDM}{Joint Spatial Division and Multiplexing}
 \acro{NN}{neural network}
 \acro{IF}{intermediate frequency}
 \acro{LoS}{line-of-sight}
 \acro{NLoS}{non-line-of-sight}
 \acro{DSP}{digital signal processing}
 \acro{AFE}{analog front end}
 \acro{SQNR}{signal-to-quantisation-noise-ratio}
 \acro{SINR}{signal-to-interference-noise-ratio}
 \acro{ENoB}{effective number of bits}
 \acro{PCB}{printed circuit board}
 \acro{EVM}{error vector mangnitude}
 \acro{CDF}{cumulative distribution function}
 \acro{MRC}{maximum ratio combining}
 \acro{MRP}{maximum ratio precoding}
 \acro{MRT}{maximum ratio transmission}
 \acro{DeepL}{deep-learning}
 \acro{DL}{downlink}
 \acro{SISO}{single-input single-output}
 \acro{SGD}{stochastic gradient descent}
 \acro{CP}{cyclic prefix}
 \acro{MISO}{Multiple Input Single Output}
 \acro{LMMSE}{linear minimum mean square error}
 \acro{ZF}{zero forcing}
 \acro{USRP}{universal software radio peripheral}
 \acro{RNN}{recurrent neural network}
 \acro{GRU}{gated recurrent unit}
 \acro{LSTM}{long short-term memory}
 \acro{NTM}{neural turing machine}
 \acro{DNC}{differentiable neural computer}
 \acro{TCN}{temporal convolutional network}
 \acro{FCL}{fully connected layer}
 \acro{MANN}{memory augmented neural network}
 \acro{RNN}{recurrent neural network}
 \acro{DNN}{deep neural network}
 \acro{FIR}{finite impulse response}
 \acro{BPTT}{back-propagation through time}
 \acro{GAN}{generative adversarial network}
 \acro{ELU}{exponential linear unit}
 \acro{tanh}{hyperbolic tangent}
 \acro{BICM}{bit-interleaved coded modulation}
 \acro{OTA}{over-the-air}
 \acro{IM}{intensity modulation}
 \acro{DD}{direct detection}
 \acro{RL}{reinforcement learning}
 \acro{SDR}{software-defined radio}
 \acro{WGAN}{Wasserstein generative adversarial network}
 \acro{BMD}{bit-metric decoding}
 \acro{BMI}{bit-wise mutual information}
 \acro{LDPC}{low-density parity-check}
 \acro{IDD}{iterative demapping and decoding}
 \acro{JSD}{Jensen-Shannon divergence}
 \acro{MMSE}{minimum mean square error}
 \acro{FFT}{fast Fourier transform}
 \acro{DFT}{discrete Fourier transform}
 \acro{IFFT}{inverse fast Fourier transform}
 \acro{QAM}{quadrature amplitude modulation}
 \acro{EMD}{earth mover's distance}
 \acro{TDL}{tapped delay line}
 \acro{KL}{Kullback-Leibler}
 \acro{PRACH}{physical random access channel}
 \acro{URLLC}{ultra-reliable low-latency communication}
 \acro{ANOMA}{asynchronous non-orthogonal multiple access}
 \acro{FEC}{forward error correction}
 \acro{PAPR}{peak-to-average power ratio}
 \acro{APP}{a posteriori probability}
 \acro{COTS}{commercial off-the-shelf}
 \acro{PLL}{phase locked loop}
 \acro{STO}{sampling time offset}
 \acro{SFO}{sampling frequency offset}
 \acro{CFO}{carrier frequency offset}
 \acro{CPO}{carrier phase offset}
 \acro{CSI}{channel state information}
 \acro{GNSS}{global navigation satellite system}
 \acro{ELAA}{extremely large aperture array}
 \acro{UE}{user equipment}
 \acro{DICHASUS}{\underline{Di}stributed \underline{Cha}nnel \underline{S}ounder by \underline{U}niversity of \underline{S}tuttgart}
 \acro{JCaS}{Joint Communication and Sensing}
 \acro{CT}{continuity}
 \acro{TW}{trustworthiness}
 \acro{KS}{Kruskal's stress}
 \acro{PCA}{principal component analysis}
 \acro{AoA}{angle of arrival}
 \acro{RD}{Rajski's distance}
 \acro{ADP}{angle-delay profile}
 \acro{MPC}{multipath component}
 \acro{CIR}{channel impulse response}
 \acro{MAE}{mean absolute error}
 \acro{MDS}{multidimensional scaling}
 \acro{t-SNE}{t-distributed stochastic neighbor embedding}
 \acro{SM}{Sammon's mapping}
 \acro{CIRA}{channel impulse response amplitude}
 \acro{CS}{cosine similarity}
 \acro{FCF}{forward charting function}
 \acro{CMD}{correlation matrix distance}

\end{acronym}

\definecolor{mittelblau}{RGB}{0, 126, 198}
\definecolor{violettblau}{cmyk}{0.9, 0.6, 0, 0}
\definecolor{rot}{RGB}{238, 28 35}
\definecolor{apfelgruen}{RGB}{140, 198, 62}
\definecolor{gelb}{RGB}{1, 221, 0}
\definecolor{orange}{RGB}{244, 111, 33}
\definecolor{pink}{RGB}{237, 0, 140}
\definecolor{lila}{RGB}{128, 10, 145}
\definecolor{hellgrau}{RGB}{224, 224, 224}
\definecolor{mittelgrau}{RGB}{128, 128, 128}
\definecolor{dunkelgrau}{RGB}{80,80,80}
\definecolor{anthrazit}{RGB}{19, 31, 31}

\DeclareMathOperator*{\argmin}{arg\,min}

\definecolor{TableGray}{gray}{0.9}

\begin{document}

\title{Angle-Delay Profile-Based and Timestamp-Aided Dissimilarity Metrics for Channel Charting}

\author{Phillip Stephan\textsuperscript{\orcidicon{0009-0007-4036-668X}}, \IEEEmembership{Member, IEEE}, Florian Euchner\textsuperscript{\orcidicon{0000-0002-8090-1188}}, \IEEEmembership{Member, IEEE}, Stephan ten Brink\textsuperscript{\orcidicon{0000-0003-1502-2571}}, \IEEEmembership{Fellow, IEEE}
\thanks{This work is supported by the German Federal Ministry of Education and Research (BMBF) within the projects Open6GHub (grant no. 16KISK019) and KOMSENS-6G (grant no. 16KISK113).}}

\markboth{IEEE Transactions on Communications}%
{Submitted paper}

\maketitle

\begin{abstract}
Channel charting is a self-supervised learning technique whose objective is to reconstruct a map of the radio environment, called channel chart, by taking advantage of similarity relationships in high-dimensional channel state information.
We provide an overview of processing steps and evaluation methods for channel charting and propose a novel dissimilarity metric that takes into account angular-domain information as well as a novel deep learning-based metric.
Furthermore, we suggest a method to fuse dissimilarity metrics such that both the time at which channels were measured as well as similarities in channel state information can be taken into consideration while learning a channel chart.
By applying both classical and deep learning-based manifold learning to a dataset containing sub-6$\,$GHz distributed massive MIMO channel measurements, we show that our metrics outperform previously proposed dissimilarity measures.
The results indicate that the new metrics improve channel charting performance, even under non-line-of-sight conditions.
\end{abstract}


\section{Introduction}
\IEEEPARstart{S}{patial} multiplexing techniques like \ac{mMIMO} are indispensable for future wide-area wireless telecommunication standards to fulfill the ever-increasing throughput demands with the limited available sub-$6\,\mathrm{GHz}$ \ac{EM} spectrum. 
The large number of \ac{BS} antennas required for \ac{mMIMO} also enable observation of the channel with high spatial resolution, facilitating additional applications including \ac{JCaS} and localization services.
These applications make use of \ac{CSI}, which has to be collected at the \ac{BS} for communication purposes, regardless.
Classical source localization techniques such as, e.g., triangulation or multilateration, assume a channel model with \ac{LoS} conditions.
They tend to be inaccurate if strong multipath components disturb the direct \ac{LoS} component and fail entirely under \ac{NLoS} conditions.
\ac{CSI} fingerprinting-based localization methods, on the other hand, do not make any model assumptions and have been applied successfully, even in more complex environments \cite{savic2015fingerprinting} \cite{vieira2017deep} \cite{cc_features_ferrand}.
However, these machine learning methods require accurate ground truth position labels, which are generally unavailable in practical systems.
Channel charting, on the other hand, learns a usually two- or three-dimensional map of the radio environment, the so-called channel chart, in a self-supervised manner \cite{studer_cc}, i.e., without relying on position labels.
While most of the early work on channel charting concentrated on reconstructing local geometrical features of the radio environment \cite{studer_cc}, later work increasingly placed focus on also learning its global topology \cite{siamese_cc} \cite{n2dcc}.
The environment reconstruction is expressed as a manifold learning problem, relying on the assumption that in a sufficiently static environment, measured \ac{CSI} data is mainly a function of the \ac{UE} position and therefore lies on a low-dimensional manifold in a high-dimensional \ac{CSI} feature space.
Some applications of channel charting, like user localization, require global spatial consistency.
For others, such as radio resource management tasks like handover prediction \cite{kazemi_cc_snr_prediction} and beam tracking \cite{kazemi_cc_beam_tracking}, the accurate representation of local spatial relationships is sufficient.
In this work, we aim at an accurate reconstruction of both local geometry and global topology of the environment, with a focus on comparing different novel and established dissimilarity metrics.

\subsection{Related Work}

In their seminal work on channel charting \cite{studer_cc}, C. Studer et al. suggest multiple manifold learning methods for obtaining a low-dimensional representation of high-dimensional \ac{CSI}: \ac{PCA} \cite{pca_hotelling}, \ac{SM} \cite{sammon_mapping} and a deep learning approach using an autoencoder \cite{huang2019improving}.
In contrast to the other two methods, \ac{SM} does not directly process features derived from \ac{CSI}, but relies on a matrix of pairwise dissmilarities between any two datapoints, forming the basis of what we call \emph{dissmilarity metric-based} channel charting, which is the focus of this work.
Their paper also coins the term \emph{\ac{FCF}}, denoting the mapping from \ac{CSI} feature space to low-dimensional channel chart, and establishes a distinction between two types of \acp{FCF}, namely \emph{parametric} and \emph{nonparametric} ones.
Parametric \acp{FCF} (like aforementioned autoencoder), once learned, can map previously unseen datapoints onto an existing channel chart given the additional sample's \ac{CSI} features.
Nonparametric \acp{FCF} (like \ac{SM}), on the other hand, cannot trivially predict channel chart positions for points outside their training set.
This makes makes parametric (i.e., usually deep learning-based) methods the preferred way to perform channel charting for many practical applications.
Parametric dissimilarity metric-based channel charting was only enabled later by the proposals to employ a Siamese neural network structure \cite{siamese_cc} to parameterize \ac{SM} or to use a triplet neural network based on triplet loss \cite{triplet_cc}.

Besides the choice of manifold learning technique, the definition of the dissimilarity metric is crucial for the success of dissimilarity metric-based channel charting.
The first paper on channel charting \cite{studer_cc} proposes a dissimilarity metric based on the Frobenius distance between special matrices derived from \ac{CSI}.
Later papers propose different metrics, some of them geometrically motivated and mathematically derived, others devised in a more ad-hoc manner based on empirical results.
Notable \ac{CSI} dissimilarity metric proposals include a metric derived from superresolution angular-domain information \cite{multipoint_cc}, a \ac{CMD}-based metric \cite{cc_euclidean_distance_matrix}, a \ac{CS}-based metric \cite{magoarou_efficient_cc}, a metric making use of time domain \ac{CSI} that we call \ac{CIRA} \cite{fraunhofer_cc} and, more recently, an \ac{ADP}-based metric \cite{al_tous_cc_adpp}.
In addition, \cite{fraunhofer_cc} highlights that it is often sufficient to be able to compute dissimilarities that are indicative of local distances and that global distances, which are then called \emph{geodesic distances}, can be computed as a combination of many local distances.
In \cite{triplet_cc} and in our previous work \cite{euchner_cc}, \ac{CSI}-based dissimilarities were avoided entirely by using side information, namely the absolute difference in datapoint timestamps, as a metric.

While most research in channel charting is carried out on simulated channels \cite{studer_cc}\cite{siamese_cc}\cite{multipoint_cc}\cite{cc_euclidean_distance_matrix}\cite{magoarou_efficient_cc}\cite{al_tous_cc_adpp}, some authors compute channel charts for real-world measurement datasets \cite{triplet_cc}\cite{fraunhofer_cc}\cite{euchner_cc}.
Some previous research \cite{multipoint_cc}\cite{ponnada2019out}\cite{geng2020multipoint}\cite{pihlajasalo2020absolute} specifically considers the case of \emph{multipoint} channel charting, where \ac{CSI} is available from multiple distributed \acp{BS}.

\subsection{Contributions and Outline}
After listing some limitations of our work and defining some terminology and notations, we first review our system model and present a short overview of the principles of dissimilarity metric-based channel charting in Section \ref{sec:overview}.
In Section \ref{sec:dissimilarity_metrics}, we then introduce several commonly used dissimilarity metrics and explain our own novel metrics.
Our proposals include exploiting the \ac{ADP}, learning the dissimilarity metric with a neural network and fusing information from multiple existing metrics.
Building on well-known classical and deep learning-based manifold learning methods summarized in Sections \ref{sec:classical-manifold-learning} and \ref{sec:dl-manifold-learning}, respectively, and using the customary channel charting evaluation metrics given in Section \ref{sec:evaluation-metrics}, we compare all proposed dissimilarity metrics on a real-world dataset in Section \ref{sec:results}.
Additionally, we demonstrate that, with our novel metrics, we can get good performance even if a large portion of the dataset experiences \ac{NLoS} propagation conditions.
Finally, in Section \ref{sec:conclusion}, we draw conclusions from the observed results and suggest some open questions for future work.

\subsection{Limitations}
To be able to provide an extensive performance comparison of many different dissimilarity metrics and manifold learning techniques, we had to constrain our evaluations to just one measurement dataset.
Out of the dissimilarity metrics from literature, we had to limit ourselves to those that perform best on our dataset.
Like most current channel charting work, we assume that the environment is mostly static over the duration of the data capture.

\subsection{Terminology}
In the context of channel charting, \emph{distributed \ac{mMIMO}} can be seen as an extension of \emph{multipoint} channel charting.
Whereas the latter only assumes the availability of \ac{CSI} at multiple \acp{BS}, the former also ensures some additional synchronization between them.
In our case, we use datasets with phase synchronization across all antennas and will therefore refer to ``distributed \ac{mMIMO}'', even though many of our results also apply to multipoint channel charting.

The notion of computing \emph{dissimilarity metrics} between datapoints is common in channel charting literature.
Various terms like \emph{distance}, \emph{similarity} or \emph{dissimilarity metric} have been proposed to refer to this measure, but it is yet unclear which will prevail.
We chose to use the expression \emph{dissimilarity metric}, since it conveys the distinction between physical distances to dissimilarities in \ac{CSI} space.
Note that \emph{metric} is used in a colloquial sense here, as not all \emph{dissimilarity metrics} fulfill the requirements for a metric in the sense of a metric space.

\subsection{Notations}
We use boldface uppercase letters for matrices and tensors, and comma-separated subscripts for indexing.
For example $\mathbf H_{i, j}$ refers to the entry in the $i$-th row and $j$-th column of the matrix $\mathbf H$.
Boldface lowercase letters are reserved for vectors.

\section{Overview: Principles of Channel Charting}
\label{sec:overview}

\begin{figure}
    \centering
    \begin{tikzpicture}
    
        \node (dataset) [font={\small\sffamily}, align = center] {Collect\\CSI Dataset};
    
        \node (dm) [rectangle, right = 0.3cm of dataset, draw = green!40!black, very thick, rounded corners = 1pt, inner sep = 3pt, align = center, font={\small\sffamily}, minimum height = 1.3cm] {Compute\\Dissimilarity\\Matrix};

        \node (ldr) [rectangle, right = 0.3cm of dm, draw = blue!40!black, very thick, rounded corners = 1pt, inner sep = 3pt, align = center, font={\small\sffamily}, minimum height = 1.3cm] {Perform\\Manifold\\Learning};
        
        
        
        \node (cc) [right = 0.3cm of ldr, font={\small\sffamily}, align = center] {Channel\\Chart};
        
        \draw [-latex] (dataset) -- (dm);
        \draw [-latex] (dm) -- (ldr);
        \draw [-latex] (ldr) -- (cc);
    \end{tikzpicture}
    \caption{The two central steps in dissimilarity metric-based channel charting}
    \label{fig:cc_steps_overview}
\end{figure}
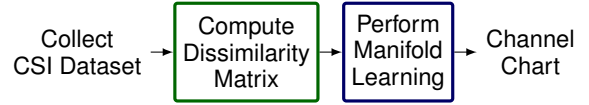

In a sufficiently static environment and neglecting hardware impairments and noise, \ac{CSI}, which is estimated at the \ac{BS} for communication purposes, is primarily determined by the location and orientation of the \ac{UE}.
This is the premise of channel charting, which assumes that \ac{CSI} in a suitable feature space lies on a single low-dimensional manifold.
The steps involved in dissimilarity metric-based channel charting are illustrated in Fig. \ref{fig:cc_steps_overview}:
First, having collected a large \ac{CSI} dataset at the \ac{BS}, a \emph{dissimilarity matrix} containing the dissimilarities between any two datapoints in the dataset is computed using a particular dissimilarity metric.
Then, manifold learning is applied to obtain the \ac{FCF}.
The success of the second step is highly dependent on the first step, namely the choice of dissimilarity metric.
Ideally, the computed dissimilarities reflect the physical distances between datapoints to a large extent.
The low-dimensional representation should minimize the discrepancy between the point-to-point distances in the channel chart and the respective distances in the dissimilarity matrix.
A suitable representation can be found with either classical or deep learning-based manifold learning methods.
The latter implement parametric \acp{FCF} which can predict channel chart positions based on \ac{CSI} input features.
These input features are computed from the raw estimated \ac{CSI} in a \emph{feature engineering} step.

\subsection{System and Data Model}
In the following, we describe an abstract system model that we will employ to define dissimilarity metrics and manifold learning techniques.
Refer to Section \ref{sec:dichasus} for information on the particular dataset and system used for our analyses.

We consider a \ac{mMIMO} \ac{BS} consisting of $B$ distributed antenna arrays equipped with $M$ antennas each, and a \ac{UE} with a single antenna.
A dataset consisting of $L$ datapoints measured at discrete time instances $l = 1, \ldots, L$ is collected, where one datapoint consists of the following parameters:
\begin{itemize}
    \item Channel coefficients between the \ac{UE} and all $B\times M$ \ac{BS} antennas for $N_\mathrm{sub}$ \ac{OFDM} subcarriers, collected in the frequency-domain \ac{CSI} tensor $\mathbf{H}^{(l)} \in \mathbb{C}^{B \times M \times N_\mathrm{sub}}$.
    We assume that $N_\mathrm{sub}$ is even and that the subcarrier channel coefficients in $\mathbf H$ are ordered such that the coefficients $\mathbf H^{(l)}_{b, m, 1}$ correspond to the subcarrier at the center of the bandwidth (customary for \ac{DFT} frequency bins).
    \item Corresponding \ac{UE} positions $\mathbf{x}^{(l)} \in \mathbb{R}^D$, measured by some reference localization system, with $D$ being the spatial dimension (we assume $D = 2$).
    This information may be used for evaluation purposes.
    \item A timestamp $t^{(l)} \in \mathbb{R}$, measured in seconds, indicating the exact (millisecond-precision) time the \ac{CSI} and position measurements were taken.
\end{itemize}

Hence, the complete dataset may be written as a set of 3-tuples $\left(\mathbf{H}^{(l)}, \mathbf x^{(l)}, t^{(l)}\right)$ of cardinality $L$:
\[
    \text{Dataset:} ~ \mathcal S = \left\{ (\mathbf{H}^{(l)}, \mathbf x^{(l)}, t^{(l)}) \right\}_{l = 1, \ldots, L}
\]

For some processing steps, it may be beneficial to use a time-domain \ac{CSI} representation $\tilde {\mathbf H}^{(l)} \in \mathbb C^{B\times M \times N_\mathrm{sub}}$.
This representation is obtained by applying the unitary inverse discrete Fourier transform to $\mathbf{H}^{(l)}$ along the frequency axis:
\begin{equation}
    \tilde {\mathbf H}_{b,m,\tau}^{(l)} = \frac{1}{\sqrt{N_\mathrm{sub}}} \sum_{n = 1}^{N_\mathrm{sub}} \mathrm e^{2 \pi \mathrm j \cdot \nicefrac{1}{N_\mathrm{sub}} \cdot (n - 1) \cdot \left(\tau - \nicefrac{N_\mathrm{sub}}{2} - 1\right)} ~ \mathbf{H}_{b, m, n}^{(l)}
    \label{eq:dft}
\end{equation}
Note that $\tau$ in Eq. (\ref{eq:dft}) is shifted such that $\tau = 1$ corresponds to the earliest and $\tau = N_\mathrm{sub}$ corresponds to the last time tap.

\subsection{Manifold Learning and Dissimilarity Metrics}
Manifold learning finds a mapping from the \ac{CSI} tensors $\mathbf{H}^{(l)} \in \mathbb{C}^{B \times M \times N_\mathrm{sub}}$ to channel chart positions $\mathbf{z}^{(l)} \in \mathbb{R}^{D'}$, expressed by the \ac{FCF}
\[
    \mathcal C: \mathbb{C}^{B \times M \times N_\mathrm{sub}} \rightarrow \mathbb{R}^{D'},
\]
with $D'$ being the dimensionality of the channel chart.
Subsequently, we assume $D' = D = 2$ and only consider manifold learning techniques which are based on dissimilarity metrics.
A good \ac{FCF} ensures that the pairwise Euclidean distance between any two mapped positions in the channel chart $\mathbf{z}^{(i)}, \mathbf{z}^{(j)} \in \mathbb R^{D'}$ is approximately proportional to the distances between the corresponding true \ac{UE} positions $\mathbf{x}^{(i)}, \mathbf{x}^{(j)}$:
\[
    \lVert \mathbf{z}^{(i)} - \mathbf{z}^{(j)}\rVert_2 \propto \lVert \mathbf{x}^{(i)} - \mathbf{x}^{(j)}\rVert_2, \quad \forall i,j \in \{1,\ldots, L\}.
\]

In case of dissimilarity metric-based channel charting, dissimilarities computed based on a dissimilarity metric $d$ are considered while learning the \ac{FCF}.
Ideally, these dissimilarity metric should also fulfill
\begin{equation}
    d_{i, j} \propto \lVert \mathbf{x}^{(i)} - \mathbf{x}^{(j)}\rVert_2, \quad \forall i,j \in \{1,\ldots, L\},
    \label{eq:proportional_dissimilarities}
\end{equation}
with $d_{i, j}$ being the computed dissimilarity between the datapoints $i$ and $j$.
Dissimilarities are calculated based on information that is available to the \ac{BS}, such as the estimated \ac{CSI} \cite{studer_cc} \cite{fraunhofer_cc} \cite{magoarou_efficient_cc} or timestamp information \cite{triplet_cc}.
A more detailed overview of different dissimilarity metrics for channel charting is given in Section \ref{sec:dissimilarity_metrics}.
The pairwise dissimilarities between all samples in the dataset are collected in the dissimilarity matrix
\[
    \mathbf{D}_\mathrm{pw}=
    \begin{pmatrix}
        d_{1,1} & \cdots & d_{1,L}\\
        \vdots & \ddots & \vdots\\
        d_{L,1} & \cdots & d_{L,L}
    \end{pmatrix},
\]
which is generally symmetric ($d_{i, j} = d_{j, i}$) and fulfills $d_{i, j} \geq 0 \;\; \forall i, j$ and $d_{i, i} = 0 \;\; \forall i$.

\subsection{Feature Engineering}\label{sec:feature_engingeering}
For deep learning-based techniques, we propose to use \ac{CSI} features based on sample autocorrelations of the time-domain \ac{CSI} matrices $\tilde {\mathbf H}^{(l)}$.
These sample autocorrelations are computed for every pair $b_1$, $b_2$ of antenna array indices, for any pair $m_1$, $m_2$ of antenna indices and for every tap index $\tau$.
This way, the feature vector also contains sample correlations of antennas that are part of different antenna arrays.
To this end, we define a set of index tuples $\mathcal J$ as the cartesian product
\[
    \mathcal J = \left\{ 1, \ldots, B \right\}^2 \times \left\{ 1, \ldots, M \right\}^2 \times \left\{ \tau_\mathrm{min}, \ldots, \tau_\mathrm{max} \right\}.
\]

Note that only time tap indices $\tau$ from $\tau_\mathrm{min}$ to $\tau_\mathrm{max}$ are taken into account, where $1 \ll \tau_\mathrm{min} < \nicefrac{N_\mathrm{sub}}{2} < \tau_\mathrm{max} \ll N_\mathrm{sub}$, i.e., only a few \ac{CIR} taps containing the \ac{LoS} path (if it exists) as well as the first few multipath components are even considered.
The choice of $\tau_\mathrm{min}$ and $\tau_\mathrm{max}$ depends on the expected maximum delay spread.
Since \ac{OFDM} is usually operated such that the number of subcarriers $N_\mathrm{sub}$ is high enough to ensure sufficiently frequency-flat subcarriers despite the channel's frequency selectivity, the considered interval of \ac{CIR} taps $\tau_\mathrm{min}, \ldots, \tau_\mathrm{max}$ is small compared to the overall number of taps $N_\mathrm{sub}$.

For each combination of indices, we compute a sample autocorrelation and collect the autocorrelations in the vector $\mathbf c^{(l)} \in \mathbb C^{B^2 M^2 (\tau_\mathrm{max} - \tau_\mathrm{min} + 1)}$:
\[
    \mathbf c^{(l)} = \left( \tilde {\mathbf H}^{(l)}_{b_1 m_1 \tau} \left ( \tilde {\mathbf H}^{(l)}_{b_2 m_2 \tau} \right)^* \right)_{(b_1, b_2, m_1, m_2, \tau) \in \mathcal J}
\]

Since we use a neural network implementation that can only handle real-valued numbers, we provide $\mathbf c^{(l)}$ to the network in a vectorized form with real and imaginary part separated, yielding the feature vectors
\[
    \mathbf{f}^{(l)} = \begin{pmatrix}
        \mathrm{Re} \left\{ \mathbf c^{(l)} \right\} & \mathrm{Im} \left\{ \mathbf c^{(l)} \right\}
    \end{pmatrix} \in \mathbb C^{2 B^2 M^2 (\tau_\mathrm{max} - \tau_\mathrm{min} + 1)}.
\]
Intuitively, this means that the neural network is provided information about relative phases and powers between any two antennas in the system.
Obviously, our choice of feature engineering produces very large input vectors $\mathbf{f}^{(l)}$.
The feature engineering stage may be further improved, e.g., by compressing useful information in a more sparse set of features.

\newcommand\Star[3][]{%
    \path[#1] (0  :#3) -- ( 36:#2) 
        -- (72 :#3) -- (108:#2)
        -- (144:#3) -- (180:#2)
        -- (216:#3) -- (252:#2)
        -- (288:#3) -- (324:#2)--cycle;}

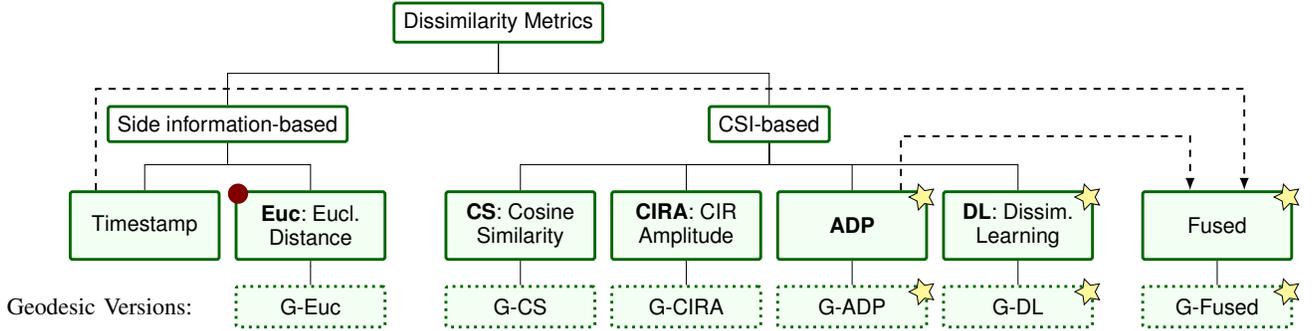
\begin{figure*}
    \centering

    \scalebox{0.9}{
        \begin{tikzpicture}[
        	metricgroup/.style={rectangle, draw = green!40!black, very thick, rounded corners = 1pt, inner sep = 4pt, align = center, font={\small\sffamily}},
        	metric/.style={rectangle, draw = green!40!black, very thick, rounded corners = 1pt, inner sep = 4pt, align = center, fill = green!5!white, minimum height = 1.0cm, minimum width = 2.2cm, font={\small\sffamily}},
        	ghostmetric/.style={rectangle, draw = green!40!black, dotted, very thick, rounded corners = 1pt, inner sep = 4pt, align = center, fill = green!5!white, minimum height = 0.6cm, minimum width = 2.2cm, font={\small\sffamily}}
        ]
        
        \node [metricgroup] at (0, 0) {Dissimilarity Metrics}
        	[edge from parent fork down, sibling distance=8cm, level distance = 1.5cm]
        	child { node [metricgroup, align = center] {Side information-based}
        		[edge from parent fork down, sibling distance=2.45cm, level distance = 1.5cm]
        		child { node (timestamp) [metric] {Timestamp}
        		    [level distance = 1.2cm]
        			child [opacity = 0, white] { node [draw = none] {} }
        		}
        		child { node (rs) [metric] {\textbf{Euc}: Eucl.\\Distance}
        		    [level distance = 1.2cm]
        			child { node (geuc) [ghostmetric] {G-Euc} }
        		}
        	}
        	child { node [metricgroup, align = center] {CSI-based}
        		[edge from parent fork down, sibling distance=2.45cm, level distance = 1.5cm]
        		child { node [metric] {\textbf{CS}: Cosine\\Similarity}
        		    [level distance = 1.2cm]
        			child { node (gcs) [ghostmetric] {G-CS} }
        		}
        		child { node [metric] {\textbf{CIRA}: CIR\\Amplitude}
        		    [level distance = 1.2cm]
        			child { node [ghostmetric] {G-CIRA} }
        		}
        		child { node (pdpcs) [metric] {\textbf{ADP}}
        		    [level distance = 1.2cm]
        			child { node (gpdpcs) [ghostmetric] {G-ADP} }
        		}
        		child { node (dl) [metric] {\textbf{DL}: Dissim.\\Learning}
        		    [level distance = 1.2cm]
        			child { node (gdl) [ghostmetric] {G-DL} }
        		}
        	};
        	
        	\node [left = 0.5cm of geuc] {Geodesic Versions:};
        	
        	\node [xshift = -0.1cm, yshift = -0.1cm] at (dl.north east) { \begin{tikzpicture} \Star[fill=yellow!50,draw]{0.2}{0.1} \end{tikzpicture} };
        	\node [xshift = -0.1cm, yshift = -0.1cm] at (gdl.north east) { \begin{tikzpicture} \Star[fill=yellow!50,draw]{0.2}{0.1} \end{tikzpicture} };
        	\node [xshift = -0.1cm, yshift = -0.1cm] at (pdpcs.north east) { \begin{tikzpicture} \Star[fill=yellow!50,draw]{0.2}{0.1} \end{tikzpicture} };
        	\node [xshift = -0.1cm, yshift = -0.1cm] at (gpdpcs.north east) { \begin{tikzpicture} \Star[fill=yellow!50,draw]{0.2}{0.1} \end{tikzpicture} };

        	\node [xshift = 0.05cm, yshift = -0.05cm] at (rs.north west) { \begin{tikzpicture} \node[circle, fill = red!50!black, inner sep = 3pt] {}; \end{tikzpicture} };
        	
        	\node (fused) [metric, right = 0.7cm of dl] {Fused};
        	\node (gfused) [ghostmetric, right = 0.7cm of gdl] {G-Fused};
        	\node [xshift = -0.1cm, yshift = -0.1cm] at (fused.north east) { \begin{tikzpicture} \Star[fill=yellow!50,draw]{0.2}{0.1} \end{tikzpicture} };
        	\node [xshift = -0.1cm, yshift = -0.1cm] at (gfused.north east) { \begin{tikzpicture} \Star[fill=yellow!50,draw]{0.2}{0.1} \end{tikzpicture} };
        	\draw (fused) -- (gfused);

        	\draw [dashed, thick, -latex] ($(timestamp.north west) + (0.4, 0)$) -- ($(timestamp.north west) + (0.4, 1.5)$) -| ($(fused.north) + (0.4, 0)$);
        	\draw [dashed, thick, -latex] ($(pdpcs.north east) + (-0.4, 0)$) -- ($(pdpcs.north east) + (-0.4, 0.8)$) -| ($(fused.north) + (-0.4, 0)$);;

        \end{tikzpicture}
    }
    \caption{Overview of dissimilarity metrics compared in this paper. Metrics marked with a star (\protect\tikz{\protect\Star[fill=yellow!50,draw]{0.15}{0.075}}) are newly introduced. Metrics marked with a red circle (\protect\tikz{\protect\node[circle, fill = red!50!black, inner sep = 2.5pt] {}}) use ground truth positioning data and are only meant as baselines, not for practical implementation of a channel charting system. Note that we are considering a metric that fuses information from both \ac{ADP} and timestamp metrics here, but, in general, any set of metrics may be fused into one.}
    \label{fig:dissimilarity_metrics_overview}
\end{figure*}

\section{Dissimilarity Metrics}
\label{sec:dissimilarity_metrics}
A variety of dissimilarity metrics that compute pseudo-distances between measured datapoints have been proposed for channel charting.
The choice of metric depends on properties of the available \ac{CSI} data (e.g., number and distribution of antennas, bandwidth, phase-coherence) and the availability of side information.
In general, dissimilarity metrics can be grouped into side information-based and \ac{CSI}-based metrics, as illustrated in Fig. \ref{fig:dissimilarity_metrics_overview}.
Side information-based methods assume that some additional information is available at the \ac{BS} in addition to wireless channel estimates.
Other dissimilarity metrics are computed solely based on measured \ac{CSI}.
We compare the suitability of different \ac{CSI}-based dissimilarity metrics from literature, namely a \ac{CS}-based metric \cite{magoarou_efficient_cc} and a \ac{CIRA}-based metric \cite{fraunhofer_cc}, to our newly developed metric, which we refer to as \ac{ADP} dissimilarity.
In addition, we describe an approach where a \ac{DNN} learns a suitable dissimilarity metric between measured \ac{CSI}, which we call dissimilarity metric learning, and we propose a method to combine information from multiple metrics.

\subsection{Euclidean Distance (Reference System)}
Under the assumption that some reference system (e.g., a \ac{GNSS}) provides ``ground truth'' position labels $\{ \mathbf x^{(l)} \}_{l = 1, \ldots, L}$, the Euclidean distance
\[
    d_{\mathrm{Euc}, i, j} = \lVert \mathbf{x}^{(i)}-\mathbf{x}^{(j)} \rVert_2
\]
between these labels can be used directly as the dissimilarity metric.
Obviously, this metric is unsuitable for most practical uses of channel charting, where only channel measurements at the \ac{BS} are available, but is nevertheless included here as a baseline for benchmarking subsequent processing steps.

\subsection{Absolute Time Difference}
\label{sec:time-distance}
As first suggested in \cite{triplet_cc}, timestamps may also reveal some information about spatial relationships between datapoints:
Since the velocity of a \ac{UE} is bounded, two channel measurements between \ac{BS} and the same \ac{UE} which are measured in quick succession are likely to be also spatially close to each other.
It is reasonable to assume that timestamps are available alongside \ac{CSI} measurements in practical systems.

The timestamp-based dissimilarity metric is defined as the absolute time difference between two measurements:
\begin{equation}
    d_{\mathrm{time},i, j} = \lvert t^{(i)}-t^{(j)} \rvert
    \label{eq:time_distance}
\end{equation}

Obviously, this dissimilarity metric fails to predict meaningful distances for large absolute time differences, but its usefulness has been demonstrated before \cite{triplet_cc} \cite{euchner_cc} and it can be especially powerful in combination with \ac{CSI}-based metrics.

\subsection{Channel Impulse Response Amplitude (CIRA)}
A dissimilarity metric based on the \ac{CIRA} was introduced in \cite{fraunhofer_cc}.
An important advantage of this metric is that phase synchronization between receivers is not required, since it only operates on \ac{CIR} amplitudes, not phases.
However, the metric assumes a setup with distributed receivers and is less suitable for setups with one single or few \ac{mMIMO} arrays.
We can apply the \ac{CIRA} metric to our time-domain \ac{CIR} tensors $\tilde {\mathbf H}^{(l)}$ as follows:
\begin{equation}
    d_{\mathrm{CIRA},i, j}=\sum\limits_{b=1}^{B}\sum\limits_{m=1}^{M}\sum\limits_{\tau=\tau_\mathrm{min}}^{\tau_\mathrm{max}} \left| \left|\tilde {\mathbf H}_{b,m,\tau}^{(i)}\right| - \left|\tilde {\mathbf H}_{b,m,\tau}^{(j)}\right|\right|
    \label{eq:cira}
\end{equation}

As previously explained in Section \ref{sec:feature_engingeering}, $\tau_\mathrm{min}$ and $\tau_\mathrm{max}$ should be chosen appropriately.

\subsection{Cosine Similarity (CS)}
The use of a \ac{CS}-based metric for channel charting was first proposed in \cite{magoarou_efficient_cc}, which assumes a narrowband system with just one \ac{mMIMO} antenna array at the \ac{BS}.
In \cite{magoarou_efficient_cc}, the dissimilarity between two channel coefficient vectors $\mathbf h^{(i)} \in \mathbb C^M$ and $\mathbf h^{(j)} \in \mathbb C^M$, where $M$ is the number of antennas in the \ac{mMIMO} array, is defined as
\[
    d_{\mathrm{Magoarou},i, j} = \sqrt{2 - 2 \frac{\left| \left(\mathbf h^{(i)}\right)^\mathrm{H} \mathbf h^{(j)} \right|}{\left\lVert \mathbf h^{(i)} \right\rVert_2 \left\lVert \mathbf h^{(j)} \right\rVert_2}}.
\]

Empirically we determined that a slightly modified version of the metric considering the squared \ac{CS}, which can be interpreted as a normalized power value, performs better with respect to all evaluation metrics:
\[
    d_{\mathrm{CS-vec},i, j} = 1 - \frac{\left| \left(\mathbf h^{(i)}\right)^\mathrm{H} \mathbf h^{(j)} \right|^2}{\left\lVert \mathbf h^{(i)} \right\rVert_2^2 \left\lVert \mathbf h^{(j)} \right\rVert_2^2}
\]
Note that $0 \leq d_{\mathrm{CS-vec},i, j} \leq 1$, with $d_{\mathrm{CS-vec},i, j} = 0$ if and only if $\mathbf h^{(i)}$ and $\mathbf h^{(j)}$ are linearly dependent.
We generalize this metric to wideband \ac{OFDM} channels with $N_\mathrm{sub}$ subcarriers and distributed antenna setups with $B$ antenna arrays.
We achieve this by computing the summation over the dissimilarities calculated separately for each antenna array and each subcarrier:
\begin{equation}
    d_{\mathrm{CS}, i, j} = \sum_{b=1}^B \sum_{n=1}^{N_\mathrm{sub}} \left(1 - \frac{\left\lvert\sum_{m=1}^{M} \left({\mathbf H}_{b, m, n}^{(i)}\right)^* {\mathbf H}_{b, m, n}^{(j)}\right\rvert^2} {\left(\sum_{m=1}^{M}\left\lvert {\mathbf H}_{b, m, n}^{(i)}\right\rvert^2 \right) \left(\sum_{m=1}^{M} \left\lvert {\mathbf H}_{b, m, n}^{(j)}\right\rvert^2\right)}\right)
    \label{eq:cs}
\end{equation}

\subsection{Angle-Delay Profile (ADP)}
\label{sec:adp_distance}
Taking inspiration from the \ac{CIRA} metric in Eq. (\ref{eq:cira}), we propose an enhanced version of a \ac{CS}-based metric by exploiting the sparsity of time-domain \ac{CSI}.
Most signal power in time domain is concentrated in the \ac{LoS} path (if it exists) and a few strong \acp{MPC}.
As opposed to Eq. (\ref{eq:cs}), which computes cosine similarities for frequency-domain \ac{CSI} data, we suggest to perform the summation over time-domain \ac{CSI} matrices $\tilde {\mathbf H}$ for taps $\tau = \tau_\mathrm{min}, \ldots, \tau_\mathrm{max}$, where $\tau_\mathrm{min}$ and $\tau_\mathrm{max}$ must be adapted to the radio environment such that all relevant \acp{MPC} are accounted for:

\begin{equation}
    \begin{split}
        d_{\mathrm{ADP}, i, j} =& \sum_{b=1}^B \sum_{\tau=\tau_\mathrm{min}}^{\tau_\mathrm{max}} \left( 1 - \frac{\left\lvert \sum_{m=1}^{M} \left(\tilde {\mathbf H}_{b, m, \tau}^{(i)}\right)^* \tilde {\mathbf H}_{b, m, \tau}^{(j)}\right\rvert^2} { \left(\sum_{m=1}^{M}\left\lvert \tilde {\mathbf H}_{b, m, \tau}^{(i)}\right\rvert^2 \right) \left(\sum_{m=1}^{M} \left\lvert \tilde {\mathbf H}_{b, m, \tau}^{(j)}\right\rvert^2\right)} \right)
    \end{split}
    \label{eq:adp}
\end{equation}

Eq. (\ref{eq:adp}) can be interpreted as using the angular-domain similarity in terms of power measured across multiple time-domain taps and \acp{MPC}, hence we suggest the name \ac{ADP} for our novel metric.
The restriction to $\tau_\mathrm{min} \leq \tau \leq \tau_\mathrm{max}$ excludes taps with weak signal and strong noise contributions (refer to Section \ref{sec:feature_engingeering}).
In contrast to previously proposed angular domain metrics \cite{multipoint_cc} \cite{studer_earthmover}, we deliberately forgo the use of superresolution techniques or the computation of \acp{EMD}, which makes computation practical for large datasets.

\subsection{Dissimilarity Metric Learning (DL)}
\label{sec:dissimilarity_metric_learning}
\begin{figure}
    \centering
    \begin{tikzpicture}
        \node (featureext) [align = center, draw, minimum width = 3.3cm, minimum height = 0.6cm, inner sep = 0mm] {Feature Engineering};
        \node (Hi) at ($(featureext.north) + (-1, 0.6)$) {$\tilde {\mathbf H}^{(i)}$};
        \node (Hj) at ($(featureext.north) + (1, 0.6)$) {$\tilde {\mathbf H}^{(j)}$};
        \node (dnn) [draw, below = 0.6cm of featureext, minimum width = 3.3cm, minimum height = 0.6cm, inner sep = 0mm] {\ac{DNN} $\mathcal D_\Theta$};
        \node (lossfunc) [draw, below = 0.4cm of dnn, minimum width = 3.3cm, minimum height = 0.6cm, inner sep = 0mm] {Loss Function $\mathcal L_\mathrm{DL}$};
        
        \draw [thick, -latex] (Hi) -- ($(featureext.north) + (-1, 0.0)$);
        \draw [thick, -latex] (Hj) -- ($(featureext.north) + ( 1, 0.0)$);
        \draw [thick, -latex] ($(featureext.south) + (-1, 0)$) -- ($(dnn.north) + (-1, 0)$) node[midway, right] {$\mathbf f^{(i)}$};
        \draw [thick, -latex] ($(featureext.south) + (1, 0)$) -- ($(dnn.north) + (1, 0)$) node[midway, left] {$\mathbf f^{(j)}$};
        \draw [thick, -latex] (dnn) -- (lossfunc);
        
        \node [draw, minimum width = 3.3cm, minimum height = 0.6cm, inner sep = 0mm] (timedist) [right = 0.3cm of featureext] {$d_{\mathrm{time}, i, j} = \left| t^{(i)} - t^{(j)} \right|$};

        \node (ti) at ($(timedist.north) + (-1, 0.6)$) {$t^{(i)}$};
        \node (tj) at ($(timedist.north) + (1, 0.6)$) {$t^{(j)}$};
        \draw [thick, -latex] (ti) -- ($(timedist.north) + (-1, 0.0)$);
        \draw [thick, -latex] (tj) -- ($(timedist.north) + ( 1, 0.0)$);
        
        \draw [thick, -latex] (timedist.south) |- (lossfunc.east);
    \end{tikzpicture}
    \caption{Dissimilarity Metric Learning: The \ac{DNN} $\mathcal D_\Theta$ is trained to estimate the absolute time difference between samples $i$, $j$ based on their \ac{CSI}.}
    \label{fig:distancelearning}
\end{figure}
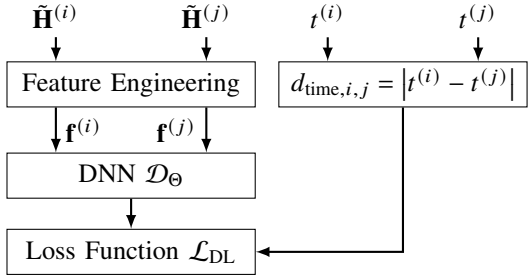

The previous dissimilarity metrics were explicitly defined and motivated based on assumptions grounded in the movement of \acp{UE} or in \ac{EM} wave propagation.
However, modeling wave propagation is complicated, which makes it difficult to formulate dissimilarity metrics that adapt to all environments.
In an entirely different approach, we suggest to not only use neural networks for manifold learning, but to use them to learn the dissimilarity metric itself.
The architecture of the dissimilarity metric learning process is shown Fig. \ref{fig:distancelearning}.

The challenge for this approach is the definition of a suitable loss function that does not rely on the ``ground truth'' \ac{UE} positions $\mathbf x^{(l)}$, which must not be used.
Instead, we suggest to train a \ac{DNN} $\mathcal{D}_\Theta$ to estimate the absolute time difference $d_{\mathrm{time}, i, j}$ (as defined in Eq. (\ref{eq:time_distance})) between pairs of \ac{CSI} features $\mathbf{f}^{(i)}, \mathbf{f}^{(j)}$ (as defined in Section \ref{sec:feature_engingeering}).
The hope is that the predictions of $\mathcal{D}_\Theta$ may also be a useful indicator of physical distances (compare Section \ref{sec:time-distance}).
As loss function, we employ a modified form of \ac{NMSE}:
\begin{equation}
    \mathcal{L}_\mathrm{DL} = \sum\limits_{i,j} \left(\frac{\mathcal D_\Theta\left( \mathbf f^{(i)}, \mathbf f^{(j)} \right) - d_{\mathrm{time}, i, j}}{d_{\mathrm{time},i, j} + \beta}\right)^2
    \label{eq:customnmse}
\end{equation}

In contrast to \ac{MSE}, \ac{NMSE} places a higher importance on predicting small time differences correctly by normalizing the error to $d_{\mathrm{time}, i, j}$.
The additional hyperparameter $\beta > 0$ in Eq. (\ref{eq:customnmse}) has the following impact on the behavior of $\mathcal L_\mathrm{DL}$:
For $d_\mathrm{time,i,j} \ll \beta$, Eq. (\ref{eq:customnmse}) behaves more like the regular \ac{MSE} loss function, i.e., normalization is applied less aggressively to very small time differences.
For $d_\mathrm{time,i,j} \gg \beta$, Eq. (\ref{eq:customnmse}) approaches an \ac{NMSE} loss function, implying that prediction errors for large true time differences $d_{\mathrm{time}, i, j}$ are weighted low.
Intuitively, this makes sense since we cannot expect $\mathcal D_\Theta$ to generate accurate distance predictions for large time differences due to unknown \ac{UE} trajectories.

For training the \ac{DNN}, we only select pairs of datapoints $i, j$ for which $d_\mathrm{time,i,j} \leq \alpha$, i.e., for which the absolute time difference is smaller than some threshold $\alpha$. 
The \ac{DNN} $\mathcal D_\Theta$ consists of four hidden layers with 1024, 512, 256 and 128 neurons each, all of them with ReLU activation.
In our experiments, we choose $\beta = 1\,\mathrm{s}$ and $\alpha = 500\,\mathrm{s}$.
Since subsequent processing steps may assume perfectly symmetric dissimilarity matrices, we symmetrize the metric during inference to obtain
\[
    d_{\mathrm{DL}, i, j} = \frac{1}{2} \left(\mathcal D_\Theta(\mathbf f^{(i)}, \mathbf f^{(j)}) + \mathcal D_\Theta(\mathbf f^{(j)}, \mathbf f^{(i)})\right).
\]

\subsection{Fused Dissimilarity Metrics}
\label{sec:combined_metrics}
\begin{figure*}
    \centering
    \begin{minipage}[b]{.64\textwidth}
        \begin{subfigure}{0.45\textwidth}
            \centering
            \includegraphics[width=0.9\textwidth]{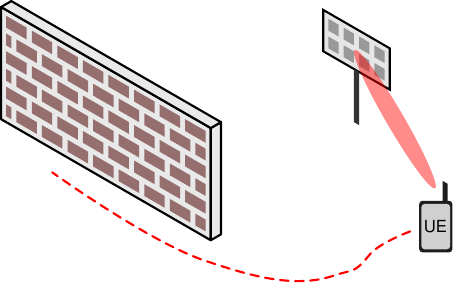}
            \caption{Vastly different channel conditions in physical proximity: CSI-based metrics fail}
            \label{fig:timestamps_wall}
        \end{subfigure}
        \quad
        \begin{subfigure}{0.5\textwidth}
            \centering
            \includegraphics[width=0.75\textwidth]{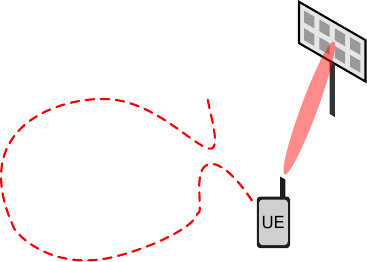}
            \caption{\ac{UE} locations are close in space, but far away in time: Timestamp-based metric fails}
            \label{fig:omega_path}
        \end{subfigure}
        \caption{Pathological cases for \ac{CSI}- or timestamp-based dissimilarity metrics}
        \label{fig:pathological_cases}
    \end{minipage}
    \begin{minipage}[b]{.32\textwidth}
        \centering
        \begin{tikzpicture}
            \tikzset{isomapgraphnode/.style={circle, inner sep = 1.5pt, fill}}
            \node (A) at(0.4, 0.2) [isomapgraphnode] {};
            \node (B) at(0.1, 0.85) [isomapgraphnode] {};
            \node (C) at(1.5, 0) [isomapgraphnode] {};
            \node (D) at(1, 0.7) [isomapgraphnode] {};
            \node (E) at(2.2, -0.2) [isomapgraphnode] {};
            \node (F) at(2, 1.1) [isomapgraphnode] {};
            \node (G) at(2.9, -0.25) [isomapgraphnode] {};
            \node (H) at(2.8, 0.2) [isomapgraphnode] {};
            \node (I) at(4, -0.1) [isomapgraphnode] {};
            \node (J) at(4.2, 0.5) [isomapgraphnode] {};
            \node (K) at(4.8, 0.2) [isomapgraphnode] {};
            \node (L) at(5, 0.7) [isomapgraphnode] {};
            
            \node [left = 0.02cm of B] {A};
            \node [right = 0.02cm of L] {B};
            
            \draw [dashed, red!80!black, thick, -latex] (B) to[bend left] node[above] {Dissimilarity $d_{\mathrm{ADP}, \mathrm{A,B}}$: Inaccurate} (L);
            
            \draw [blue!80!black, thick] (B.center) -- (D.center) -- (C.center) -- (E.center) -- (G.center) -- (I.center) -- (J.center) -- (L.center);
            \node [below = 0.1cm of G, blue!80!black] {Geodesic dissimilarity $d_{\mathrm{G-ADP}, \mathrm{A,B}}$};
        \end{tikzpicture}
        \caption{Illustration of geodesic dissimilaritiy}
        \label{fig:geodesic_distances}
    \end{minipage}
\end{figure*}

Observing some pathological situations, illustrated in Fig. \ref{fig:pathological_cases}, leads to the realization that different metrics may fail or  perform well under different sets of circumstances:
If channel conditions change dramatically over a short spatial distance, as is the case in Fig. \ref{fig:timestamps_wall}, the channel charting algorithm may fail to place neighboring datapoints in proximity to each other in the channel chart if using a \ac{CSI}-based dissimilarity metric.
On the other hand, if a \ac{UE} takes the $\Omega$-shaped trajectory shown in Fig. \ref{fig:omega_path}, any timestamp-based metric will fail to correctly indicate the closeness of the two turning points.

These two examples, which can occur in real-world datasets, lead us to suggest \emph{fused} dissimilarity metrics that combine information from multiple sources.
A fused dissimilarity metric $d_\mathrm{fuse}$ is computed separately for every combination of two datapoints $i$, $j$ and takes into account two or more other dissimilarity metrics $d_\mathrm{A}$, $d_\mathrm{B}$, \ldots:
\begin{equation}
    d_{\mathrm{fuse}, i, j} = f_\mathrm{fuse}(\gamma_\mathrm{A} d_{\mathrm{A}, i, j}, \gamma_\mathrm{B} d_{\mathrm{B}, i, j}, \ldots)
    \label{eq:combined_distmatrix}
\end{equation}

The function $f_\mathrm{fuse}$ in Eq. (\ref{eq:combined_distmatrix}) defines how the combined metric is calculated.
For example, if the dissimilarity metrics are prone to sometimes overestimating the true distance, but immune from underestimating, then $f_\mathrm{fuse}$ may simply choose the minimum of all metrics.
The constants $\gamma_\mathrm{A}$, $\gamma_\mathrm{B}$, $\ldots$ in Eq. (\ref{eq:combined_distmatrix}) are scaling factors which may be required to scale all metrics to a common reference.
For instance, all metrics may be scaled such that they approximate the true distance in meters.
In practice, these scaling factors can often be estimated from the dataset itself.
As will be shown in Section \ref{sec:results}, the fusion of a \ac{CSI}-based metric like $d_\mathrm{ADP}$ with $d_\mathrm{time}$ as a timestamp-based dissimilarity metric is of particular interest and leads to significant performance improvements.

\subsection{Global Geodesic Dissimilarities}
\label{sec:geodesic_distances}
As \cite{fraunhofer_cc} points out, dissimilarity metrics are most reliable if the true distance between two datapoints is small.
On the other hand, if two datapoints are separated widely in physical space, \ac{CSI}-based dissimilarity metrics are rarely indicative of the true physical distance.
A channel chart learned directly from \ac{CSI}-based dissimilarity metrics without any further processing will usually preserve local neighborhood relationships, but fail at capturing the global topology of the environment.
As in \cite{fraunhofer_cc} and inspired by the idea behind Isomap \cite{isomap}, we compute \emph{geodesic dissimilarities}, which more accurately predict true distances between far-away datapoints.
The basic concept is illustrated in Fig. \ref{fig:geodesic_distances}:
The dissimilarity between two far-away points A and B is computed as the sum of smaller dissimilarities (which are more accurate) between intermediate points that are closer together.

In practice, geodesic distances are computed as follows:
First, we find the pairwise dissimilarities between any two datapoints in the dataset using one of the aforementioned metrics and collect them in the dissimilarity matrix $\mathbf D_\mathrm{pw} \in \mathbb R^{L \times L}$.
If, as explained in Section \ref{sec:combined_metrics}, a fused dissimilarity metric shall be used, $\mathbf D_\mathrm{pw}$ must contain the fused dissimilarities.
$\mathbf D_\mathrm{pw}$ is interpreted as a matrix containing the weights of a graph, and we find its $k$-nearest neighbor graph $G_{k\mathrm{-NN}}$, where $k$ is a tunable hyperparameter that we choose as $k = 20$.
We then apply a shortest path algorithm, such as Dijkstra's algorithm \cite{dijkstra} to $G_{k\mathrm{-NN}}$.
The length of the computed shortest paths between any two nodes of $G_{k\mathrm{-NN}}$ are the geodesic dissimilarities, which we collect in the geodesic dissimilarity matrix $\mathbf D_\mathrm{geo} \in \mathbb R^{L \times L}$.
In our notation, geodesic versions of pairwise dissimilarity metrics are denoted by a ``G-'' prefix (for example, $d_\mathrm{CIRA}$ becomes $d_\mathrm{G-CIRA}$).

\section{Classical Manifold Learning}
\label{sec:classical-manifold-learning}

\begin{figure*}
    \centering
    \begin{minipage}[b]{.65\textwidth}
        \centering
        \scalebox{0.9}{
            \begin{tikzpicture}[
            	techniquegroup/.style={rectangle, draw = blue!40!black, very thick, rounded corners = 1pt, inner sep = 4pt, align = center, font={\small\sffamily}},
            	technique/.style={rectangle, draw = blue!40!black, very thick, rounded corners = 1pt, inner sep = 4pt, align = center, fill = blue!5!white, minimum height = 1.0cm, minimum width = 2.2cm, font={\small\sffamily}},
            ]
            
            \node [techniquegroup] at (0, 0) {Manifold Learning}
            	[edge from parent fork down, sibling distance=7cm, level distance = 1.2cm]
            	child { node [techniquegroup, align = center] {Classical}
            		[edge from parent fork down, sibling distance=2.45cm, level distance = 1.4cm]
            		child { node [technique] {MDS /\\Isomap} }
            		child { node [technique] {Sammon's\\Mapping} }
            		child { node (tsne) [technique] {t-SNE} }
            	}
            	child { node [techniquegroup, align = center] {Deep Learning-based}
            		[edge from parent fork down, sibling distance=2.45cm, level distance = 1.4cm]
            		child { node [technique] {Siamese\\Network} }
            		child { node [technique] {Triplet\\Network} }
            	};
            \end{tikzpicture}
        }
        \caption{Overview of classical and deep learning-based manifold learning methods}
        \label{fig:dimensionality_reduction_overview}
    \end{minipage}\qquad
    \begin{minipage}[b]{.29\textwidth}
        \centering
        \scalebox{0.9}{
\begin{tikzpicture}
    \tikzstyle{box1} = [rectangle, draw = blue!40!black, very thick, rounded corners=1pt,inner sep = 4pt, align = center, fill = blue!5!white, minimum height=16pt,minimum width=1.5cm]
    \tikzstyle{box2} = [rectangle, draw = blue!40!black, very thick, rounded corners=1pt,inner sep = 4pt, align = center, minimum height=16pt,minimum width=3.8cm]

    \node (in_1) at (2,0){$\mathbf{f}^{(i)}$};
    \node (in_2) at (4,0){$\mathbf{f}^{(j)}$};
    
    \node (in_3) at (3,0){$d_{i,j}$};
    
    \node[box1] (dnn_1) at (2,-1.3) {DNN $\mathcal{C}_\Theta$};
    \node[box1] (dnn_2) at (4,-1.3) {DNN $\mathcal{C}_\Theta$};
    
    \node [color=blue!40!black, text width=2.5cm] (shared_weights) at (6.5,-0.6){shared\\weights $\Theta$};
    
    \node[box2] (contrastive_loss) at (3,-2.6) {Loss $\mathcal{L}_\mathrm{Siamese}$};
    
    \draw [->, thick]  (in_1.south) -- (dnn_1.north)
    node[midway,anchor=east]{};
    \draw [->, thick]  (in_2.south) -- (dnn_2.north)
    node[midway,anchor=east]{};
    
    \draw [->, thick]  (in_3.south) -- (contrastive_loss)
    node[midway,anchor=east]{};
    
    \draw [->, thick,color=blue!40!black]  (shared_weights.west) -- (2.7,-0.6) -- (dnn_1);
    \draw [->, thick,color=blue!40!black]
    (4.7,-0.6) -- (dnn_2);
    
    \draw [->, thick]  (dnn_1.south) -- (dnn_1.south|-contrastive_loss.north)
    node[midway,anchor=east]{$\mathbf{z}^{(i)}$};
    \draw [->, thick]  (dnn_2.south) -- (dnn_2.south|-contrastive_loss.north)
    node[midway,anchor=west]{$\mathbf{z}^{(j)}$};
\end{tikzpicture}
        }
        \caption{Siamese neural network structure}
        \label{fig:siamese_nn_structure}
    \end{minipage}
\end{figure*}

Classical manifold learning techniques formulate optimization problems, where the objective is to find a low-dimensional representation, which, according to some cost function, minimizes the discrepancy in point-to-point distances between channel chart and dissimilarity matrix.

\subsection{Multidimensional Scaling and Isomap}
Multidimensional scaling (MDS) \acused{MDS} \cite{kruskal_stress} places points $\{\mathbf{z}^{(l)}\}_{l=1}^L$ in the channel chart such that the mean squared distance between point-to-point distances in the channel chart $\left\lVert \mathbf{z}^{(i)} - \mathbf{z}^{(j)} \right\rVert_2$ and the dissimilarities $d_{i, j}$ in the dissimilarity matrix is minimized, which is achieved by
\begin{equation}
    \min_{\{\mathbf{z}^{(l)}\}_{l=1}^L} \sum\nolimits_{i=1}^{L-1}\sum\nolimits_{j=i+1}^L \left(d_{i, j} - \lVert \mathbf{z}^{(i)} - \mathbf{z}^{(j)} \rVert_2 \right)^2.
\end{equation}
This non-convex optimization problem is solved with gradient descent.
It is not guaranteed that the global optimum of the objective function is found.
Any orthogonal transform of the optimized point set $\{\mathbf{z}^{(l)}\}_{l=1}^L$ is also an optimum.
\ac{MDS} applied to a geodesic dissimilarity matrix is known as Isomap.

\subsection{Sammon's Mapping}
\ac{SM} \cite{sammon_mapping} is a variation of \ac{MDS} which emphasizes the importance of small point-to-point dissimilarities.
This is achieved by adding a weighting factor to the objective function:
\begin{equation}
    \min_{\{\mathbf{z}^{(l)}\}_{l=1}^L} \sum\nolimits_{i=1}^{L-1}\sum\nolimits_{j=i+1}^L \frac{1}{d_{i, j}} \left(d_{i, j}-\lVert \mathbf{z}^{(i)} - \mathbf{z}^{(j)} \rVert_2\right)^2
\end{equation}

\subsection{t-distributed Stochastic Neighbor Embedding}
\ac{t-SNE} \cite{t_sne} is another commonly used manifold learning technique.
The algorithm starts by computing what are called conditional probabilities for the distances in the dissimilarity matrix:
\begin{equation}
    p_{j|i} = \frac{\exp{\left(-d_{i, j}^2/2\sigma_i^2\right)}}{\sum_{k\neq i}\exp{\left(-d_{i, k}^2/2\sigma_i^2\right)}} \quad \text{and} \quad p_{i|i} = 0
    \label{eq:tsne-conditional}
\end{equation}
In Eq. (\ref{eq:tsne-conditional}), $\sigma_i$ is a hyperparameter.
For the computation of $\sigma_i$, we refer to \cite[Section 2]{t_sne} and note that our values of $\sigma_i$ are based on a ``perplexity'' of 120, which was empirically found to deliver good results.
\ac{t-SNE} then defines symmetrized conditional probabilities as $p_{ij} = \frac{p_{j|i} + p_{i|j}}{2L}$ and defines pairwise similarities in the channel chart as
\[
    q_{ij} = \frac{\left(1 + \lVert \mathbf{z}_i - \mathbf{z}_j\rVert_2^2\right)^{-1}}{\sum_k \sum_{\ell\neq k} \left(1 + \lVert \mathbf{z}_k - \mathbf{z}_\ell\rVert_2^2\right)^{-1}} \quad \text{and} \quad q_{ii} = 0.
\]

The location of channel chart points $\{\mathbf{z}^{(l)}\}_{l=1}^L$ is then determined by minimizing the Kullback-Leibler divergence:
\[
\min_{\{\mathbf{z}^{(l)}\}_{l=1}^L} \sum_{i \neq j} p_{ij}\log\frac{p_{ij}}{q_{ij}}
\]

\section{Deep Learning-Based Manifold Learning}
\label{sec:dl-manifold-learning}

In the case of deep-learning based manifold learning methods, the \ac{FCF} is realized as a \ac{DNN} $\mathcal{C}_\Theta$, which maps some \ac{CSI} feature vector  $\mathbf{f}^{(l)}$ (see Section \ref{sec:feature_engingeering}) to the estimated corresponding position in the channel chart $\mathbf{z}^{(l)}$, i.e.,
\[
    \mathbf{z}^{(l)} = \mathcal{C}_\Theta\left(\mathbf{f}^{(l)}\right),
\]
with $\Theta$ being the set of trainable parameters.
The challenge is that labels for $\mathbf z^{(l)}$ are unavailable, hence $\mathcal{C}_\Theta$ cannot be trained in a supervised manner.
Instead, during the training step, it has been proposed to embed $\mathcal{C}_\Theta$ into either a \emph{Siamese} \cite{siamese_cc} or a \emph{triplet} \cite{triplet_cc} neural network, as explained in the following.
For inference, this embedding is no longer required.

\subsection{Siamese Neural Networks}
The structure of a Siamese neural network, as illustrated in Fig. \ref{fig:siamese_nn_structure}, enables the use of objective functions known from classical manifold learning techniques as loss function.
During training, two weight-sharing \acp{DNN} $\mathcal C_\Theta$ map two \ac{CSI} feature vectors $\mathbf{f}^{(i)}, \mathbf{f}^{(j)}$ to channel chart positions $\mathbf{z}^{(i)}, \mathbf{z}^{(j)}$ at once.
Then, the weights are updated based on a loss function $\mathcal L_\mathrm{Siamese}$ that takes into account $\mathbf{z}^{(i)}, \mathbf{z}^{(j)}$ and the dissimilarity metric $d_{i, j}$ from the dissimilarity matrix.
Here, we assume that the \ac{MDS} cost function is used as loss function for training:
\begin{equation}
\mathcal{L}_\mathrm{Siamese}=\sum\nolimits_{i=1}^{N-1}\sum\nolimits_{j=i+1}^N \left(d_{i, j}-\Vert\mathbf{z}^{(i)}-\mathbf{z}^{(j)}\Vert_2\right)^2.
\label{eq:contrastive_loss}
\end{equation}
As for \ac{MDS}, this ensures that the \ac{DNN} learns to place the estimated points such that their distance in the channel chart match with the corresponding computed dissimilarities.
In that sense, the Siamese neural network can be seen as a parametric version of \ac{MDS}. However, other loss functions such as, e.g., Sammon's loss, could also be used \cite{siamese_cc}.

\subsection{Triplet Neural Networks}
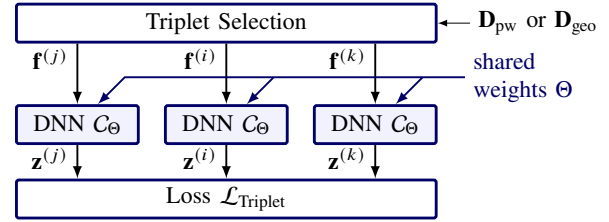
\begin{figure}
    \centering
    \scalebox{0.9}{
\begin{tikzpicture}
    \tikzstyle{box1} = [rectangle, draw = blue!40!black, very thick, rounded corners=1pt,inner sep = 2pt, align = center, fill = blue!5!white, minimum height=16pt,minimum width=1.8cm]
    \tikzstyle{box2} = [rectangle, draw = blue!40!black, very thick, rounded corners=1pt,inner sep = 2pt, align = center, minimum height=16pt,minimum width=6.2cm]
    
    \node[box2] (tripletsel) at (4.2, 0.2) {Triplet Selection};
    
    \node[box1] (dnn_1) at (2,-1.3) {DNN $\mathcal{C}_\Theta$};
    \node[box1] (dnn_2) at (4.2,-1.3) {DNN $\mathcal{C}_\Theta$};
    \node[box1] (dnn_3) at (6.4,-1.3) {DNN $\mathcal{C}_\Theta$};
    
    \node [anchor = west] (distmatrix) at (7.8, 0.2) {$\mathbf D_\mathrm{pw}$ or $\mathbf D_\mathrm{geo}$};
    \draw [->] (distmatrix) -- (tripletsel.east);
    
    \node [color=blue!40!black, text width=2.5cm] (shared_weights) at (9.1,-0.6){shared\\weights $\Theta$};
    
    \node[box2] (triplet_loss) at (4.2,-2.4) {Loss $\mathcal{L}_\mathrm{Triplet}$};
    
    \draw [->, thick] ($(tripletsel.south) + (-2.2, 0)$) -- (dnn_1.north) node[pos=0.25,anchor=east] {$\mathbf{f}^{(j)}$};
    \draw [->, thick] ($(tripletsel.south) + (0, 0)$) -- (dnn_2.north) node[pos=0.25,anchor=east] {$\mathbf{f}^{(i)}$};
    \draw [->, thick] ($(tripletsel.south) + (2.2, 0)$) -- (dnn_3.north) node[pos=0.25,anchor=east] {$\mathbf{f}^{(k)}$};
    
    \draw [->, thick,color=blue!40!black]  (shared_weights.west) -- (2.7,-0.6) -- (dnn_1);
    \draw [->, thick,color=blue!40!black]
    (4.9,-0.6) -- (dnn_2);
    \draw [->, thick,color=blue!40!black]
    (7.1,-0.6) -- (dnn_3);
    
    \draw [->, thick]  (dnn_1.south) -- (dnn_1.south|-triplet_loss.north)
    node[pos=0.4,anchor=east]{$\mathbf{z}^{(j)}$};
    \draw [->, thick]  (dnn_2.south) -- (dnn_2.south|-triplet_loss.north)
    node[pos=0.4,anchor=east]{$\mathbf{z}^{(i)}$};
    \draw [->, thick]  (dnn_3.south) -- (dnn_3.south|-triplet_loss.north)
    node[pos=0.4,anchor=east]{$\mathbf{z}^{(k)}$};
\end{tikzpicture}
    }
    \caption{Triplet neural network structure}
    \label{fig:triplet_nn_structure}
\end{figure}
Another approach is to embed the \ac{DNN} $\mathcal{C}_\Theta$ into a triplet neural network \cite{facenet}, as depicted in Fig. \ref{fig:triplet_nn_structure}.
Here, the loss function does not require information from the dissimilarity matrix.
Instead, information about dissimilarities is incorporated into \emph{triplet selection}.
This triplet selection step selects 3-tuples $(i, j, k) \in \{ 1, \ldots, L \}^3$ of datapoints, where the datapoint with index $i$ is called the \emph{anchor} sample, datapoint $j$ is the \emph{close} sample and datapoint $k$ is the \emph{far} sample.
In our case, we choose the anchor sample $i$ randomly among all datapoints in the dataset with uniform probability.
Then, we define the set $\mathcal N_q^{(i)}$ of suitable close samples for the anchor sample $i$ as
\[
    \mathcal N_q^{(i)} = \left\{ l ~ | ~ d_{i, l} < d_{\mathrm{thresh}, q}^{(i)} \right\},
\]
where $d_{\mathrm{thresh}, q}^{(i)}$ is the $q$-quantile of all dissimilarities $\{ d_{i, l} \}_{l = 1, \ldots, L}$.
In other words, $\mathcal N_q^{(i)}$ contains the $qL$ nearest neighbors of datapoint $i$ with respect to the dissimilarities.

The parameter $q$ is a hyperparameter that may be adjusted during training.
We suggest large values for $q$ at first (e.g., $q = 0.2$) and to reduce $q$ over the training epochs (e.g., to $q = 0.02$).
This way, the neural network first learns the global and then the local structure of the manifold.
Next, the index of the close sample $j$ is picked randomly from $\mathcal N_q^{(i)}$, and the index of the far sample $k$ is picked randomly from $\{ 1, \ldots, L \} \setminus \mathcal N_q^{(i)}$, in both cases with uniform probability.
This process is repeated until the desired number of triplets $(i, j, k)$ has been collected in the set $\mathcal T$.
Ideally, each triplet should fulfill
\begin{equation}\label{eq:triplet_inequality}
    \lVert \mathbf{x}^{(i)} - \mathbf{x}^{(j)} \rVert_2 < \lVert \mathbf{x}^{(i)} - \mathbf{x}^{(k)} \rVert_2 \quad \text{with} \quad (i, j, k) \in \mathcal T.
\end{equation}
That is, in physical space, the close sample should be nearer to the anchor sample than the far sample.
Since triplet selection guarantees that $d_{i, j} < d_{i, k}$, Eq. (\ref{eq:triplet_inequality}) should be fulfilled for most triplets as long as a good dissimilarity metric is chosen.

The three weight-sharing \acp{DNN} estimate channel chart positions $\mathbf z^{(i)}$, $\mathbf z^{(j)}$ and $\mathbf z^{(k)}$ based on the \ac{CSI} feature vectors $\mathbf f^{(i)}$, $\mathbf f^{(j)}$ and $\mathbf f^{(k)}$.
The triplet loss $\mathcal{L}_\mathrm{Triplet}$ \cite{facenet} then makes use of the knowledge that $\mathbf z^{(j)}$ should be closer to $\mathbf z^{(i)}$ than $\mathbf z^{(k)}$:
\begin{equation}
    \mathcal{L}_\mathrm{Triplet} = \sum_{\left(i,j,k\right)\in \mathcal{T}} \frac{\max \left\{\Vert\mathbf{z}^{(i)}-\mathbf{z}^{(j)}\Vert_2-\Vert\mathbf{z}^{(i)}-\mathbf{z}^{(k)}\Vert_2+M, 0\right\}}{\mathcal{|T|}},
    \label{eq:triplet_loss}
\end{equation}
In Eq. (\ref{eq:triplet_loss}), $M$ is yet another hyperparameter called \emph{margin} that we choose to be $M = 1$.
In contrast to Siamese networks, triplet neural networks only require qualitative information about dissimilarities:
It is sufficient to know that the close sample is closer to the anchor than the far sample.
Therefore, triplet neural networks only require a monotonically positive relationship between true distances $\lVert \mathbf x^{(i)} - \mathbf x^{(j)} \rVert_2$ and dissimilarities $d_{i, j}$, but not necessarily the proportional relationship stipulated in Eq. (\ref{eq:proportional_dissimilarities}).
This is a key advantage for metrics that do not approximate the proportional relationship (e.g., $d_\mathrm{time}$), but can also be a disadvantage if good absolute dissimilarities are available, but the information is not used.

\section{Evaluation Metrics}
\label{sec:evaluation-metrics}
The quality of a channel chart can be subjectively judged by comparing the learned chart to the map of true locations.
However, for a more objective assessment, well-known evaluation metrics from manifold learning like \ac{CT}, \ac{TW} \cite{trustworthiness_continuity} and \ac{KS} \cite{kruskal_stress} are commonly applied.
While \ac{CT} and \ac{TW} indicate how well local neighborhood relationships are reflected in the channel chart, \ac{KS} measures to what extent the global structure is preserved.

Instead of applying these metrics to the resulting channel chart, and thereby evaluating dissimilarity metric and manifold learning technique jointly, \ac{CT}, \ac{TW} and \ac{KS} may also be used to evaluate dissimilarity metrics separately \cite{studer_cc}.
We also consider \ac{RD} \cite{rajski_distance} for dissimilarity metrics and, for channel charts, we compute the \ac{MAE} between the ground truth positions and the channel chart positions assuming an optimal affine coordinate transform.

\subsection{Continuity and Trustworthiness}
\ac{CT} and \ac{TW} compare neighborhood relationships in physical space to neighborhood relationships in the representation space.
Let $r_{l, i}$ denote the rank of the neighbor with index $i$ among all neighbors of datapoint $l$ in physical space.
That is, $r_{l, i} = 1$ if $i$ is the closest neighbor of $l$, $r_{l, i} = 2$ if it is the second closest neighbor and so on.
We define the set of $K$ nearest neighbors of $l$ in physical space as $\mathcal{V}_{K}^{(l)} = \left\{ i ~ | ~ r_{l, i} \leq K \right\}$.

In a similar way, let $\hat r_{l, i}$ denote the rank of the neighbor with index $i$ among all neighbors of datapoint $l$ in the representation space.
When evaluating \ac{CT} / \ac{TW} of a dissimilarity metric, this rank is calculated over the entries in the dissimilarity matrix, i.e., over $\left\{d_{l, i}\right\}_{i = 1, \ldots, L}$.
If \ac{CT} / \ac{TW} are applied to a channel chart, the rank is calculated over the Euclidean distances of the channel chart locations, i.e., over $\left\{\lVert \mathbf z^{(l)} - \mathbf z^{(i)} \rVert_2\right\}_{i = 1, \ldots, L}$.
We define the set of $K$ nearest neighbors of $l$ in the representation space as $\mathcal{U}_{K}^{(l)} = \left\{ i ~ | ~ \hat r_{l, i} \leq K \right\}$.
\ac{CT} is then defined as
\[
    \mathrm{CT}\left(K\right) = 1 - \frac{2}{LK \left(2L - 3K - 1\right)} \sum^L_{l=1}
    \sum_{i \in \mathcal{V}_{K}^{(l)}} \max\left\{0, \hat{r}_{l, i} - K\right\},
\]
and, intuitively speaking, indicates whether the nearest neighbors in physical space are among those of the representation space.
Analogously, \ac{TW} is defined as
\[
    \mathrm{TW}\left(K\right) = 1 - \frac{2}{LK \left(2L - 3K - 1\right)} \sum^L_{l=1}
    \sum_{i \in \mathcal{U}_{K}^{(l)}} \max \left\{0, r_{l, i} - K\right\},
\]
and indicates whether the neighbors in representation space are among those in physical space, or, conversely, if the representation contains false neighbors.
The normalization factor ensures that both \ac{CT} and \ac{TW} are in the range $\left[0,1\right]$, where values close to $1$ are better.
As already proposed in \cite{studer_cc}, we choose a neighborhood size of $K = 0.05 L$.

\subsection{Kruskal's Stress}
\ac{KS} \cite{kruskal_stress} is a performance measure for the preservation of the global structure of the channel chart.
It compares pairwise distances between points in the original space to those between points in the representation space:
\[
\begin{split}
    \mathrm{KS} = \sqrt{\frac{\sum_{i=1}^{L-1}\sum_{j=i+1}^L\left(\Vert\mathbf{x}^{(i)}-\mathbf{x}^{(j)}\Vert_2-\beta\Vert\mathbf{z}^{(i)}-\mathbf{z}^{(j)}\Vert_2\right)^2}{\sum_{i=1}^{L-1}\sum_{j=i+1}^L\Vert\mathbf{x}^{(i)}-\mathbf{x}^{(j)}\Vert_2^2}} \\
    \text{with} \quad
    \beta = \frac{\sum_{i=1}^{L-1}\sum_{j=i+1}^L\Vert \mathbf{x}^{(i)}-\mathbf{x}^{(j)}\Vert_2 \Vert \mathbf{z}^{(i)}-\mathbf{z}^{(j)}\Vert_2}{\sum_{i=1}^{L-1}\sum_{j=i+1}^L\Vert \mathbf{z}^{(i)}-\mathbf{z}^{(j)} \Vert_2^2}.
\end{split}
\]
The scaling factor $\beta$ ensures that the value of \ac{KS} is bounded to the range $\left[0,1\right]$, where $0$ indicates the best preservation of the global structure.
If \ac{KS} is evaluated for a dissimilarity metric (as opposed to a channel chart), $\Vert\mathbf{z}^{(i)}-\mathbf{z}^{(j)}\Vert_2$ is replaced by the corresponding dissimilarity $d_{i, j}$.

\subsection{Rajski's Distance}
\ac{RD}, which was proposed for use in channel charting in \cite{studer_earthmover}, is an indicator for the mutual information between the true distances and the dissimilarities.
To this end, all pairwise true distances $\lVert \mathbf{x}^{(l)} - \mathbf{x}^{(i)} \rVert_2$, as well as all dissimilarities $d_{l, i}$ ($l, i \in \{ 1, \ldots, L \}$) are quantized into $100$ bins each such that they can be interpreted as distributions modeled by the random variables $V$ and $Q$, respectively.
\ac{RD} is then defined as \cite{rajski_distance}
\[
    \begin{split}
    \mathrm{RD}\left(V,Q\right)=1-\frac{I\left(V,Q\right)}{H\left(V,Q\right)}, \quad \text{for} \quad H\left(V,Q\right)\neq0, \\
    \text{with} \quad I\left(V,Q\right)=\sum_{v\in V, q\in Q} P_{V,Q}\left(v,q\right) \mathrm{log}_2 \frac{P_{V,Q}\left(v,q\right)}{P_V\left(v\right)P_Q\left(q\right)}
    \end{split}
\]
denoting the mutual information between the distributions of $V$ and $Q$,
where $P_{V,Q}\left(v,q\right)$ is the joint probability distribution of $V$ and $Q$ and $P_V\left(v\right)$ and $P_Q\left(q\right)$ are the respective marginal distributions.
The joint entropy of $V$ and $Q$ is given by
\[
    H\left(V,Q\right)=-\sum_{v\in V, q\in Q} P_{V,Q}\left(v,q\right) \mathrm{log}_2 P_{V,Q}\left(v,q\right).
\]
\ac{RD} is between 0 and 1, and 0 indicates the best performance.

\subsection{Mean Absolute Error}
\label{sec:mae}
A learned channel chart should reconstruct both local and global geometry, but may be rotated, scaled, flipped and / or translated compared to the true map.
As previously proposed by Stahlke et al. \cite{fraunhofer_cc}, we find the optimal affine transformation from channel chart to the true locations and compute the \ac{MAE} under the assumption that this transformation has been found and applied.
The resulting \ac{MAE} thereby provides a lower bound to the absolute localization error.
We anticipate that it will be possible to achieve performance close to this bound by applying proposals like \cite{pihlajasalo2020absolute}.
To find the optimal affine transformation, we solve the least squares problem
\[
    (\mathbf{\hat A}, \mathbf{\hat b}) = \argmin\limits_{(\mathbf{A}, \mathbf{b})} \sum_{l = 1}^L \lVert\mathbf{A}\mathbf{z}^{(l)} + \mathbf b - \mathbf{x}^{(l)}\rVert_2^2
\]
and compute the \ac{MAE} as $\mathrm{MAE} = \frac{1}{L}\sum_{l=1}^L\lVert \mathbf{\hat{A}}\mathbf{z}^{(l)} + \mathbf{\hat b} - \mathbf{x}^{(l)}\rVert_2.$

\section{Experimental Results}
\label{sec:results}

\subsection{DICHASUS: Channel Measurement Datasets}
\label{sec:dichasus}
\begin{figure*}
    \centering
    \begin{subfigure}[b]{0.33\textwidth}
        \centering
        \includegraphics[width=0.95\textwidth, trim = 30 200 30 0, clip]{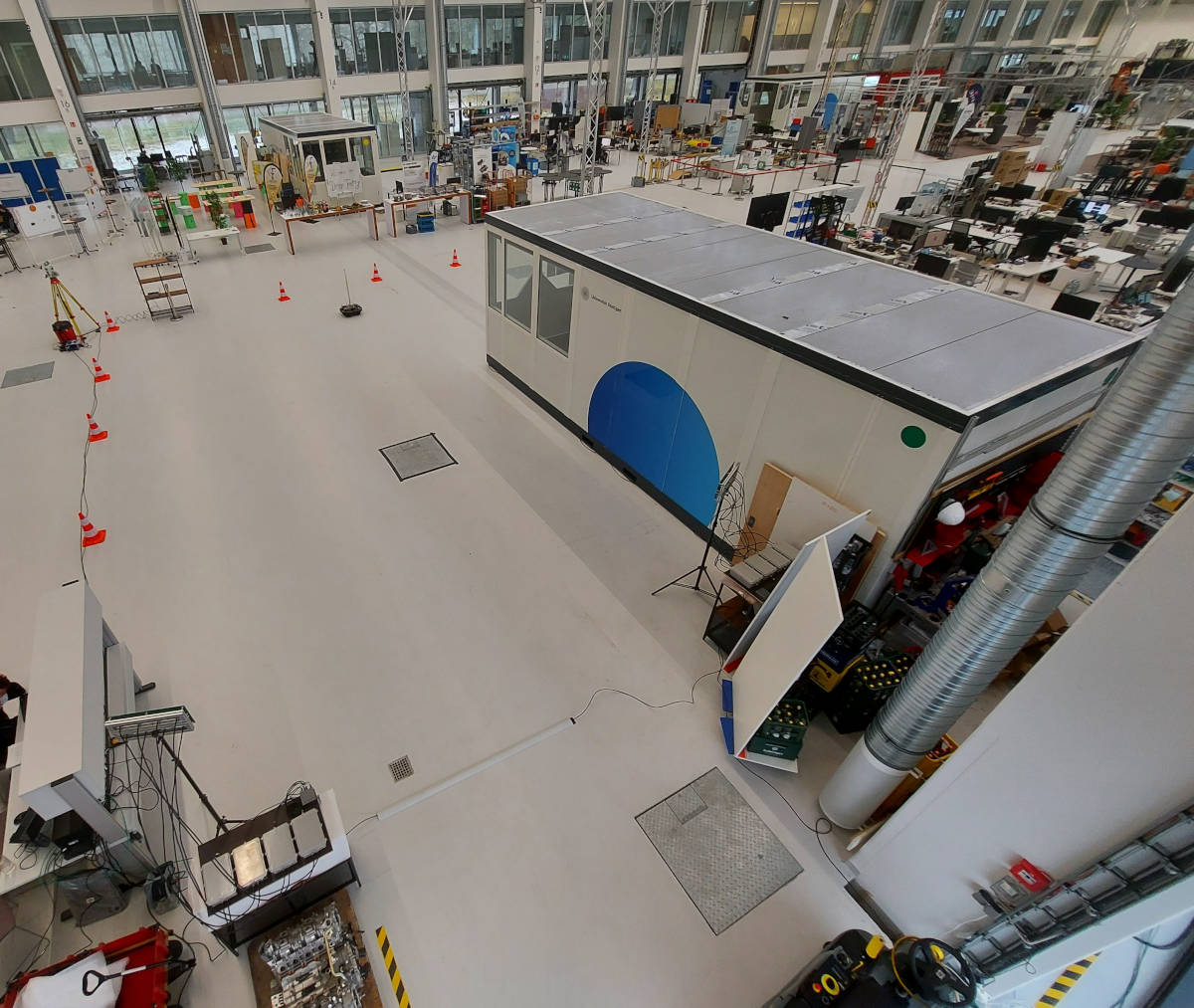}
        \vspace{0.4cm}
        \vspace{-0.1cm}
        \caption{}
    \end{subfigure}
    \begin{subfigure}[b]{0.32\textwidth}
        \centering
        \begin{tikzpicture}
            \begin{axis}[
                width=0.729\columnwidth,
                height=0.6\columnwidth,
                scale only axis,
                xmin=-15.5,
                xmax=6.1,
                ymin=-18.06,
                ymax=-1.5,
                xlabel = {Coordinate $x ~ [\mathrm{m}]$},
                ylabel = {Coordinate $y ~ [\mathrm{m}]$},
                ylabel shift = -8 pt,
                xlabel shift = -4 pt,
                xtick={-10, -6, -2, 2}
            ]
                \addplot[thick,blue] graphics[xmin=-14.5,ymin=-17.06,xmax=4.1,ymax=-1.5] {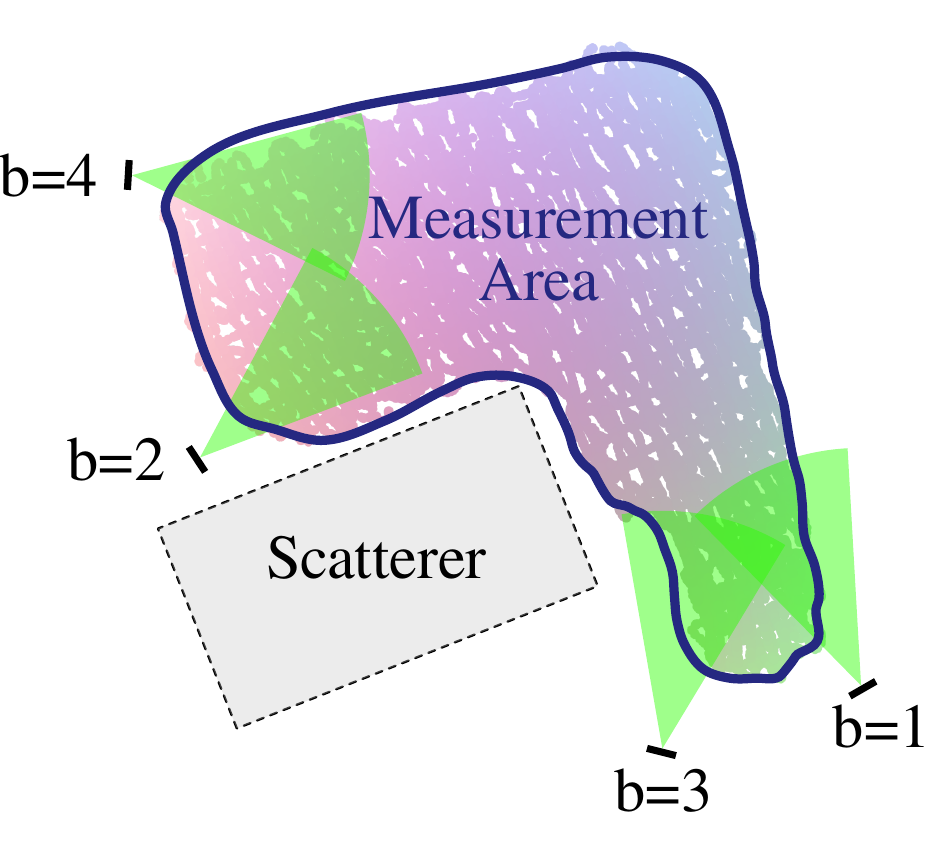};
            \end{axis}
        \end{tikzpicture}
        \vspace{-0.1cm}
        \caption{}
        \label{fig:labelled-area}
    \end{subfigure}
    \begin{subfigure}[b]{0.32\textwidth}
        \centering
        \begin{tikzpicture}
            \begin{axis}[
                width=0.6\columnwidth,
                height=0.6\columnwidth,
                scale only axis,
                xmin=-12.5,
                xmax=2.5,
                ymin=-14.5,
                ymax=-1.5,
                xlabel = {Coordinate $x ~ [\mathrm{m}]$},
                ylabel = {Coordinate $y ~ [\mathrm{m}]$},
                ylabel shift = -8 pt,
                xlabel shift = -4 pt,
                xtick={-10, -6, -2, 2}
            ]
                \addplot[thick,blue] graphics[xmin=-12.5,ymin=-14.5,xmax=2.5,ymax=-1.5] {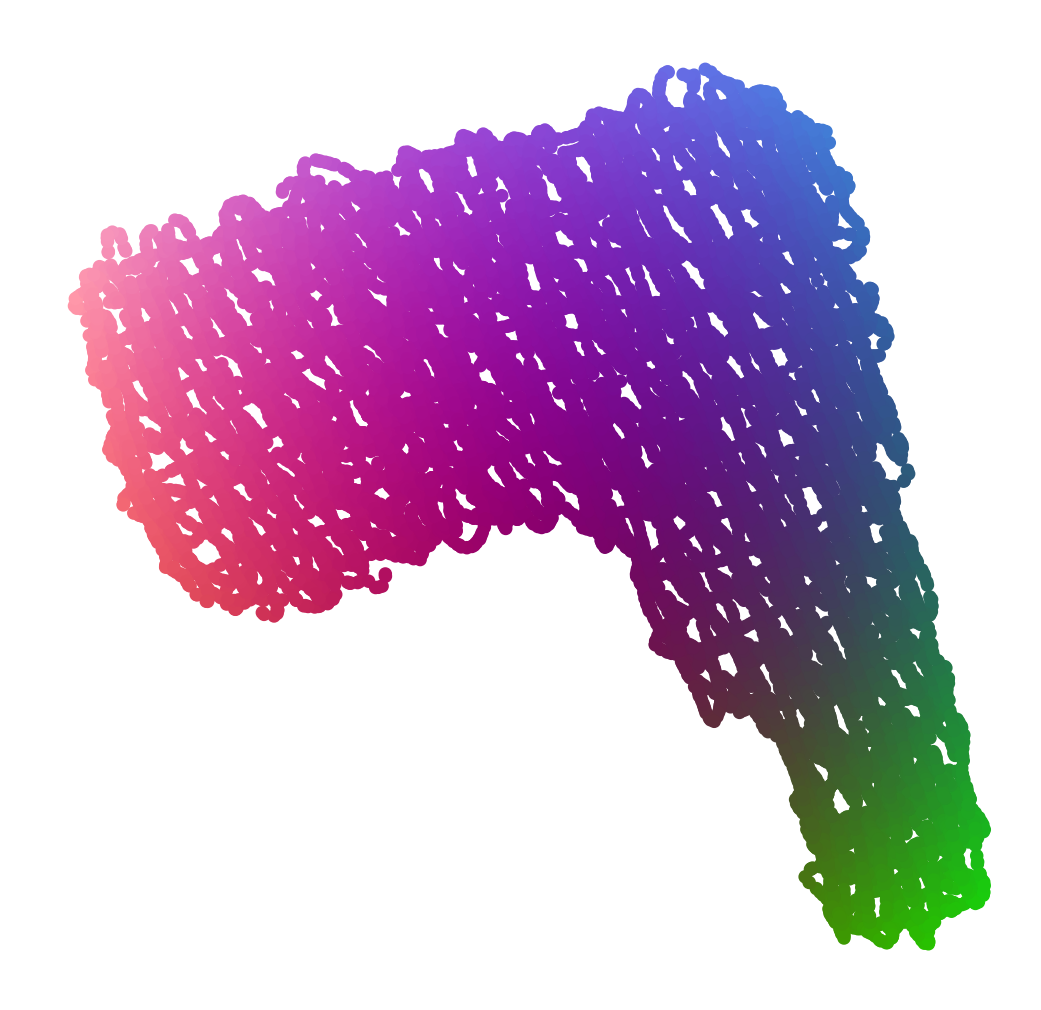};
            \end{axis}
        \end{tikzpicture}
        \vspace{-0.1cm}
        \caption{}
        \label{fig:groundtruth-map}
    \end{subfigure}
    \vspace{-0.3cm}
    \caption{Information about the industrial environment the measurement was conducted in. The figure shows (a) a photograph of the environment, (b) a top view map and (c) a scatter plot of colorized ``ground truth'' positions of datapoints in $\mathcal S_\mathrm{full}$, also in top view. The antenna arrays in the map are drawn to scale as black rectangles and their viewing direction is indicated by the green sectors.}
    \label{fig:industrial_environment}
\end{figure*}

All dissimilarity metrics and manifold learning techniques were tested on a dataset generated by \emph{\ac{DICHASUS}}, our distributed \ac{mMIMO} channel sounder, whose architecture is thoroughly described in \cite{dichasus2021}.
In brief, \ac{DICHASUS} measures the propagation channel between a single transmitter and many receive antennas.
It achieves long-term phase-coherence, even if antennas are distributed over a wide area.
\ac{DICHASUS} provides large datasets containing frequency-domain \ac{CSI} tensors $\mathbf H^{(l)}$, alongside side information like timestamps and accurate information about the positions of all antennas.

The dataset chosen for the following analyses is called \emph{dichasus-cf0x} \cite{dataset-dichasus-cf0x}, and was captured in an industrial environment with $B = 4$ separate antenna arrays made up of $M = 2 \times 4$ antennas each.
$N_\mathrm{sub} = 1024$ \ac{OFDM} channel coefficients were measured at a carrier frequency of $1.272\,\mathrm{GHz}$ and with a channel bandwidth of $50\,\mathrm{MHz}$.
The single dipole transmit antenna is mounted on top of a robot, which travels along a set of trajectories inside a defined, L-shaped area, with an overall size of approximately $14\,\mathrm{m} \times 14\,\mathrm{m}$.
A photo and a top view map of the environment are shown in Fig. \ref{fig:industrial_environment}.
A large metal container is located at the inner corner of the L-shape, blocking the \ac{LoS}.
Every position in the measurement area has at least one \ac{LoS} path to one of the four arrays, and some positions in the dataset even have a \ac{LoS} path to all arrays.

To reduce the total amount of data, we choose three subsets of \emph{dichasus-cf0x} called \emph{dichasus-cf02}, \emph{dichasus-cf03} and \emph{dichasus-cf04} and then collect only every 5\textsuperscript{th} datapoint into a smaller dataset that we call $\mathcal S_\mathrm{full}$, now containing $\left|\mathcal S_\mathrm{full}\right| = 16797$ datapoints.
The true datapoint positions $\mathbf x^{(l)}$ are shown in Fig. \ref{fig:industrial_environment}.
The points have been colorized and the datapoints will retain their color even as the \ac{FCF} maps them to a position in the channel chart.
This allows for a visual evaluation of the generated chart: If the global topology is preserved, a similar color gradient should appear in the chart.

\begin{figure}
    \centering
\begin{tikzpicture}

\definecolor{crimson2143940}{RGB}{214,39,40}
\definecolor{darkgray176}{RGB}{176,176,176}
\definecolor{darkorange25512714}{RGB}{255,127,14}
\definecolor{forestgreen4416044}{RGB}{44,160,44}
\definecolor{gray127}{RGB}{127,127,127}
\definecolor{mediumpurple148103189}{RGB}{148,103,189}
\definecolor{orchid227119194}{RGB}{227,119,194}
\definecolor{sienna1408675}{RGB}{140,86,75}
\definecolor{steelblue31119180}{RGB}{31,119,180}

\begin{axis}[
width=\columnwidth,
height=.5\columnwidth,
tick align=outside,
tick pos=left,
x grid style={darkgray176},
xlabel={$\tau$},
xmajorgrids,
xmin=504, xmax=527,
xtick style={color=black},
y grid style={darkgray176},
ylabel={$|\tilde {\mathbf H}_{1,m,\tau}^{(l)}|$},
ymajorgrids,
ymin=-0.15, ymax=3.22,
ytick style={color=black},
ticklabel style={fill=white},
legend cell align={left},
legend style={
  at={(0.74,0.94)},
  anchor=north,
},
legend columns=2,
clip = false]
\path [draw=none, fill=red, fill opacity=0.15]
(axis cs:507,-0.15)
--(axis cs:507,3.22) node (taumintop) {}
--(axis cs:520,3.22) node (taumaxtop) {}
--(axis cs:520,-0.15)
--cycle;

\addplot [semithick, mittelblau]
table {%
504 0.0103528667241335
505 0.0111632393673062
506 0.0113612022250891
507 0.0161797441542149
508 0.0537847653031349
509 0.253659516572952
510 1.6299991607666
511 0.364338994026184
512 0.180298626422882
513 0.182507187128067
514 0.371507734060287
515 0.0866622477769852
516 0.0944439321756363
517 0.205963924527168
518 0.179692849516869
519 0.32526758313179
520 0.247379690408707
521 0.208555892109871
522 0.237336829304695
523 0.119296237826347
524 0.105581969022751
525 0.0523029528558254
526 0.0824050009250641
527 0.0513913035392761
};
\addlegendentry{$m = 1$}
\addplot [semithick, darkorange25512714]
table {%
504 0.0268455017358065
505 0.0318318605422974
506 0.0350442044436932
507 0.0419967658817768
508 0.0449570119380951
509 0.0747723579406738
510 1.61510109901428
511 0.989263832569122
512 0.274097681045532
513 0.620353162288666
514 0.186997696757317
515 0.183664992451668
516 0.0402505174279213
517 0.198869839310646
518 0.120992079377174
519 0.0921802967786789
520 0.146016702055931
521 0.0302751157432795
522 0.0533650852739811
523 0.103227272629738
524 0.167705044150352
525 0.100765399634838
526 0.0645769611001015
527 0.072935163974762
};
\addlegendentry{$m = 2$}
\addplot [semithick, forestgreen4416044]
table {%
504 0.029728539288044
505 0.0362099297344685
506 0.040357518941164
507 0.049402441829443
508 0.0569900348782539
509 0.155370965600014
510 2.18984603881836
511 1.49561381340027
512 0.482259333133698
513 0.445671856403351
514 0.22804482281208
515 0.340645223855972
516 0.0237672217190266
517 0.214470028877258
518 0.0975789576768875
519 0.10644906014204
520 0.210441306233406
521 0.251645356416702
522 0.324277549982071
523 0.199505925178528
524 0.230835855007172
525 0.0957760736346245
526 0.147033527493477
527 0.035957720130682
};
\addlegendentry{$m = 3$}
\addplot [semithick, crimson2143940]
table {%
504 0.02880834415555
505 0.0353155359625816
506 0.0512269139289856
507 0.0780953094363213
508 0.180096700787544
509 0.873564779758453
510 3.06944608688354
511 0.672043859958649
512 0.277742147445679
513 0.485104382038116
514 0.085481658577919
515 0.145617991685867
516 0.327041000127792
517 0.122862055897713
518 0.321855932474136
519 0.0509163439273834
520 0.214505434036255
521 0.138341039419174
522 0.175600335001945
523 0.101853184401989
524 0.167477741837502
525 0.0398262478411198
526 0.127994954586029
527 0.111697137355804
};
\addlegendentry{$m = 4$}
\addplot [semithick, mediumpurple148103189]
table {%
504 0.00812591332942247
505 0.0100526846945286
506 0.016503756865859
507 0.0289446040987968
508 0.0788524895906448
509 0.326440066099167
510 1.44285535812378
511 0.302280128002167
512 0.241887629032135
513 0.0537204705178738
514 0.302305281162262
515 0.177190884947777
516 0.256053775548935
517 0.392749786376953
518 0.111598685383797
519 0.280430138111115
520 0.351891577243805
521 0.12333345413208
522 0.154770165681839
523 0.174174293875694
524 0.0455650240182877
525 0.132184311747551
526 0.0936159715056419
527 0.10612715035677
};
\addlegendentry{$m = 5$}
\addplot [semithick, sienna1408675]
table {%
504 0.0216435547918081
505 0.0257453620433807
506 0.0301795359700918
507 0.039658185094595
508 0.053691390901804
509 0.14266075193882
510 1.52022743225098
511 0.852588772773743
512 0.187465310096741
513 0.830275177955627
514 0.238002210855484
515 0.162579163908958
516 0.339133322238922
517 0.31077840924263
518 0.159670427441597
519 0.0450350269675255
520 0.323084563016891
521 0.0321344584226608
522 0.096453420817852
523 0.0462092943489552
524 0.115544281899929
525 0.106010183691978
526 0.0332904532551765
527 0.0762070119380951
};
\addlegendentry{$m = 6$}
\addplot [semithick, orchid227119194]
table {%
504 0.0154235316440463
505 0.0198191814124584
506 0.0232229325920343
507 0.0325905121862888
508 0.0674420371651649
509 0.304871320724487
510 2.24849057197571
511 1.04776406288147
512 0.480736434459686
513 0.324414908885956
514 0.22627130150795
515 0.312449932098389
516 0.0662628933787346
517 0.124514371156693
518 0.130337506532669
519 0.106767065823078
520 0.147359177470207
521 0.247018605470657
522 0.226406127214432
523 0.205203756690025
524 0.151094377040863
525 0.111207783222198
526 0.174613013863564
527 0.0416953861713409
};
\addlegendentry{$m = 7$}
\addplot [semithick, gray127]
table {%
504 0.0405595079064369
505 0.0475164838135242
506 0.0658112987875938
507 0.0941096916794777
508 0.217966228723526
509 1.64028179645538
510 2.6294801235199
511 0.230025336146355
512 0.271033227443695
513 0.217806711792946
514 0.01800093986094
515 0.107974514365196
516 0.294531226158142
517 0.105678655207157
518 0.14122448861599
519 0.0797541365027428
520 0.128566950559616
521 0.152358815073967
522 0.218004047870636
523 0.0613677948713303
524 0.0621242746710777
525 0.0307801831513643
526 0.115684106945992
527 0.0718737989664078
};
\addlegendentry{$m = 8$}

\draw [thick, dotted] (axis cs: 507, 3.22) -- (axis cs: 507, -0.15) node[pos = 0.79, anchor = center, fill = white, opacity = 0.7, text opacity = 1.0, inner sep = 1pt, red!80!black, fill = white] {$\tau_\mathrm{min}$};
\draw [thick, dotted] (axis cs: 520, 3.22) -- (axis cs: 520, -0.15) node[pos = 0.79, anchor = center, fill = white, opacity = 0.7, text opacity = 1.0, inner sep = 1pt, red!80!black, fill = white] {$\tau_\mathrm{max}$};
\end{axis}

\end{tikzpicture}
    \vspace{-0.2cm}
    \caption{Examplary time-domain \ac{CSI} amplitudes $|\tilde {\mathbf H}_{1,m,\tau}^{(l)}|$ at some time instant $l$. Each line corresponds to one antenna $m$ at antenna array $b = 1$ .}
    \label{fig:cir}
\end{figure}
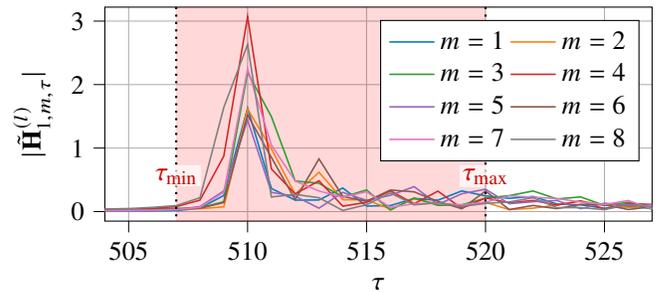

Exemplary \ac{CSI} data measured by \ac{DICHASUS} is visualized in Fig. \ref{fig:cir}, which shows the \acp{CIR} of all antennas in antenna array $b = 1$.
Note that phase information is also available, but not displayed in the figure.
It is easy to see that much of the received signal power is concentrated within a relatively short interval of taps (area shaded red in Fig. \ref{fig:cir}), leading us to choose $\tau_\mathrm{min} = 507$ and $\tau_\mathrm{max} = 520$ for all subsequent analyses.

\subsection{Fusion of $d_\mathrm{ADP}$ and $d_\mathrm{time}$}
\begin{figure}
    \centering
    \begin{tikzpicture}
        \begin{axis}[
            tick align=outside,
            tick pos=left,
            xmin=-1.5, xmax=31.5,
            xtick style={color=black},
            ymin=0, ymax=4930.8,
            ytick style={color=black},
            width=0.8\columnwidth,
            height=0.4\columnwidth,
            ylabel style={align=center},
            ylabel={Occurences for\\$d_{\mathrm{time}, i, j} < t_\mathrm{thresh}$},
            xlabel={$\nicefrac{d_{\mathrm{ADP}, i, j}}{d_{\mathrm{time}, i, j}}$}
        ]
            \draw[draw=none,fill=mittelgrau] (axis cs:0,0) rectangle (axis cs:0.375,428);
            \draw[draw=none,fill=mittelgrau] (axis cs:0.375,0) rectangle (axis cs:0.75,1502);
            \draw[draw=none,fill=mittelgrau] (axis cs:0.75,0) rectangle (axis cs:1.125,1378);
            \draw[draw=none,fill=mittelgrau] (axis cs:1.125,0) rectangle (axis cs:1.5,1150);
            \draw[draw=none,fill=mittelgrau] (axis cs:1.5,0) rectangle (axis cs:1.875,966);
            \draw[draw=none,fill=mittelgrau] (axis cs:1.875,0) rectangle (axis cs:2.25,790);
            \draw[draw=none,fill=mittelgrau] (axis cs:2.25,0) rectangle (axis cs:2.625,644);
            \draw[draw=none,fill=mittelgrau] (axis cs:2.625,0) rectangle (axis cs:3,620);
            \draw[draw=none,fill=mittelgrau] (axis cs:3,0) rectangle (axis cs:3.375,562);
            \draw[draw=none,fill=mittelgrau] (axis cs:3.375,0) rectangle (axis cs:3.75,484);
            \draw[draw=none,fill=mittelgrau] (axis cs:3.75,0) rectangle (axis cs:4.125,456);
            \draw[draw=none,fill=mittelgrau] (axis cs:4.125,0) rectangle (axis cs:4.5,458);
            \draw[draw=none,fill=mittelgrau] (axis cs:4.5,0) rectangle (axis cs:4.875,436);
            \draw[draw=none,fill=mittelgrau] (axis cs:4.875,0) rectangle (axis cs:5.25,422);
            \draw[draw=none,fill=mittelgrau] (axis cs:5.25,0) rectangle (axis cs:5.625,350);
            \draw[draw=none,fill=mittelgrau] (axis cs:5.625,0) rectangle (axis cs:6,320);
            \draw[draw=none,fill=mittelgrau] (axis cs:6,0) rectangle (axis cs:6.375,338);
            \draw[draw=none,fill=mittelgrau] (axis cs:6.375,0) rectangle (axis cs:6.75,288);
            \draw[draw=none,fill=mittelgrau] (axis cs:6.75,0) rectangle (axis cs:7.125,304);
            \draw[draw=none,fill=mittelgrau] (axis cs:7.125,0) rectangle (axis cs:7.5,334);
            \draw[draw=none,fill=mittelgrau] (axis cs:7.5,0) rectangle (axis cs:7.875,292);
            \draw[draw=none,fill=mittelgrau] (axis cs:7.875,0) rectangle (axis cs:8.25,280);
            \draw[draw=none,fill=mittelgrau] (axis cs:8.25,0) rectangle (axis cs:8.625,274);
            \draw[draw=none,fill=mittelgrau] (axis cs:8.625,0) rectangle (axis cs:9,322);
            \draw[draw=none,fill=mittelgrau] (axis cs:9,0) rectangle (axis cs:9.375,290);
            \draw[draw=none,fill=mittelgrau] (axis cs:9.375,0) rectangle (axis cs:9.75,318);
            \draw[draw=none,fill=mittelgrau] (axis cs:9.75,0) rectangle (axis cs:10.125,288);
            \draw[draw=none,fill=mittelgrau] (axis cs:10.125,0) rectangle (axis cs:10.5,284);
            \draw[draw=none,fill=mittelgrau] (axis cs:10.5,0) rectangle (axis cs:10.875,316);
            \draw[draw=none,fill=mittelgrau] (axis cs:10.875,0) rectangle (axis cs:11.25,288);
            \draw[draw=none,fill=mittelgrau] (axis cs:11.25,0) rectangle (axis cs:11.625,384);
            \draw[draw=none,fill=mittelgrau] (axis cs:11.625,0) rectangle (axis cs:12,406);
            \draw[draw=none,fill=mittelgrau] (axis cs:12,0) rectangle (axis cs:12.375,392);
            \draw[draw=none,fill=mittelgrau] (axis cs:12.375,0) rectangle (axis cs:12.75,448);
            \draw[draw=none,fill=mittelgrau] (axis cs:12.75,0) rectangle (axis cs:13.125,668);
            \draw[draw=none,fill=mittelgrau] (axis cs:13.125,0) rectangle (axis cs:13.5,774);
            \draw[draw=none,fill=mittelgrau] (axis cs:13.5,0) rectangle (axis cs:13.875,1062);
            \draw[draw=none,fill=mittelgrau] (axis cs:13.875,0) rectangle (axis cs:14.25,1382);
            \draw[draw=none,fill=mittelgrau] (axis cs:14.25,0) rectangle (axis cs:14.625,1690);
            \draw[draw=none,fill=mittelgrau] (axis cs:14.625,0) rectangle (axis cs:15,2018);
            \draw[draw=none,fill=mittelgrau] (axis cs:15,0) rectangle (axis cs:15.375,2540);
            \draw[draw=none,fill=mittelgrau] (axis cs:15.375,0) rectangle (axis cs:15.75,2810);
            \draw[draw=none,fill=mittelgrau] (axis cs:15.75,0) rectangle (axis cs:16.125,3350);
            \draw[draw=none,fill=mittelgrau] (axis cs:16.125,0) rectangle (axis cs:16.5,3664);
            \draw[draw=none,fill=mittelgrau] (axis cs:16.5,0) rectangle (axis cs:16.875,4038);
            \draw[draw=none,fill=mittelgrau] (axis cs:16.875,0) rectangle (axis cs:17.25,4252);
            \draw[draw=none,fill=mittelgrau] (axis cs:17.25,0) rectangle (axis cs:17.625,4444);
            \draw[draw=none,fill=mittelgrau] (axis cs:17.625,0) rectangle (axis cs:18,4696);
            \draw[draw=none,fill=mittelgrau] (axis cs:18,0) rectangle (axis cs:18.375,4620);
            \draw[draw=none,fill=mittelgrau] (axis cs:18.375,0) rectangle (axis cs:18.75,4630);
            \draw[draw=none,fill=mittelgrau] (axis cs:18.75,0) rectangle (axis cs:19.125,4668);
            \draw[draw=none,fill=mittelgrau] (axis cs:19.125,0) rectangle (axis cs:19.5,4208);
            \draw[draw=none,fill=mittelgrau] (axis cs:19.5,0) rectangle (axis cs:19.875,4332);
            \draw[draw=none,fill=mittelgrau] (axis cs:19.875,0) rectangle (axis cs:20.25,4012);
            \draw[draw=none,fill=mittelgrau] (axis cs:20.25,0) rectangle (axis cs:20.625,3980);
            \draw[draw=none,fill=mittelgrau] (axis cs:20.625,0) rectangle (axis cs:21,3688);
            \draw[draw=none,fill=mittelgrau] (axis cs:21,0) rectangle (axis cs:21.375,3616);
            \draw[draw=none,fill=mittelgrau] (axis cs:21.375,0) rectangle (axis cs:21.75,3560);
            \draw[draw=none,fill=mittelgrau] (axis cs:21.75,0) rectangle (axis cs:22.125,3306);
            \draw[draw=none,fill=mittelgrau] (axis cs:22.125,0) rectangle (axis cs:22.5,3344);
            \draw[draw=none,fill=mittelgrau] (axis cs:22.5,0) rectangle (axis cs:22.875,3076);
            \draw[draw=none,fill=mittelgrau] (axis cs:22.875,0) rectangle (axis cs:23.25,3068);
            \draw[draw=none,fill=mittelgrau] (axis cs:23.25,0) rectangle (axis cs:23.625,2990);
            \draw[draw=none,fill=mittelgrau] (axis cs:23.625,0) rectangle (axis cs:24,2782);
            \draw[draw=none,fill=mittelgrau] (axis cs:24,0) rectangle (axis cs:24.375,2696);
            \draw[draw=none,fill=mittelgrau] (axis cs:24.375,0) rectangle (axis cs:24.75,2666);
            \draw[draw=none,fill=mittelgrau] (axis cs:24.75,0) rectangle (axis cs:25.125,2590);
            \draw[draw=none,fill=mittelgrau] (axis cs:25.125,0) rectangle (axis cs:25.5,2578);
            \draw[draw=none,fill=mittelgrau] (axis cs:25.5,0) rectangle (axis cs:25.875,2552);
            \draw[draw=none,fill=mittelgrau] (axis cs:25.875,0) rectangle (axis cs:26.25,2244);
            \draw[draw=none,fill=mittelgrau] (axis cs:26.25,0) rectangle (axis cs:26.625,2338);
            \draw[draw=none,fill=mittelgrau] (axis cs:26.625,0) rectangle (axis cs:27,2274);
            \draw[draw=none,fill=mittelgrau] (axis cs:27,0) rectangle (axis cs:27.375,2240);
            \draw[draw=none,fill=mittelgrau] (axis cs:27.375,0) rectangle (axis cs:27.75,2164);
            \draw[draw=none,fill=mittelgrau] (axis cs:27.75,0) rectangle (axis cs:28.125,2078);
            \draw[draw=none,fill=mittelgrau] (axis cs:28.125,0) rectangle (axis cs:28.5,2068);
            \draw[draw=none,fill=mittelgrau] (axis cs:28.5,0) rectangle (axis cs:28.875,1894);
            \draw[draw=none,fill=mittelgrau] (axis cs:28.875,0) rectangle (axis cs:29.25,1946);
            \draw[draw=none,fill=mittelgrau] (axis cs:29.25,0) rectangle (axis cs:29.625,1886);
            \draw[draw=none,fill=mittelgrau] (axis cs:29.625,0) rectangle (axis cs:30,1808);

            \draw [very thick, red!80!black] (axis cs: 12, 0) -- (axis cs: 12, 5000) node[midway, right] {$\gamma$};
            \node [align = center, red!80!black, inner sep = 5pt, fill = white, opacity = 0.65, text opacity = 1.0] at (axis cs: 5, 2800) {Rely on\\$d_{\mathrm{ADP}, i, j}$};
            \node [align = center, red!80!black, inner sep = 5pt, fill = white, opacity = 0.65, text opacity = 1.0] at (axis cs: 22, 2800) {Rely on\\$d_{\mathrm{time}, i, j}$};
        \end{axis}
    \end{tikzpicture}
    \vspace{-0.2cm}
    \caption{Histogram depiciting the distribution of the ratios $\nicefrac{d_{\mathrm{ADP}, i, j}}{d_{\mathrm{time}, i, j}}$ for over all pairs of datapoints $i, j$ in $\mathcal S_\mathrm{full}$ with $d_{\mathrm{time}, i, j} < t_\mathrm{thresh}$.}
    \label{fig:metric_ratio_histogram}
\end{figure}
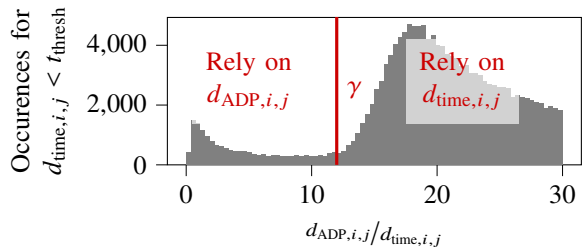

As described in Section \ref{sec:combined_metrics}, multiple dissimilarity metrics may be combined into a fused metric.
We choose to fuse $d_\mathrm{ADP}$ as a \ac{CSI}-based dissimilarity metric and $d_\mathrm{time}$ as a side information-based metric and show that using the fused metric $d_\mathrm{fuse}$ significantly improves the channel chart.
In \emph{dichasus-cf0x}, the robot either moves at approximately constant velocity $v_\mathrm{robot}$, or remains at a standstill.
The combination of $d_\mathrm{ADP}$ and $d_\mathrm{time}$ is based on a simple assumption:
For short time intervals $d_\mathrm{time} < t_\mathrm{thresh}$, we assume that the transmitter either moves along an approximately straight line (due to its inertia) with velocity $v_\mathrm{robot}$ or does not move at all.
For such short intervals, if the robot is moving, the time-based metric $d_\mathrm{time}$ predicts the true distance between points much better than any \ac{CSI}-based metric could.
Therefore, we propose to fuse $d_\mathrm{time}$ and $d_\mathrm{ADP}$ by taking their minimum after appropriate scaling of $d_\mathrm{time}$:
\begin{equation}
    d_{\mathrm{fuse}, i, j} = \mathrm{min} \left\{ d_{\mathrm{ADP}, i, j}, \gamma \cdot d_{\mathrm{time}, i, j} \right\}
    \label{eq:combine_time_adp}
\end{equation}
This way, whenever $d_{\mathrm{time}, i, j}$ is sufficiently small, it predicts the dissimilarity between points $i$ and $j$.
The challenge remains to find the scaling factor $\gamma$ in Eq. (\ref{eq:combine_time_adp}):
To determine $\gamma$, we need to observe the distribution of ratios $\nicefrac{d_{\mathrm{ADP}, i, j}}{d_{\mathrm{time}, i, j}}$ in Fig. \ref{fig:metric_ratio_histogram}, there shown for all pairs of datapoints $i$, $j$ in $\mathcal S_\mathrm{full}$ with $d_{\mathrm{time}, i, j} < t_\mathrm{thresh} = 2\,\mathrm{s}$.
The diagram clearly shows a bimodal distribution for the ratios, where the right mode can be attributed to the datapoints where the robot is moving, and the left mode can be attributed to the cases where it remains static.
We choose a scaling factor that separates the two modes, e.g., $\gamma = 12$, such that $\gamma d_{\mathrm{time}, i, j} < d_{\mathrm{ADP}, i, j}$ if the robot is moving.
For other real-world datasets, the assumption of constant transmitter velocity may not hold.
In that case, it might be necessary to estimate the relative velocities of transmitters over time and between transmitters based on \ac{CSI} data.
This remains a topic for future research that we do not consider within the scope of this work.
After computing the pairwise fused metric $d_{\mathrm{fuse}, i, j}$, we find geodesic distances as explained in Section \ref{sec:geodesic_distances} to obtain $d_{\mathrm{G-fuse}, i, j}$.

\subsection{Evaluation of Dissimilarity Metrics}\label{sec:dissimilarity_metric_evaluation}

\begin{table}
	\caption{Evaluation of dissimilarity metrics on $\mathcal S_\mathrm{full}$.}
	\centering
	\begin{tabular}{ccccc}
	\toprule
 	Dissimilarity & \ac{CT} $\uparrow$ & \ac{TW} $\uparrow$ & \ac{KS} $\downarrow$ & \ac{RD} $\downarrow$\\
 	\midrule
 	\rowcolor{TableGray}
 	$d_\mathrm{Euc}$ & $1.0000$ & $1.0000$ & $0.0000$ & $0.0000$ \\
 	\rowcolor{TableGray}
 	$d_\mathrm{G-Euc}$ & $0.9999$ & $1.0000$ & $0.0231$ & $0.3617$ \\
 	\midrule
 	$d_\mathrm{CS}$ & $0.9529$ & $0.9542$ & $0.4281$ & $0.9595$ \\
 	$d_\mathrm{G-CS}$ & $0.9799$ & $0.9774$ & $0.2676$ & $0.9112$ \\
 	$d_\mathrm{CIRA}$ & $0.9278$ & $0.9477$ & $0.3173$ & $0.9406$ \\
 	$d_\mathrm{G-CIRA}$ & $0.9685$ & $0.9684$ & $0.2197$ & $0.8972$ \\
 	\midrule
 	$d_\mathrm{DL}$ & $0.9762$ & $0.9742$ & $0.4267$ & $0.9499$ \\
 	$d_\mathrm{G-DL}$ & $0.9800$ & $0.9772$ & $0.2250$ & $0.8953$ \\
 	\midrule
	$d_\mathrm{ADP}$ & $0.9653$ & $0.9689$ & $0.4350$ & $0.9431$ \\
	$d_\mathrm{G-ADP}$ & $0.9885$ & $0.9866$ & $0.1738$ & $0.8384$ \\
	\midrule
	$d_\mathrm{fuse}$ & $0.9655$ & $0.9691$ & $0.4347$ & $0.9429$ \\
	$d_\mathrm{G-fuse}$ & $\mathbf{0.9957}$ & $\mathbf{0.9948}$ & $\mathbf{0.1098}$ & $\mathbf{0.7619}$ \\
	\bottomrule
	\end{tabular}
	\label{tab:distance_evaluation}
	\vspace*{-0.2cm}
\end{table}

In Tab. \ref{tab:distance_evaluation}, all dissimilarity metrics presented in Section \ref{sec:dissimilarity_metrics}, as well as previously described fused dissimilarity metric $d_{\mathrm{fuse},i, j}$ are evaluated using \ac{CT}, \ac{TW}, \ac{KS} and \ac{RD}.
The true distance $d_{\mathrm{Euc},i, j}$ is only included for reference and it obviously achieves the optimal performance metric in every category.
In all tables, results based on true distances are only included for comparison and are shaded in grey.
The results indicate that our newly proposed dissimilarity metrics $d_\mathrm{ADP}$ and $d_\mathrm{DL}$ outperform the baseline dissimilarities $d_\mathrm{CS}$ and $d_\mathrm{CIRA}$ regarding most performance metrics.
Overall, the geodesic fused dissimilarity $d_{\mathrm{G-fuse},i, j}$ is the best-performing one.

\begin{figure}[h]
    \centering
    \input{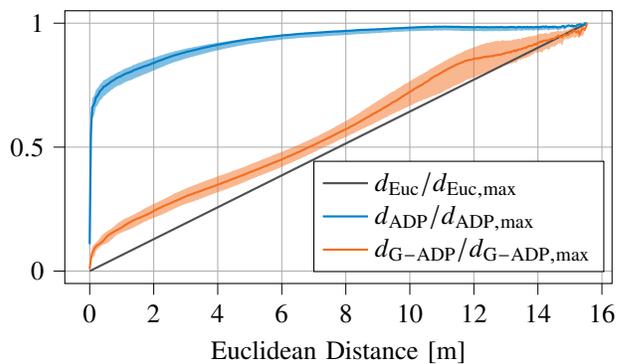}
    \vspace{-0.2cm}
    \caption{Dissimilarity metrics $d_\mathrm{ADP}$ and $d_\mathrm{G-ADP}$ normalized to a range from 0 to 1 as a function of the true distance. To generate the diagram, the dissimilarities are grouped into 100 bins according to the true distances. The lines represent the medians of these bins, and the shaded areas represent the range from the 25\textsuperscript{th} to the 75\textsuperscript{th} percentile. See also \cite[Fig. 6 and 7]{fraunhofer_cc}.}
\label{fig:adp_euclidean}
\end{figure}

In general, it is apparent that geodesic dissimilarities are better than the respective non-geodesic versions regarding \ac{CT}, \ac{TW}, \ac{KS} and \ac{RD}.
This is also illustrated by Fig. \ref{fig:adp_euclidean}, which, inspired by \cite{fraunhofer_cc}, plots $d_{\mathrm{ADP},i, j}$ and $d_{\mathrm{G-ADP},i, j}$ as a function of the Euclidean distance $d_{\mathrm{Euc},i, j}$.
Unsurprisingly, the geodesic dissimilarity approximates the Euclidean distance better (down to a scaling factor) than the non-geodesic dissimilarity.

\subsection{Channel Charting Results}
All channel charts in this section are learned on the training set $\mathcal S_\mathrm{full}$, as specified in Section \ref{sec:dichasus}.
While classical manifold learning methods directly produce channel charts, deep learning-based (parametric) methods are evaluated on a different test set $\mathcal S_\mathrm{full}'$.
This test set also contains $\left|\mathcal S_\mathrm{full}'\right| = 16797$ datapoints measured in the same area, and is generated by selecting a different subset of points from \emph{dichasus-cf02}, \emph{dichasus-cf03} and \emph{dichasus-cf04}.
This ensures that deep learning-based methods actually work with previously unseen feature vectors, and, in a performance comparison, puts them at a slight disadvantage compared to classical methods.

In our experiments, the \ac{DNN} $\mathcal C_\Theta$ consists of six hidden layers with 1024, 512, 512, 256, 128, 64 neurons and ReLU activation each, and an output layer with 2 neurons and linear activation.
Batch normalization is applied between the layers to enhance training performance.

Tab. \ref{tab:cc_performance_classical} shows the channel charting performance for different classical methods.
The results for the distance $d_{\mathrm{Euc}}$ serve as benchmark for the respective manifold learning technique.
As expected, the channel charts learned by \ac{SM} often represent the local geometry better than those learned by \ac{MDS} (higher \ac{CT} / \ac{TW}).
Overall, the best performance is achieved by Isomap, which applies \ac{MDS} to geodesic distances.
As already concluded in Section \ref{sec:dissimilarity_metric_evaluation}, the newly introduced dissimilarity metrics, and especially the fused metric appear to be most suitable, as they result in the best channel charts.
It is remarkable that the \ac{t-SNE} algorithm produces only slightly worse channel charts than Isomap without relying on geodesic distances.

The performance metrics for deep learning-based methods are shown in Tab. \ref{tab:cc_performance_deeplearning}.
Unsurprisingly, the Siamese network produces better channel charts for most geodesic dissimilarity metrics, whereas the triplet neural network performs better for non-geodesic metrics.
This can be explained by the fact that the triplet neural network only uses qualitative information about dissimilarity, and does not require the metrics to be proportional to true distances, whereas the Siamese network can also make use of absolute dissimilarity information.

The channel chart produced by a triplet neural network using purely time-based triplet selection ($d_{\mathrm{time}}$) is surprisingly good with respect to all metrics, but is still outperformed by the Siamese network:
The best overall performance is achieved by the Siamese network in combination with $d_\mathrm{G-fuse}$.

\begin{figure}[h]
    \centering
    \input{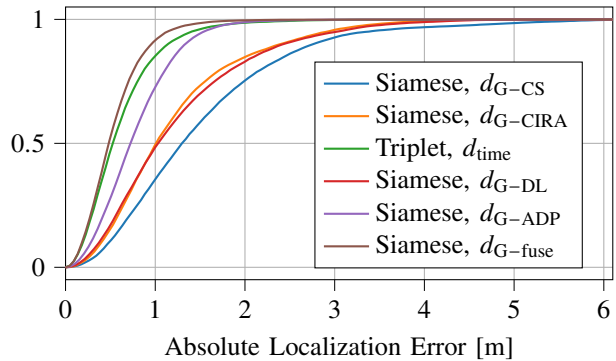}
    \vspace{-0.2cm}
    \caption{CDF of absolute localization error for different manifold learning methods and dissimilarity metrics, computed on $\mathcal S_\mathrm{full}$.}
    \label{fig:mae_cdf}
\end{figure}

\begin{figure*}
    \centering
    \begin{subfigure}[b]{0.24\textwidth}
        \begin{tikzpicture}
            \begin{axis}[
                width=0.6\columnwidth,
                height=0.6\columnwidth,
                scale only axis,
                xmin=-604,
                xmax=846,
                ymin=-611,
                ymax=865,
                xlabel = {Latent variable $x$},
                ylabel = {Latent variable $y$},
                tick label style={font=\small},
                label style={font=\small},
                ylabel shift = -8 pt,
                xlabel shift = -4 pt,
            ]
                \addplot[thick,blue] graphics[xmin=-604,ymin=-611,xmax=846,ymax=865] {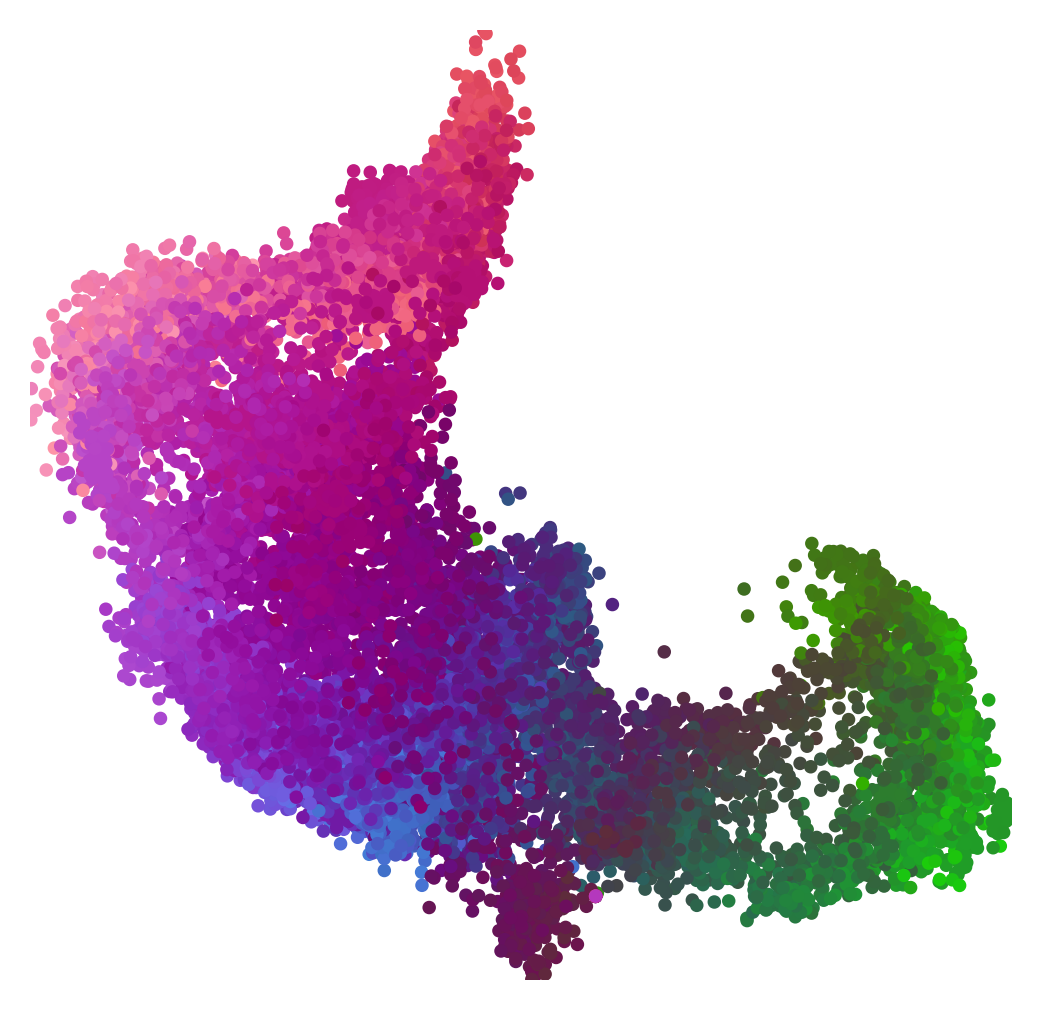};
            \end{axis}
        \end{tikzpicture}
        \vspace{-0.2cm}
        \caption{Siamese, $d_\mathrm{G-CIRA}$, $\mathcal S_\mathrm{full}$}
        \label{fig:siamese_cira_geo}
        \vspace{0.2cm}
    \end{subfigure}
    \begin{subfigure}[b]{0.24\textwidth}
        \begin{tikzpicture}
            \begin{axis}[
                width=0.6\columnwidth,
                height=0.6\columnwidth,
                scale only axis,
                xmin=-6,
                xmax=7.5,
                ymin=-11,
                ymax=11,
                xlabel = {Latent variable $x$},
                ylabel = {Latent variable $y$},
                tick label style={font=\small},
                label style={font=\small},
                ylabel shift = -8 pt,
                xlabel shift = -4 pt,
            ]
                \addplot[thick,blue] graphics[xmin=-6,ymin=-11,xmax=7.5,ymax=11] {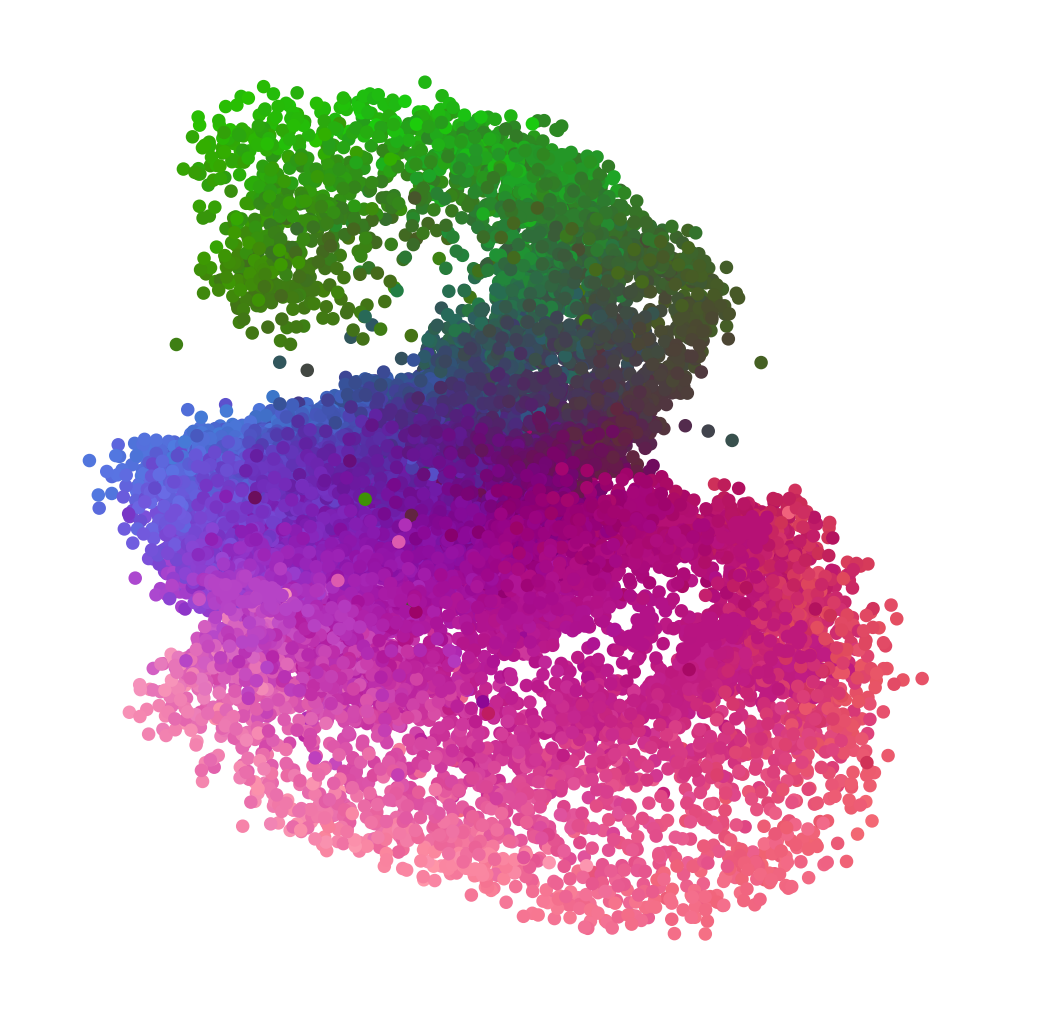};
            \end{axis}
        \end{tikzpicture}
        \vspace{-0.2cm}
        \caption{Siamese, $d_\mathrm{G-CS}$, $\mathcal S_\mathrm{full}$}
        \label{fig:siamese_cs_geo}
        \vspace{0.2cm}
    \end{subfigure}
    \begin{subfigure}[b]{0.24\textwidth}
        \begin{tikzpicture}
            \begin{axis}[
                width=0.6\columnwidth,
                height=0.6\columnwidth,
                scale only axis,
                xmin=-37,
                xmax=39.5,
                ymin=-38.5,
                ymax=36.5,
                xlabel = {Latent variable $x$},
                ylabel = {Latent variable $y$},
                tick label style={font=\small},
                label style={font=\small},
                ylabel shift = -8 pt,
                xlabel shift = -4 pt,
            ]
                \addplot[thick,blue] graphics[xmin=-37,ymin=-38.5,xmax=39.5,ymax=36.5] {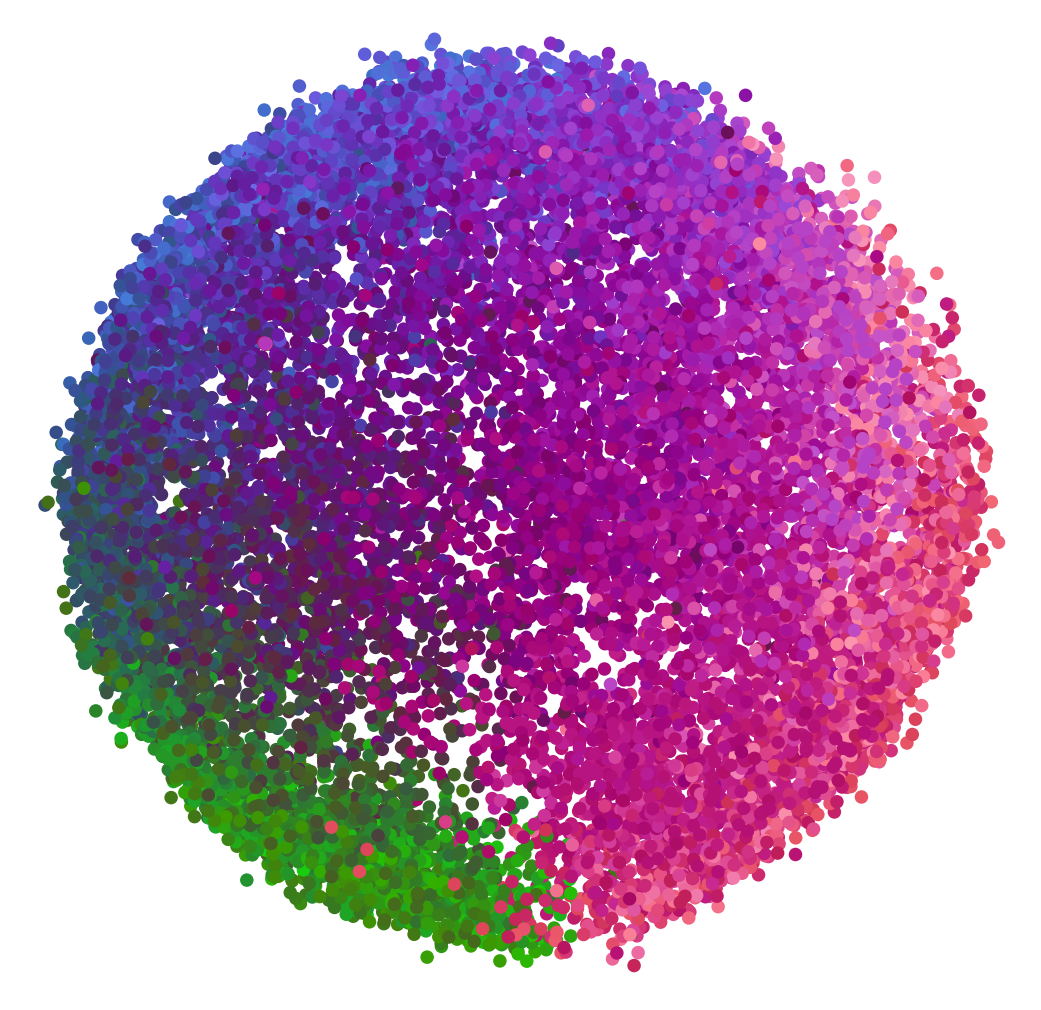};
            \end{axis}
        \end{tikzpicture}
        \vspace{-0.2cm}
        \caption{Siamese, $d_\mathrm{ADP}$, $\mathcal S_\mathrm{full}$}
        \label{fig:siamese_adp}
        \vspace{0.2cm}
    \end{subfigure}
    \begin{subfigure}[b]{0.24\textwidth}
        \begin{tikzpicture}
            \begin{axis}[
                width=0.6\columnwidth,
                height=0.6\columnwidth,
                scale only axis,
                xmin=-9,
                xmax=11.5,
                ymin=-12,
                ymax=14.5,
                xlabel = {Latent variable $x$},
                ylabel = {Latent variable $y$},
                tick label style={font=\small},
                label style={font=\small},
                ylabel shift = -8 pt,
                xlabel shift = -4 pt,
            ]
                \addplot[thick,blue] graphics[xmin=-9,ymin=-12,xmax=11.5,ymax=14.5] {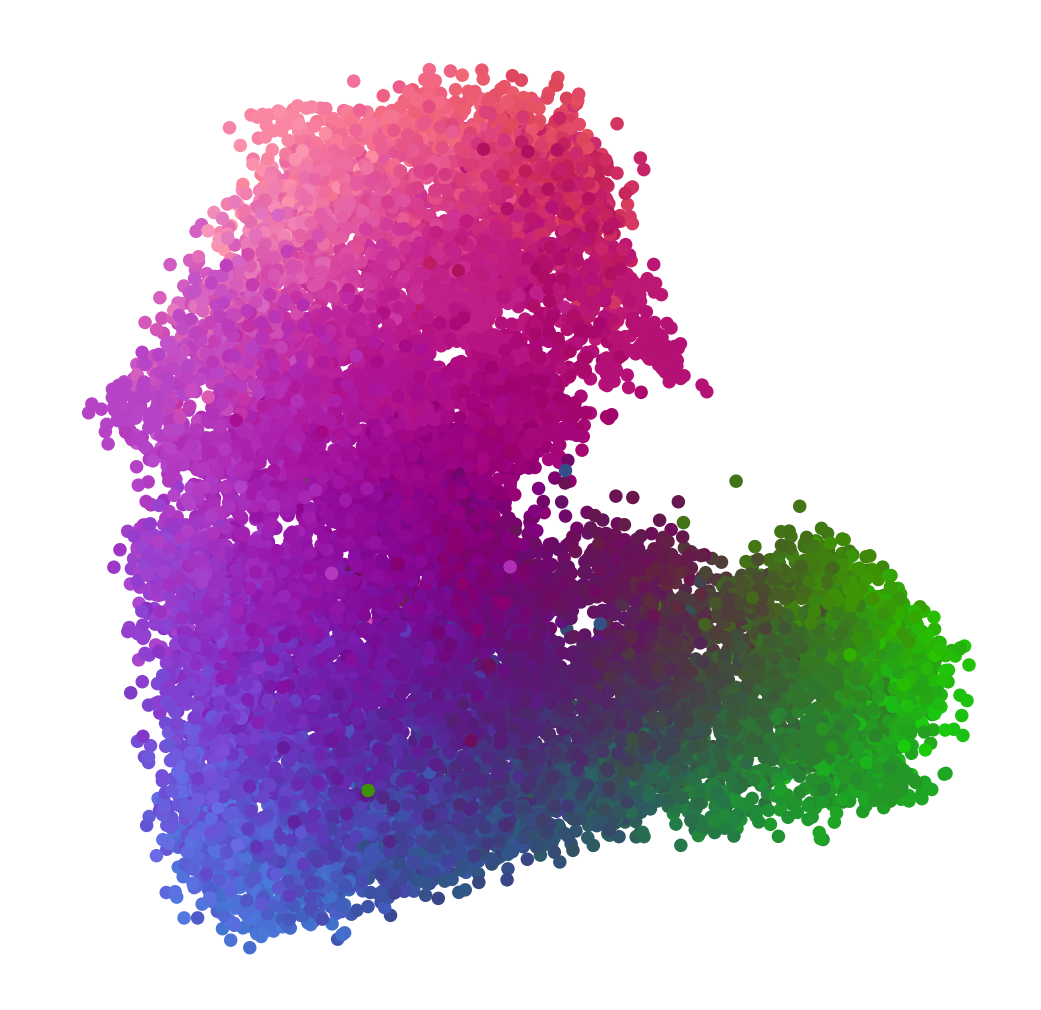};
            \end{axis}
        \end{tikzpicture}
        \vspace{-0.2cm}
        \caption{Triplet, $d_\mathrm{time}$, $\mathcal S_\mathrm{full}$}
        \label{fig:triplet_time}
        \vspace{0.2cm}
    \end{subfigure}
    \begin{subfigure}[b]{0.24\textwidth}
        \begin{tikzpicture}
            \begin{axis}[
                width=0.6\columnwidth,
                height=0.6\columnwidth,
                scale only axis,
                xmin=-247,
                xmax=221.5,
                ymin=-329.5,
                ymax=222.5,
                xlabel = {Latent variable $x$},
                ylabel = {Latent variable $y$},
                tick label style={font=\small},
                label style={font=\small},
                ylabel shift = -8 pt,
                xlabel shift = -4 pt,
            ]
                \addplot[thick,blue] graphics[xmin=-247,ymin=-329.5,xmax=221.5,ymax=222.5] {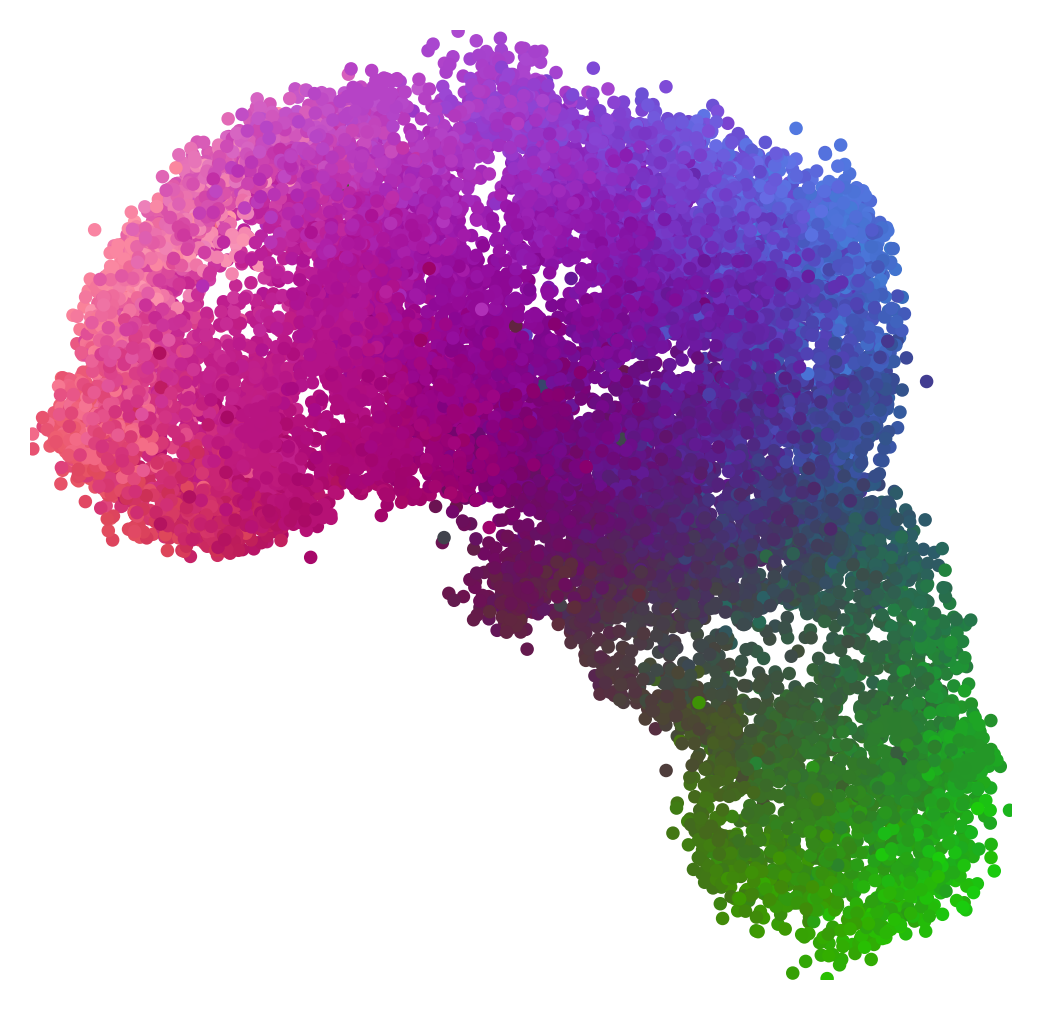};
            \end{axis}
        \end{tikzpicture}
        \vspace{-0.2cm}
        \caption{Siamese, $d_\mathrm{G-ADP}$, $\mathcal S_\mathrm{full}$}
        \label{fig:siamese_adp_geo}
        \vspace{0.2cm}
    \end{subfigure}
    \begin{subfigure}[b]{0.24\textwidth}
        \begin{tikzpicture}
            \begin{axis}[
                width=0.6\columnwidth,
                height=0.6\columnwidth,
                scale only axis,
                xmin=-6,
                xmax=7.5,
                ymin=-7,
                ymax=7.5,
                xlabel = {Latent variable $x$},
                ylabel = {Latent variable $y$},
                tick label style={font=\small},
                label style={font=\small},
                ylabel shift = -8 pt,
                xlabel shift = -4 pt,
            ]
                \addplot[thick,blue] graphics[xmin=-6,ymin=-7,xmax=7.5,ymax=7.5] {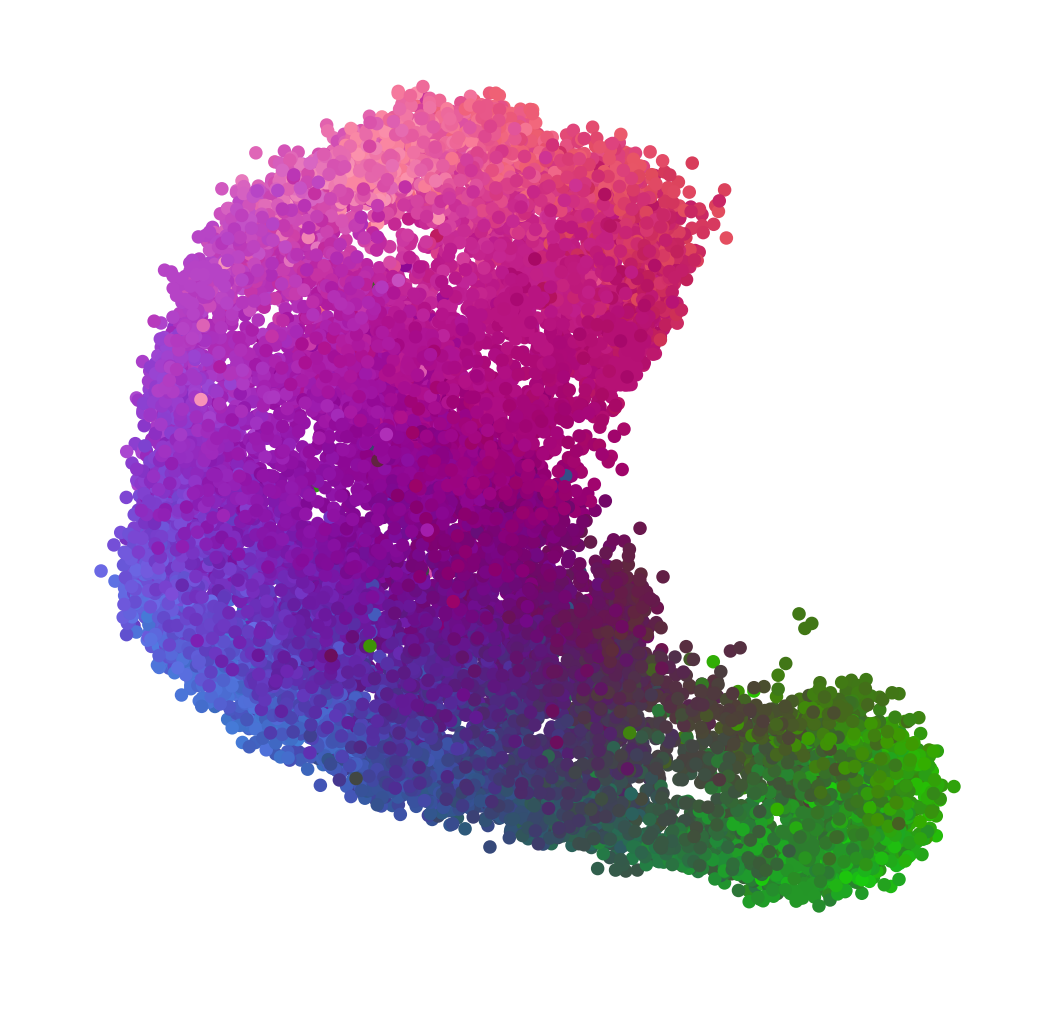};
            \end{axis}
        \end{tikzpicture}
        \vspace{-0.2cm}
        \caption{Triplet, $d_\mathrm{G-ADP}$, $\mathcal S_\mathrm{full}$}
        \label{fig:triplet_adp_geo}
        \vspace{0.2cm}
    \end{subfigure}
    \begin{subfigure}[b]{0.24\textwidth}
        \begin{tikzpicture}
            \begin{axis}[
                width=0.6\columnwidth,
                height=0.6\columnwidth,
                scale only axis,
                xmin=-281,
                xmax=274,
                ymin=-316.5,
                ymax=378,
                xlabel = {Latent variable $x$},
                ylabel = {Latent variable $y$},
                tick label style={font=\small},
                label style={font=\small},
                ylabel shift = -8 pt,
                xlabel shift = -4 pt,
            ]
                \addplot[thick,blue] graphics[xmin=-281,ymin=-316.5,xmax=274,ymax=378] {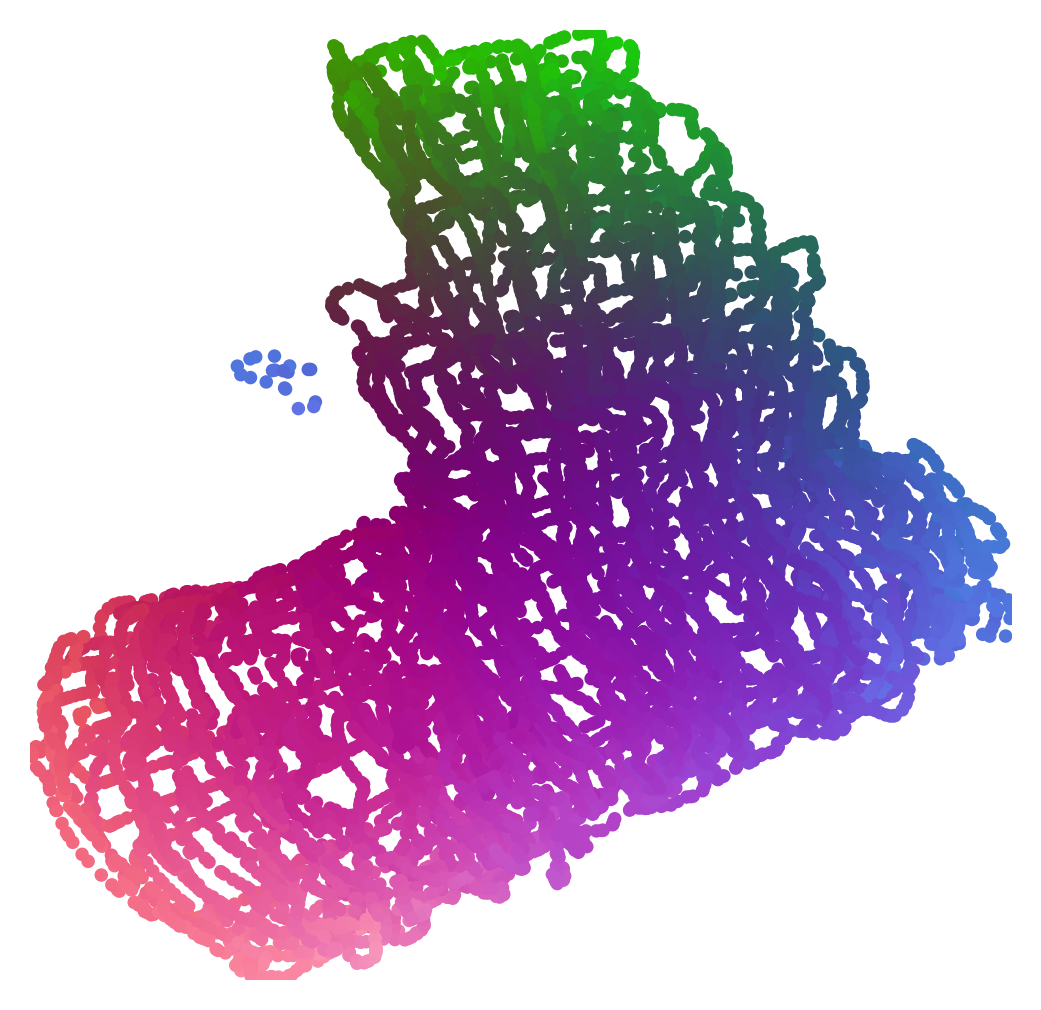};
            \end{axis}
        \end{tikzpicture}
        \vspace{-0.2cm}
        \caption{Isomap, $d_\mathrm{G-fuse}$, $\mathcal S_\mathrm{full}$}
        \label{fig:isomap_fuse_geo}
        \vspace{0.2cm}
    \end{subfigure}
    \begin{subfigure}[b]{0.24\textwidth}
        \begin{tikzpicture}
            \begin{axis}[
                width=0.6\columnwidth,
                height=0.6\columnwidth,
                scale only axis,
                xmin=-376.5,
                xmax=255,
                ymin=-320,
                ymax=246.5,
                xlabel = {Latent variable $x$},
                ylabel = {Latent variable $y$},
                tick label style={font=\small},
                label style={font=\small},
                ylabel shift = -8 pt,
                xlabel shift = -4 pt,
            ]
                \addplot[thick,blue] graphics[xmin=-376.5,ymin=-320,xmax=255,ymax=246.5] {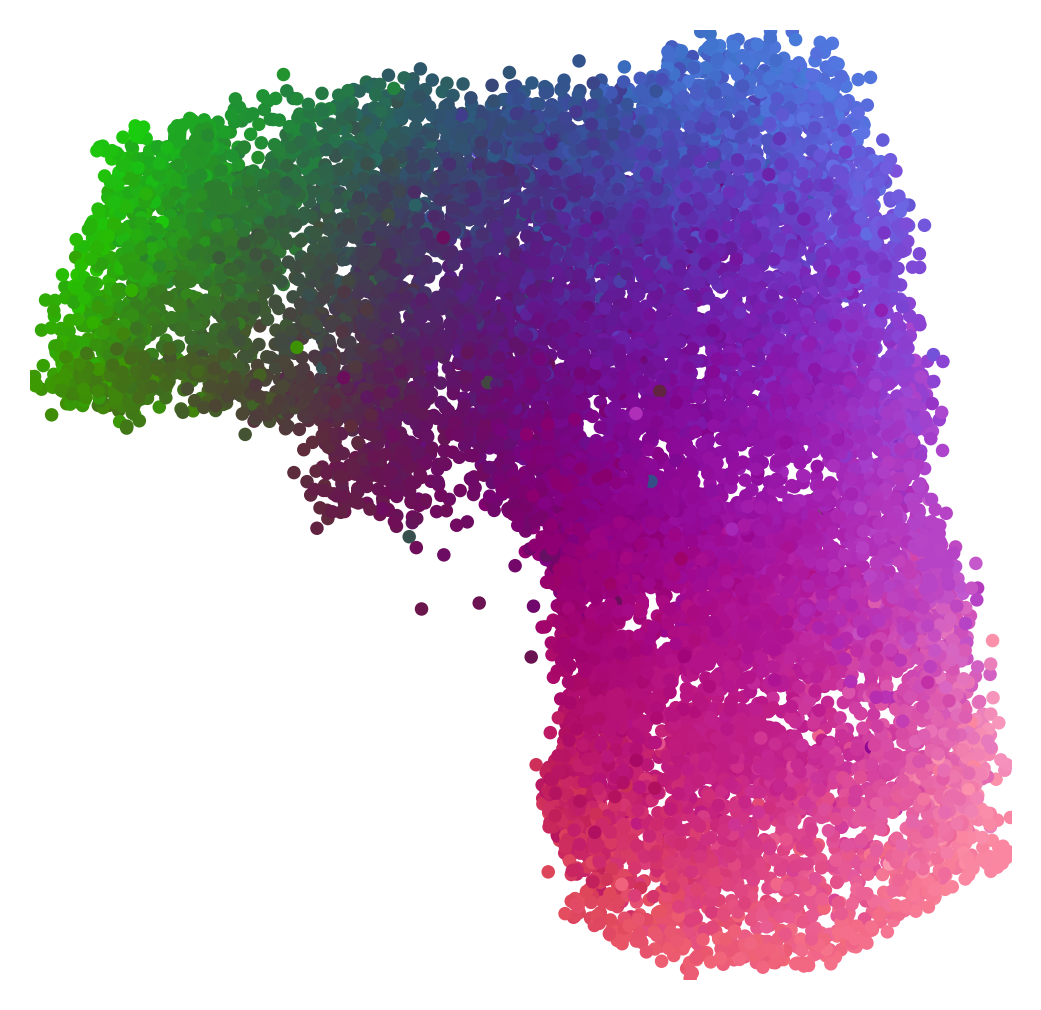};
            \end{axis}
        \end{tikzpicture}
        \vspace{-0.2cm}
        \caption{Siamese, $d_{\mathrm{G-fuse}}$, $\mathcal S_\mathrm{full}$}
        \label{fig:siamese_fuse_geo}
        \vspace{0.2cm}
    \end{subfigure}
    \begin{subfigure}[b]{0.24\textwidth}
        \begin{tikzpicture}
            \begin{axis}[
                width=0.6\columnwidth,
                height=0.6\columnwidth,
                scale only axis,
                xmin=-71.5,
                xmax=57.5,
                ymin=-57,
                ymax=92.5,
                xlabel = {Latent variable $x$},
                ylabel = {Latent variable $y$},
                tick label style={font=\small},
                label style={font=\small},
                ylabel shift = -8 pt,
                xlabel shift = -4 pt,
            ]
                \addplot[thick,blue] graphics[xmin=-71.5,ymin=-57,xmax=57.5,ymax=92.5] {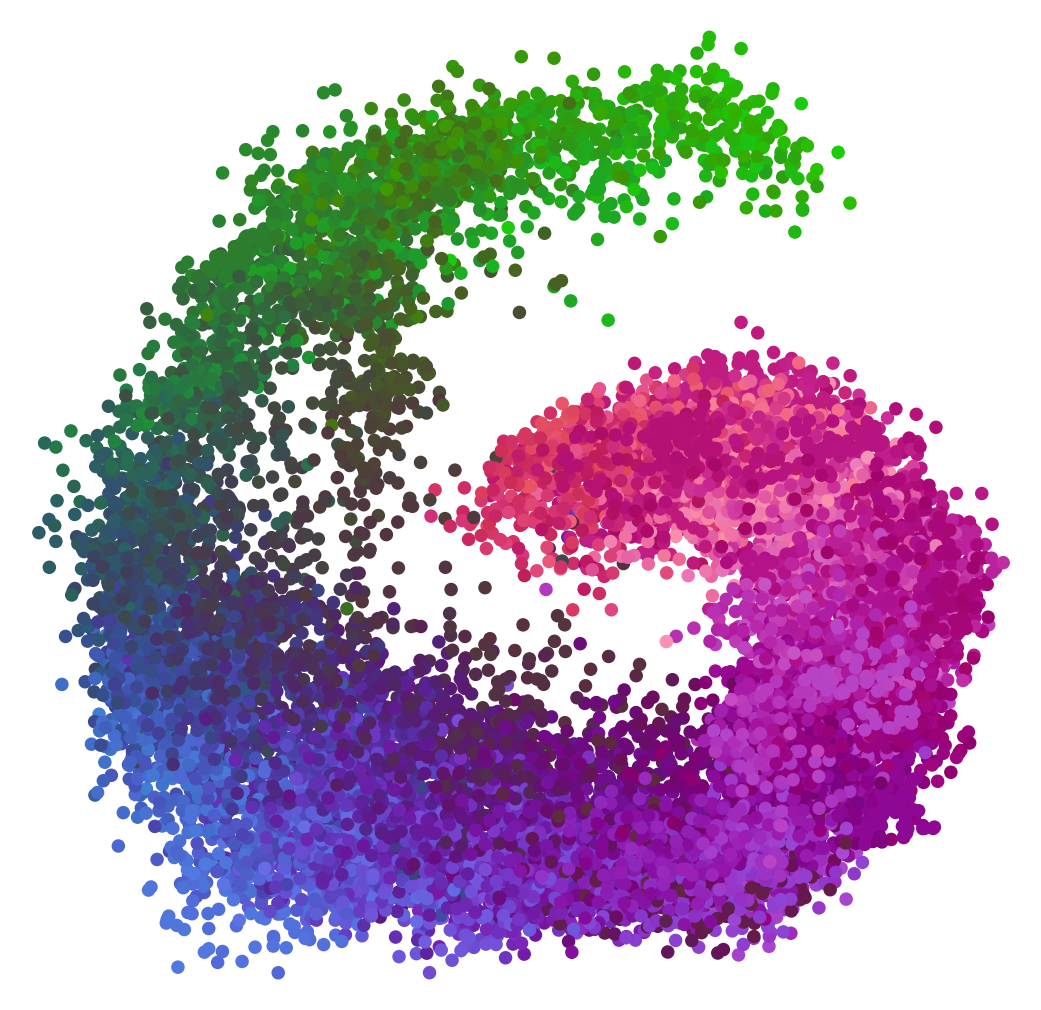};
            \end{axis}
        \end{tikzpicture}
        \vspace{-0.2cm}
        \caption{Siamese, $d_\mathrm{G-ADP}$, $\mathcal S_\mathrm{NLoS}$}
        \label{fig:siamese_adp_geo_nlos}
        \vspace{0.2cm}
    \end{subfigure}
    \begin{subfigure}[b]{0.24\textwidth}
        \begin{tikzpicture}
            \begin{axis}[
                width=0.6\columnwidth,
                height=0.6\columnwidth,
                scale only axis,
                xmin=-23.5,
                xmax=23,
                ymin=-28,
                ymax=14.5,
                xlabel = {Latent variable $x$},
                ylabel = {Latent variable $y$},
                tick label style={font=\small},
                label style={font=\small},
                ylabel shift = -8 pt,
                xlabel shift = -4 pt,
            ]
                \addplot[thick,blue] graphics[xmin=-23.5,ymin=-28,xmax=23,ymax=14.5] {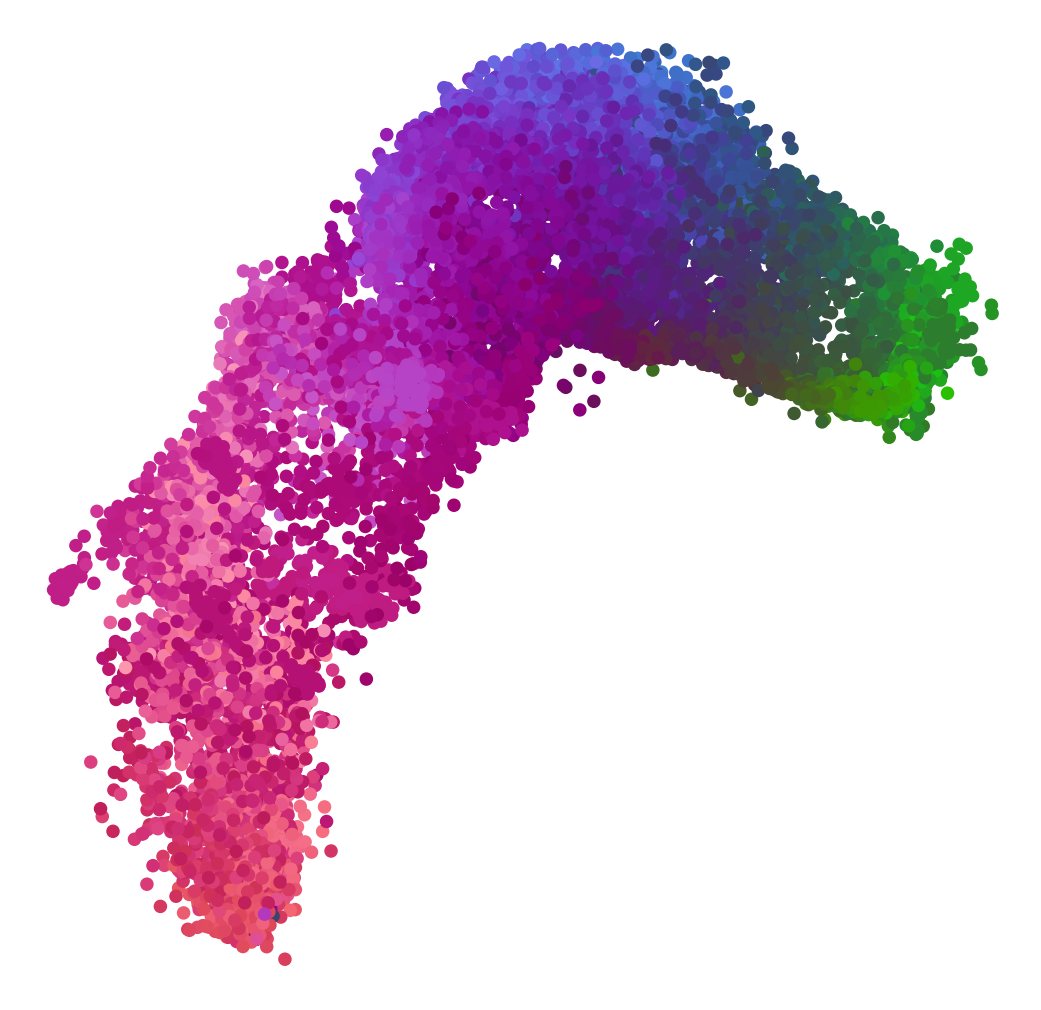};
            \end{axis}
        \end{tikzpicture}
        \vspace{-0.2cm}
        \caption{Siamese, $d_\mathrm{G-DL}$, $\mathcal S_\mathrm{NLoS}$}
        \label{fig:siamese_dl_geo_nlos}
        \vspace{0.2cm}
    \end{subfigure}
    \begin{subfigure}[b]{0.24\textwidth}
        \begin{tikzpicture}
            \begin{axis}[
                width=0.6\columnwidth,
                height=0.6\columnwidth,
                scale only axis,
                xmin=-11,
                xmax=11,
                ymin=-11.5,
                ymax=13,
                xlabel = {Latent variable $x$},
                ylabel = {Latent variable $y$},
                tick label style={font=\small},
                label style={font=\small},
                ylabel shift = -8 pt,
                xlabel shift = -4 pt,
            ]
                \addplot[thick,blue] graphics[xmin=-11,ymin=-11.5,xmax=11,ymax=13] {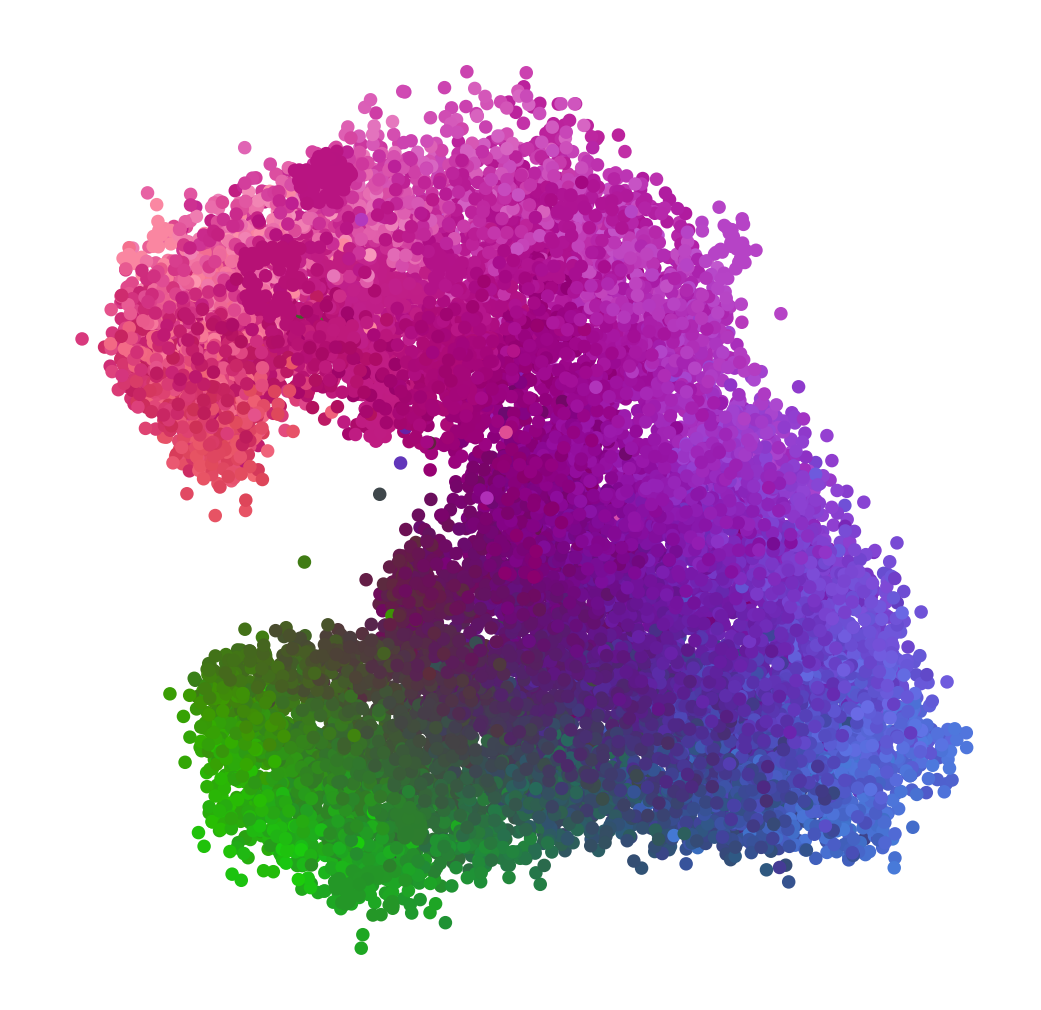};
            \end{axis}
        \end{tikzpicture}
        \vspace{-0.2cm}
        \caption{Triplet, $d_\mathrm{time}$, $\mathcal S_\mathrm{NLoS}$}
        \label{fig:triplet_time_nlos}
        \vspace{0.2cm}
    \end{subfigure}
    \begin{subfigure}[b]{0.24\textwidth}
        \begin{tikzpicture}
            \begin{axis}[
                width=0.6\columnwidth,
                height=0.6\columnwidth,
                scale only axis,
                xmin=-88,
                xmax=96.5,
                ymin=-133.5,
                ymax=120,
                xlabel = {Latent variable $x$},
                ylabel = {Latent variable $y$},
                tick label style={font=\small},
                label style={font=\small},
                ylabel shift = -8 pt,
                xlabel shift = -4 pt,
            ]
                \addplot[thick,blue] graphics[xmin=-88,ymin=-133.5,xmax=96.5,ymax=120] {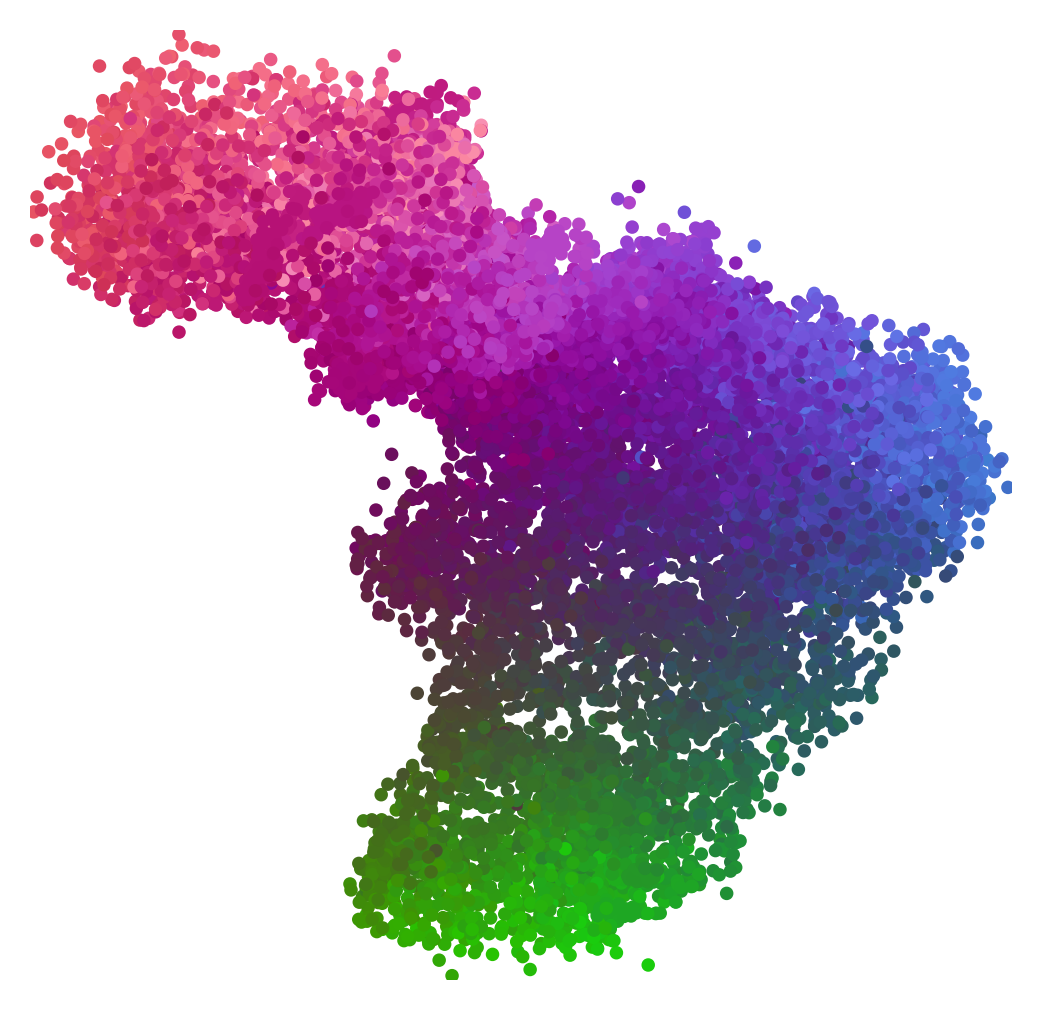};
            \end{axis}
        \end{tikzpicture}
        \vspace{-0.2cm}
        \caption{Siamese, $d_{\mathrm{G-fuse}}$, $\mathcal S_\mathrm{NLoS}$}
        \label{fig:siamese_fuse_geo_nlos}
        \vspace{0.2cm}
    \end{subfigure}
    \vspace{-0.3cm}
    \caption{Comparison of channel charts generated by various combinations of manifold learning technique, dissimilarity metric and dataset (see captions).}
    \label{fig:baseline_channel_charts_colorized}
\end{figure*}

After the optimal affine coordinate transformation has been applied as in Section \ref{sec:mae}, the \ac{CDF} of the absolute localization errors can be visualized.
Various such \acp{CDF}, belonging to channel charts produced with different dissimilarities and manifold learning techniques, are depicted in Fig. \ref{fig:mae_cdf}.
Fig. \ref{fig:baseline_channel_charts_colorized} shows several channel charts obtained using different dissimilarity metrics and manifold learning techniques.
The channel charts can be directly compared to the true map in Fig. \ref{fig:groundtruth-map}.
The first row of Fig. \ref{fig:baseline_channel_charts_colorized} shows the channel charts produced with the baseline metrics $d_\mathrm{G-CIRA}$ and $d_\mathrm{G-CS}$ as well as a chart learned from non-geodesic $d_\mathrm{ADP}$ by Siamese network.
It also contains Fig. \ref{fig:triplet_time}, which is the surprisingly good channel chart obtained through time-based triplet selection, i.e., the combination of a triplet network with $d_\mathrm{time}$.
The second row shows channel charts for $d_\mathrm{G-ADP}$, demonstrating that using the geodesic version of the \ac{ADP} metric greatly improves performance.
What is more, the fused metric $d_\mathrm{G-fuse}$, here with Isomap or a Siamese network, generates even better channel charts.

\begin{table}
	\caption{Performance of Classical Methods on $\mathcal S_\mathrm{full}$.}
	\centering
	\begin{tabular}{ccccccc}
	\toprule
 	Method & Metric & \ac{CT} $\uparrow$ & \ac{TW} $\uparrow$ & \ac{KS} $\downarrow$ & \ac{MAE} $\downarrow$ & Fig. \\
 	\midrule
 	 \rowcolor{TableGray}
 	MDS & $d_\mathrm{Euc}$ & $1.0000$ & $1.0000$ & $0.0001$ & $0.000\, \mathrm m$\\
 	MDS & $d_\mathrm{CS}$ & $0.9106$ & $0.8888$ & $0.3673$ & $1.913\, \mathrm m$\\
 	MDS & $d_\mathrm{CIRA}$ & $0.9281$ & $0.9200$ & $0.3400$ & $1.861\, \mathrm m$\\
 	MDS & $d_\mathrm{DL}$ & $0.9289$ & $0.9011$ & $0.4015$ & $2.349\, \mathrm m$\\
 	MDS & $d_\mathrm{ADP}$ & $0.8902$ & $0.9007$ & $0.3437$ & $2.037\, \mathrm m$\\
 	MDS & $d_\mathrm{fuse}$ & $0.9254$ & $0.9255$ & $0.3262$ & $1.782\, \mathrm m$\\
 	\midrule
 	Isomap & $d_\mathrm{G-CS}$ & $0.9712$ & $0.9721$ & $0.2557$ & $1.533\, \mathrm m$\\
 	Isomap & $d_\mathrm{G-CIRA}$ & $0.9690$ & $0.9656$ & $0.2340$ & $1.185\, \mathrm m$\\
 	Isomap & $d_\mathrm{G-DL}$ & $0.9689$ & $0.9617$ & $0.2237$ & $1.292\, \mathrm m$\\
 	Isomap & $d_\mathrm{G-ADP}$ & $0.9893$ & $0.9878$ & $0.1371$ & $0.705\, \mathrm m$\\
 	Isomap & $d_\mathrm{G-fuse}$ & $\mathbf{0.9956}$ & $\mathbf{0.9960}$ & $\mathbf{0.0859}$ & $\mathbf{0.438\, m}$ & \ref{fig:isomap_fuse_geo}\\
 	\midrule
  	\rowcolor{TableGray}
 	SM & $d_\mathrm{Euc}$ & $0.9833$ & $0.9339$ & $0.1853$ & $1.274\, \mathrm m$\\
 	SM & $d_\mathrm{CS}$ & $0.9311$ & $0.8984$ & $0.3482$ & $1.877\, \mathrm m$\\
 	SM & $d_\mathrm{CIRA}$ & $0.9124$ & $0.9167$ & $0.3242$ & $1.787\, \mathrm m$\\
 	SM & $d_\mathrm{DL}$ & $0.9231$ & $0.8572$ & $0.4249$ & $2.454\, \mathrm m$\\
 	SM & $d_\mathrm{ADP}$ & $0.9493$ & $0.9383$ & $0.3159$ & $1.518\, \mathrm m$\\
 	SM & $d_\mathrm{fuse}$ & $0.9529$ & $0.9413$ & $0.3153$ & $1.505\, \mathrm m$\\
 	\midrule
 	\rowcolor{TableGray}
 	t-SNE & $d_\mathrm{Euc}$ & $0.9980$ & $0.9969$ & $0.1025$ & $0.894\, \mathrm m$\\
 	t-SNE & $d_\mathrm{CS}$ & $0.9811$ & $0.9819$ & $0.1800$ & $1.030\, \mathrm m$\\
 	t-SNE & $d_\mathrm{CIRA}$ & $0.9579$ & $0.9514$ & $0.2392$ & $1.261\, \mathrm m$\\
 	t-SNE & $d_\mathrm{DL}$ & $0.9793$ & $0.9778$ & $0.1637$ & $0.914\, \mathrm m$\\
 	t-SNE & $d_\mathrm{ADP}$ & $0.9851$ & $0.9874$ & $0.1776$ & $0.775\, \mathrm m$\\
 	t-SNE & $d_\mathrm{fuse}$ & $0.9888$ & $0.9902$ & $0.1509$ & $0.697\, \mathrm m$\\
	\bottomrule
	\end{tabular}
	\label{tab:cc_performance_classical}
	\vspace*{-0.2cm}
\end{table}

\begin{table}
	\caption{Performance of Deep Learning-based Methods on $\mathcal S_\mathrm{full}$}
	\centering
	\begin{tabular}{ccccccc}
	\toprule
 	Method & Metric & \ac{CT} $\uparrow$ & \ac{TW} $\uparrow$ & \ac{KS} $\downarrow$ & \ac{MAE} $\downarrow$ & Fig. \\
 	\midrule
 	\rowcolor{TableGray}
 	Siamese & $d_\mathrm{Euc}$ & $0.9907$ & $0.9902$ & $0.0801$ & $0.364\, \mathrm m$\\
 	Siamese & $d_\mathrm{CS}$ & $0.8928$ & $0.8869$ & $0.3394$ & $1.712\, \mathrm m$\\
 	Siamese & $d_\mathrm{CIRA}$ & $0.9195$ & $0.9040$ & $0.3344$ & $1.926\, \mathrm m$\\
 	Siamese & $d_\mathrm{DL}$ & $0.8982$ & $0.8681$ & $0.4020$ & $2.368\, \mathrm m$\\
 	Siamese & $d_\mathrm{ADP}$ & $0.8882$ & $0.8845$ & $0.3496$ & $1.817\, \mathrm m$ & \ref{fig:siamese_adp}\\
 	Siamese & $d_\mathrm{fuse}$ & $0.8907$ & $0.8845$ & $0.3466$ & $1.771\, \mathrm m$\\
 	\midrule
 	Siamese & $d_\mathrm{G-CS}$ & $0.9710$ & $0.9718$ & $0.2517$ & $1.519\, \mathrm m$ & \ref{fig:siamese_cs_geo}\\
 	Siamese & $d_\mathrm{G-CIRA}$ & $0.9524$ & $0.9466$ & $0.2266$ & $1.216\, \mathrm m$ & \ref{fig:siamese_cira_geo}\\
 	Siamese & $d_\mathrm{G-DL}$ & $0.9638$ & $0.9530$ & $0.2269$ & $1.250\, \mathrm m$\\
 	Siamese & $d_\mathrm{G-ADP}$ & $0.9807$ & $0.9798$ & $0.1515$ & $0.783\, \mathrm m$ & \ref{fig:siamese_adp_geo}\\
 	Siamese & $d_\mathrm{G-fuse}$ & $\mathbf{0.9879}$ & $\mathbf{0.9880}$ & $\mathbf{0.1067}$ & $\mathbf{0.542\, m}$ & \ref{fig:siamese_fuse_geo}\\
 	\midrule
 	 \rowcolor{TableGray}
 	Triplet & $d_\mathrm{Euc}$ & $0.9825$ & $0.9803$ & $0.1338$ & $0.671\, \mathrm m$\\
 	Triplet & $d_\mathrm{CS}$ & $0.9470$ & $0.9362$ & $0.3011$ & $1.675\, \mathrm m$\\
 	Triplet & $d_\mathrm{CIRA}$ & $0.9299$ & $0.9265$ & $0.2770$ & $1.429\, \mathrm m$\\
 	Triplet & $d_\mathrm{DL}$ & $0.9487$ & $0.9246$ & $0.3198$ & $2.178\, \mathrm m$\\
 	Triplet & $d_\mathrm{ADP}$ & $0.9656$ & $0.9594$ & $0.2810$ & $1.420\, \mathrm m$\\
 	Triplet & $d_\mathrm{fuse}$ & $0.9657$ & $0.9601$ & $0.2781$ & $1.438\, \mathrm m$\\
 	Triplet & $d_\mathrm{time}$ & $0.9823$ & $0.9822$ & $0.2155$ & $0.621\, \mathrm m$ & \ref{fig:triplet_time}\\
	\midrule
 	Triplet & $d_\mathrm{G-CS}$ & $0.9702$ & $0.9669$ & $0.1733$ & $0.951\, \mathrm m$\\
 	Triplet & $d_\mathrm{G-CIRA}$ & $0.9533$ & $0.9430$ & $0.2215$ & $1.239\, \mathrm m$\\
 	Triplet & $d_\mathrm{G-DL}$ & $0.9640$ & $0.9538$ & $0.2241$ & $1.199\, \mathrm m$\\
 	Triplet & $d_\mathrm{G-ADP}$ & $0.9769$ & $0.9743$ & $0.1506$ & $0.747\, \mathrm m$ & \ref{fig:triplet_adp_geo}\\
 	Triplet & $d_\mathrm{G-fuse}$ & $0.9805$ & $0.9785$ & $0.1429$ & $0.640\, \mathrm m$\\
	\bottomrule
	\end{tabular}
	\label{tab:cc_performance_deeplearning}
	\vspace*{-0.2cm}
\end{table}

\subsection{Results for significant \ac{NLoS} propagation}
To figure out whether our metrics $d_\mathrm{G-DL}$, $d_\mathrm{G-ADP}$ and $d_\mathrm{G-fuse}$ are useful for environments with significant \ac{NLoS} propagation, we define a dataset $\mathcal S_\mathrm{NLoS}$, which we obtain by removing \ac{CSI} data for antenna arrays $b=2$ and $b=4$ from $\mathcal S_\mathrm{full}$.
This way, the metal container is blocking the \ac{LoS} path for a significant share of the area (compare Fig. \ref{fig:labelled-area}).
Tab. \ref{tab:distance_evaluation_ac} shows the performance of the respective dissimilarities for this scenario.
All metrics perform at least slightly worse than in the \ac{LoS} scenario (compare Tab. \ref{tab:distance_evaluation}), but the \ac{CSI}-based dissimilarity metrics suffer most from the \ac{NLoS} conditions.
$d_\mathrm{CS}$ and $d_\mathrm{CIRA}$ are especially hard hit, but the performance metrics for $d_\mathrm{ADP}$ also deteriorate.
Fusing $d_\mathrm{ADP}$ with $d_\mathrm{time}$ and using geodesic dissimilarities significantly improves performance, as indicated by the analyses for $d_\mathrm{G-fuse}$.
This underlines the importance of taking into account timestamp information, especially in \ac{NLoS} conditions.
The results for Isomap, Siamese networks and triplet networks on the \ac{NLoS} dataset are listed in Tab. \ref{tab:cc_performance_ac}.
As expected, the channel charts produced with metrics that exploit timestamp information perform best.

\begin{figure}
    \centering
    \input{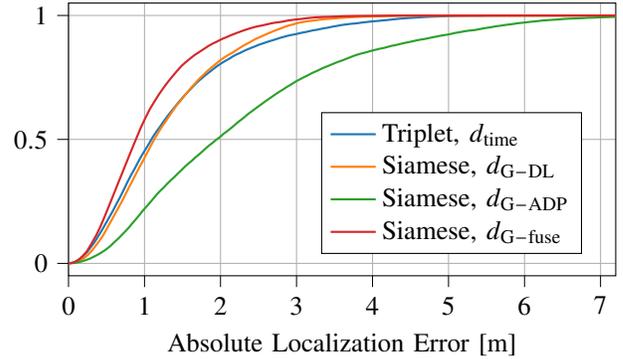}
    \vspace{-0.2cm}
    \caption{CDF of absolute localization error for different dissimilarity metrics and manifold learning techniques computed on $\mathcal S_\mathrm{NLoS}$.}
    \label{fig:mae_cdf_ac}
\end{figure}

This claim is supported by the final row of Fig. \ref{fig:baseline_channel_charts_colorized}:
The global geometry is distorted for the channel chart produced with $d_\mathrm{G-ADP}$, whereas the channel charts using timestamp-aided dissimilarity metrics ($d_\mathrm{G-DL}$, $d_\mathrm{time}$ and $d_\mathrm{G-fuse}$) are globally more meaningful.
The best localization performance is achieved by the Siamese network combined with the fused dissimilarity $d_\mathrm{G-fuse}$, where a mean localization error of $0.672\, \mathrm m$ is reached.
As before, \acp{CDF} of localization errors for different metrics and techniques are shown in Fig. \ref{fig:mae_cdf_ac}.

\begin{table}
	\caption{Evaluation of dissimilarity metrics on $\mathcal S_\mathrm{NLoS}$}
	\centering
	\begin{tabular}{ccccc}
	\toprule
 	Distance & \ac{CT} $\uparrow$ & \ac{TW} $\uparrow$ & \ac{KS} $\downarrow$ & \ac{RD} $\downarrow$\\
 	\midrule
 	$d_\mathrm{CS}$ & $0.8362$ & $0.8221$ & $0.4338$ & $0.9781$ \\
 	$d_\mathrm{G-CS}$ & $0.8285$ & $0.8141$ & $0.4349$ & $0.9752$ \\
 	$d_\mathrm{CIRA}$ & $0.7599$ & $0.8546$ & $0.4174$ & $0.9691$ \\
 	$d_\mathrm{G-CIRA}$ & $0.6025$ & $0.6066$ & $0.4693$ & $0.9869$ \\
 	\midrule
 	$d_\mathrm{DL}$ & $0.9649$ & $0.9497$ & $0.4450$ & $0.9611$ \\
 	$d_\mathrm{G-DL}$ & $0.9679$ & $0.9508$ & $0.3210$ & $0.9295$ \\
 	\midrule
	$d_\mathrm{ADP}$ & $0.9072$ & $0.9027$ & $0.4297$ & $0.9671$ \\
	$d_\mathrm{G-ADP}$ & $0.9516$ & $0.9299$ & $0.3521$ & $0.9490$ \\
	\midrule
	$d_\mathrm{fuse}$ & $0.9081$ & $0.9038$ & $0.4294$ & $0.9668$ \\
	$d_\mathrm{G-fuse}$ & $\mathbf{0.9823}$ & $\mathbf{0.9699}$ & $\mathbf{0.2204}$ & $\mathbf{0.8889}$ \\
	\bottomrule
	\end{tabular}
	\label{tab:distance_evaluation_ac}
	\vspace*{-0.2cm}
\end{table}

\begin{table}
	\caption{Channel Charting Performance on $\mathcal S_\mathrm{NLoS}$.}
	\centering
	\begin{tabular}{ccccccc}
	\toprule
 	Method & Metric & \ac{CT} $\uparrow$ & \ac{TW} $\uparrow$ & \ac{KS} $\downarrow$ & \ac{MAE} $\downarrow$ & Fig.\\
 	\midrule
 	Isomap & $d_\mathrm{G-DL}$ & $0.9623$ & $0.9485$ & $0.3300$ & $1.955\, \mathrm m$\\
 	Isomap & $d_\mathrm{G-ADP}$ & $0.9387$ & $0.8904$ & $0.3759$ & $2.298\, \mathrm m$\\
 	Isomap & $d_\mathrm{G-fuse}$ & $\mathbf{0.9778}$ & $\mathbf{0.9664}$ & $\mathbf{0.2041}$ & $\mathbf{0.935\, m}$\\
 	\midrule
 	\rowcolor{TableGray}
 	Siamese & $d_\mathrm{Euc}$ & $0.9866$ & $0.9855$ & $0.0961$ & $0.419\, \mathrm m$\\
 	Siamese & $d_\mathrm{G-DL}$ & $0.9581$ & $0.9420$ & $0.3233$ & $1.297\, \mathrm m$ & \ref{fig:siamese_dl_geo_nlos}\\
 	Siamese & $d_\mathrm{G-ADP}$ & $0.9352$ & $0.8864$ & $0.3760$ & $2.291\, \mathrm m$ & \ref{fig:siamese_adp_geo_nlos}\\
 	Siamese & $d_\mathrm{G-fuse}$ & $0.9656$ & $0.9512$ & $0.2150$ & $1.042\, \mathrm m$ & \ref{fig:siamese_fuse_geo_nlos}\\
 	\midrule
 	\rowcolor{TableGray}
 	Triplet & $d_\mathrm{Euc}$ & $0.9769$ & $0.9728$ & $0.1486$ & $0.701\, \mathrm m$\\
 	Triplet & $d_\mathrm{G-DL}$ & $0.9591$ & $0.9391$ & $0.2743$ & $1.867\, \mathrm m$\\
 	Triplet & $d_\mathrm{G-ADP}$ & $0.9477$ & $0.9162$ & $0.3894$ & $2.292\, \mathrm m$\\
 	Triplet & $d_\mathrm{G-fuse}$ & $0.9633$ & $0.9423$ & $0.2768$ & $1.323\, \mathrm m$\\
 	Triplet & $d_\mathrm{time}$ & $0.9704$ & $0.9576$ & $0.2795$ & $1.341\, \mathrm m$ & \ref{fig:triplet_time_nlos}\\
	\bottomrule
	\end{tabular}
	\label{tab:cc_performance_ac}
	\vspace*{-0.2cm}
\end{table}

\section{Conclusion}
\label{sec:conclusion}
With our \ac{ADP}-based metric, dissimilarity learning and the concept of fused metrics, we propose three novel metrics for channel charting.
An extensive study on a publicly available dataset demonstrates that we are capable of reconstructing the global geometry of the environment, outperforming previously proposed metrics.
Using our fused metric and a Siamese neural network, we achieve a mean absolute localization error of $0.542\,\mathrm{m}$ on the whole dataset when assuming an ideal coordinate transformation from channel chart to spatial coordinates.
With a mean absolute error of $1.042\,\mathrm{m}$, performance is acceptable even for the \ac{NLoS} dataset with two missing antenna arrays.
Improving methods to find the ideal transformation from channel chart to spatial coordinates, enabling absolute localization, is one potential topic for future research.
This work highlights the significance of the dissimilarity metric, illustrating that the quality of dissimilarities is at least as, if not more impactful than the applied manifold learning technique.
We encourage others to experiment on our publicly available dataset and to propose even better metrics or techniques.

\appendices

\bibliographystyle{IEEEtran}
\bibliography{IEEEabrv,references}

\begin{thebibliography}{10}
\providecommand{\url}[1]{#1}
\csname url@samestyle\endcsname
\providecommand{\newblock}{\relax}
\providecommand{\bibinfo}[2]{#2}
\providecommand{\BIBentrySTDinterwordspacing}{\spaceskip=0pt\relax}
\providecommand{\BIBentryALTinterwordstretchfactor}{4}
\providecommand{\BIBentryALTinterwordspacing}{\spaceskip=\fontdimen2\font plus
\BIBentryALTinterwordstretchfactor\fontdimen3\font minus
  \fontdimen4\font\relax}
\providecommand{\BIBforeignlanguage}[2]{{%
\expandafter\ifx\csname l@#1\endcsname\relax
\typeout{** WARNING: IEEEtran.bst: No hyphenation pattern has been}%
\typeout{** loaded for the language `#1'. Using the pattern for}%
\typeout{** the default language instead.}%
\else
\language=\csname l@#1\endcsname
\fi
#2}}
\providecommand{\BIBdecl}{\relax}
\BIBdecl

\bibitem{savic2015fingerprinting}
V.~Savic and E.~G. Larsson, ``{Fingerprinting-based positioning in distributed
  massive MIMO systems},'' in \emph{2015 IEEE 82nd vehicular technology
  conference (VTC2015-Fall)}.\hskip 1em plus 0.5em minus 0.4em\relax IEEE,
  2015.

\bibitem{vieira2017deep}
J.~Vieira, E.~Leitinger, M.~Sarajlic, X.~Li, and F.~Tufvesson, ``{Deep
  convolutional neural networks for massive MIMO fingerprint-based
  positioning},'' in \emph{2017 IEEE 28th Annual International Symposium on
  Personal, Indoor, and Mobile Radio Communications (PIMRC)}, 2017.

\bibitem{cc_features_ferrand}
P.~Ferrand, A.~Decurninge, and M.~Guillaud, ``{DNN-based Localization from
  Channel Estimates: Feature Design and Experimental Results},'' \emph{CoRR},
  vol. abs/2004.00363, 2020.

\bibitem{studer_cc}
C.~Studer, S.~Medjkouh, E.~G{\"{o}}n{\"{u}}ltas, T.~Goldstein, and
  O.~Tirkkonen, ``{Channel Charting: Locating Users within the Radio
  Environment using Channel State Information},'' \emph{CoRR}, vol.
  abs/1807.05247, 2018.

\bibitem{siamese_cc}
E.~Lei, O.~Castañeda, O.~Tirkkonen, T.~Goldstein, and C.~Studer, ``{Siamese
  Neural Networks for Wireless Positioning and Channel Charting},'' in
  \emph{57th Annual Allerton Conference on Communication, Control, and
  Computing}, 2019.

\bibitem{n2dcc}
P.~Agostini, Z.~Utkovski, S.~Stańczak, A.~A. Memon, B.~Zafar, and M.~Haardt,
  ``Not-too-deep channel charting (n2d-cc),'' in \emph{2022 IEEE Wireless
  Communications and Networking Conference (WCNC)}, 2022.

\bibitem{kazemi_cc_snr_prediction}
P.~Kazemi, H.~Al-Tous, C.~Studer, and O.~Tirkkonen, ``{SNR Prediction in
  Cellular Systems based on Channel Charting},'' in \emph{2020 IEEE Eighth
  International Conference on Communications and Networking (ComNet)}.

\bibitem{kazemi_cc_beam_tracking}
------, ``Channel charting assisted beam tracking,'' in \emph{2022 IEEE 95th
  Vehicular Technology Conference: (VTC2022-Spring)}, 2022.

\bibitem{pca_hotelling}
H.~Hotelling, ``Analysis of a complex of statistical variables into principal
  components,'' \emph{Journal of Educational Psychology}, vol.~24, no.~6, 1933.

\bibitem{sammon_mapping}
J.~Sammon, ``A nonlinear mapping for data structure analysis,'' \emph{IEEE
  Transactions on Computers}, vol. C-18, no.~5, pp. 401--409, 1969.

\bibitem{huang2019improving}
P.~Huang, O.~Casta{\~n}eda, E.~G{\"o}n{\"u}lta{\c{s}}, S.~Medjkouh,
  O.~Tirkkonen, T.~Goldstein, and C.~Studer, ``{Improving channel charting with
  representation-constrained autoencoders},'' in \emph{2019 IEEE 20th
  International Workshop on Signal Processing Advances in Wireless
  Communications (SPAWC)}.\hskip 1em plus 0.5em minus 0.4em\relax IEEE, 2019,
  pp. 1--5.

\bibitem{triplet_cc}
P.~Ferrand, A.~Decurninge, L.~G. Ordo\~nez, and M.~Guillaud, ``{Triplet-Based
  Wireless Channel Charting: Architecture and Experiments},'' \emph{IEEE
  Journal on Selected Areas in Communications}, vol.~39, no.~8, 2021.

\bibitem{multipoint_cc}
J.~Deng, S.~Medjkouh, N.~Malm, O.~Tirkkonen, and C.~Studer, ``{Multipoint
  Channel Charting for Wireless Networks},'' in \emph{2018 52nd Asilomar
  Conference on Signals, Systems, and Computers}, 2018, pp. 286--290.

\bibitem{cc_euclidean_distance_matrix}
P.~Agostini, Z.~Utkovski, and S.~Stańczak, ``{Channel Charting: an Euclidean
  Distance Matrix Completion Perspective},'' in \emph{ICASSP 2020 - 2020 IEEE
  International Conference on Acoustics, Speech and Signal Processing
  (ICASSP)}, 2020, pp. 5010--5014.

\bibitem{magoarou_efficient_cc}
L.~Le~Magoarou, ``Efficient channel charting via phase-insensitive distance
  computation,'' 2021.

\bibitem{fraunhofer_cc}
M.~Stahlke, G.~Yammine, T.~Feigl, B.~M. Eskofier, and C.~Mutschler, ``Indoor
  localization with robust global channel charting: A time-distance-based
  approach,'' \emph{IEEE Transactions on Machine Learning in Communications and
  Networking}, pp. 1--1, 2023.

\bibitem{al_tous_cc_adpp}
H.~AL–Tous, P.~Kazemi, C.~Studer, and O.~Tirkkonen, ``{Channel Charting with
  Angle-Delay-Power-Profile Features and Earth-Mover Distance},'' in \emph{2022
  56th Asilomar Conference on Signals, Systems, and Computers}, 2022, pp.
  1195--1201.

\bibitem{euchner_cc}
F.~Euchner, P.~Stephan, M.~Gauger, S.~Dörner, and S.~Ten~Brink, ``{Improving
  Triplet-Based Channel Charting on Distributed Massive MIMO Measurements},''
  in \emph{2022 IEEE 23rd International Workshop on Signal Processing Advances
  in Wireless Communication (SPAWC)}, 2022.

\bibitem{ponnada2019out}
T.~Ponnada, H.~Al-Tous, O.~Tirkkonen, and C.~Studer, ``{An Out-of-Sample
  Extension for Wireless Multipoint Channel Charting},'' in \emph{Cognitive
  Radio-Oriented Wireless Networks: 14th EAI International Conference, CrownCom
  2019, Poznan, Poland, June 11--12, 2019, Proceedings 14}.\hskip 1em plus
  0.5em minus 0.4em\relax Springer, 2019, pp. 208--217.

\bibitem{geng2020multipoint}
C.~Geng, H.~Huang, and J.~Langerman, ``{Multipoint Channel Charting with
  Multiple-Input Multiple-Output Convolutional Autoencoder},'' in \emph{2020
  IEEE/ION Position, Location and Navigation Symposium (PLANS)}.

\bibitem{pihlajasalo2020absolute}
J.~Pihlajasalo, M.~Koivisto, J.~Talvitie, S.~Ali-L{\"o}ytty, and M.~Valkama,
  ``{Absolute Positioning with Unsupervised Multipoint Channel Charting for 5G
  Networks},'' in \emph{2020 IEEE 92nd Vehicular Technology Conference
  (VTC2020-Fall)}.\hskip 1em plus 0.5em minus 0.4em\relax IEEE, 2020, pp. 1--5.

\bibitem{studer_earthmover}
H.~Al–Tous, P.~Kazemi, C.~Studer, and O.~Tirkkonen, ``{Channel Charting with
  Angle-Delay-Power-Profile Features and Earth-Mover Distance},'' in \emph{2022
  56th Asilomar Conference on Signals, Systems, and Computers}, 2022, pp.
  1195--1201.

\bibitem{isomap}
J.~B. Tenenbaum, V.~d. Silva, and J.~C. Langford, ``{A Global Geometric
  Framework for Nonlinear Dimensionality Reduction},'' \emph{science}, vol.
  290, no. 5500, pp. 2319--2323, 2000.

\bibitem{dijkstra}
E.~W. Dijkstra, ``{A Note on Two Problems in Connexion with Graphs},''
  \emph{Numer. Math.}, vol.~1, no.~1, p. 269–271, dec 1959.

\bibitem{kruskal_stress}
J.~B. Kruskal, ``Multidimensional scaling by optimizing goodness of fit to a
  nonmetric hypothesis,'' \emph{Psychometrika}, vol.~29, no.~1, 1964.

\bibitem{t_sne}
L.~van~der Maaten and G.~Hinton, ``{Visualizing Data using t-SNE},''
  \emph{Journal of Machine Learning Research}, vol.~9, no.~86, 2008.

\bibitem{facenet}
F.~Schroff, D.~Kalenichenko, and J.~Philbin, ``Facenet: {A} unified embedding
  for face recognition and clustering,'' \emph{CoRR}, vol. abs/1503.03832,
  2015.

\bibitem{trustworthiness_continuity}
J.~Venna and S.~Kaski, ``{Neighborhood preservation in nonlinear projection
  methods: An experimental study},'' in \emph{International Conference on
  Artificial Neural Networks}.\hskip 1em plus 0.5em minus 0.4em\relax Springer,
  2001, pp. 485--491.

\bibitem{rajski_distance}
C.~Rajski, ``A metric space of discrete probability distributions,''
  \emph{Information and Control}, vol.~4, no.~4, pp. 371--377, 1961.

\bibitem{dichasus2021}
F.~Euchner, M.~Gauger, S.~D\"orner, and S.~ten Brink, ``{A Distributed Massive
  MIMO Channel Sounder for "Big CSI Data"-driven Machine Learning},'' in
  \emph{WSA 2021; 25th International ITG Workshop on Smart Antennas}, 2021.

\bibitem{dataset-dichasus-cf0x}
\BIBentryALTinterwordspacing
F.~Euchner and M.~Gauger, ``{CSI Dataset dichasus-cf0x: Distributed Antenna
  Setup in Industrial Environment, Day 1},'' 2022. [Online]. Available:
  \url{https://doi.org/doi:10.18419/darus-2854}
\BIBentrySTDinterwordspacing

\end{thebibliography}

\end{document}